\documentclass[jcp,twocolumn,amsmath,amssymb]{revtex4-1}

\usepackage{graphicx}
\usepackage{bm}

\begin{document}

\title{Two-Dimensional Electronic Spectroscopy Using Incoherent Light: \\ Theoretical Analysis}

\author{Daniel B. Turner}
\affiliation{Department of Chemistry and Centre for Quantum Information and Quantum Control, 80 Saint George Street, University of Toronto, Toronto, Ontario M5S 3H6 Canada}

\author{Dylan J. Howey}
\affiliation{Department of Physics, Concordia College, Moorhead, MN, 56562 USA}

\author{Erika J. Sutor}
\affiliation{Department of Chemistry, Concordia College, Moorhead, MN, 56562 USA}

\author{Rebecca A. Hendrickson}
\affiliation{Department of Physics, Concordia College, Moorhead, MN, 56562 USA}

\author{M. W. Gealy}
\affiliation{Department of Physics, Concordia College, Moorhead, MN, 56562 USA}

\author{Darin J. Ulness}
\email{ulnessd@cord.edu}
\affiliation{Department of Chemistry, Concordia College, Moorhead, MN, 56562 USA}

\date{\today}

\begin{abstract}
Electronic energy transfer in photosynthesis occurs over a range of time scales and under a variety of intermolecular coupling conditions. Recent work has shown that electronic coupling between chromophores can lead to coherent oscillations in two-dimensional electronic spectroscopy measurements of pigment-protein complexes measured with femtosecond laser pulses. A persistent issue in the field is to reconcile the results of measurements performed using femtosecond laser pulses with physiological illumination conditions. Noisy-light spectroscopy can begin to address this question. In this work we present the theoretical analysis of incoherent two-dimensional electronic spectroscopy, I$^{(4)}$ 2D ES. Simulations reveal diagonal peaks, cross peaks, and coherent oscillations similar to those observed in femtosecond two-dimensional electronic spectroscopy experiments. The results also expose fundamental differences between the femtosecond-pulse and noisy-light techniques; the differences lead to new challenges and new opportunities. 
\end{abstract} 

\maketitle

%
%
%



\section{Introduction}
In many optoelectronic devices, synthetic macromolecules, and biological processes, electronic excitation energy is transported from one location to another. Electronic energy transfer is thus an important process in physics, chemistry, biology, and engineering. A considerable body of recent work has focused on electronic energy transfer in photosynthesis \cite{Blankenship:2002aa}. The initial steps of photosynthesis involve energy transfer over a range of distances and timescales: energy is transferred among the chromophores inside individual antenna complexes; from one antenna complex to another; and from the antenna complexes to the reaction center, where the excitation energy is used to separate charges and eventually split water. Two regimes for energy transfer in photosynthesis have been well understood since about the 1950s based on the balance between inter-chromophore electronic coupling ($J$) and the coupling of any individual chromophore to its bath ($\lambda$) \cite{May:2004aa}. F\"orster theory can be used when the coupling to the bath dominates, $J << \lambda$, and Redfield theory can be used when the coupling between chromophores dominates, $J >> \lambda$. But in the intermediate-coupling regime where $J \sim \lambda$, energy-transfer processes are neither well understood nor straightforward to treat mathematically. Some theoretical progress has been made in the past two decades \cite{Tanimura:1989aa,Tanimura:2006aa,Ishizaki:2009aa,Fassioli:2009aa,Chen:2010aa,Ishizaki:2011aa,Andrews:2011aa,Huo:2011aa,Pachon:2011aa}. 

Energy-transfer theories are important, and they need to be verified by experiments. Because the events are fast---usually nanosecond timescales or faster---it is natural for measurements to use femtosecond laser pulses. Most typical are pump-probe spectroscopy measurements \cite{Fragnito:1989aa,Vos:1993aa,vanGrondelle:1994aa,Jonas:1996aa,Savikhin:1997aa,Chachisvilis:1997aa,Sundstrom:1999aa,Prokhorenko:2000aa,Doust:2004aa,Novoderezhkin:2005aa,Doust:2006aa,Mirkovic:2007aa,Romero:2010aa,Marin:2011aa,Huang:2012aa}, but technical advances \cite{Hybl:2001aa,Brixner:2004aa,Gundogdu:2007aa,Bristow:2009aa,Nemeth:2009aa,Prokhorenko:2009aa,Tekavec:2009aa,Brinks:2010aa,Turner:2011aa,Ramunas:2011aa} have made two-dimensional electronic spectroscopy (2D ES) measurements of photosynthetic proteins \cite{Engel:2007aa,Womick:2009aa,Collini:2010aa,Panitchayangkoon:2010aa,Turner:2011ab,Turner:2012ab,Dostal:2012aa,Wong:2012aa,Harel:2012aa,Fidler:2012aa} frequent as well. 

One of the most interesting observations from 2D ES measurements of photosynthetic proteins are distinct cross peaks with oscillating amplitudes \cite{Engel:2007aa,Womick:2009aa,Collini:2010aa,Panitchayangkoon:2010aa}. Cross peaks can be signatures of coupling (in several forms) or energy transfer \cite{Jonas:2003aa,Lewis:2012aa}. Cross peaks with oscillating amplitudes are most often signatures of intramolecular vibrational modes (vibrational coherence) or strong electronic coupling (electronic coherence) \cite{Cheng:2008aa,Egorova:2008aa,Christensson:2011aa,Turner:2011ab,Turner:2012ab,Mancal:2012aa}. Similar coherent oscillations have been observed in photosynthetic proteins using related techniques as well \cite{Mercer:2009aa,Hildner:2010aa,Richards:2012aa,Richards:2012ab}. The measured coherences have evoked questions about energy transfer in the intermediate-coupling regime in biological systems and even larger questions about the nature of photosynthetic complexes \cite{Scholes:2010ab,Sarovar:2010aa,Scholes:2011aa,Miller:2012aa}. 

Femtosecond spectroscopy is a powerful tool that will continue to be the primary source of new insight into the mechanisms governing energy transfer in photosynthesis. Yet a nagging question persists regarding the disconnect between the coherent excitation via femtosecond pulses used in experiments and the almost fully incoherent excitation via sunlight that photosynthetic organisms experience in natural conditions: Can incoherent excitation produce the same coherences observed in femtosecond spectroscopy measurements? 

A few theoretical studies relate to this question  \cite{Jiang:1991aa,Jiang:1996aa,Mancal:2010ab,Brumer:2012aa,Fassioli:2012aa}, but no experiment has been performed or even suggested. Fortunately there is an established, joint theoretical and experimental framework that can address the question. This is the lesser-known technique called \emph{noisy-light spectroscopy} which was developed concurrently with (one-dimensional) ultrafast laser spectroscopy. In 1984, Morita and Yajima \cite{Morita:1984aa}, Asaka et al, \cite{Asaka:1984aa}, and Beach and Hartmann \cite{Beach:1984aa} independently demonstrated that noisy light could be used to achieve femtosecond-scale time resolution in degenerate four-wave mixing experiments. During the subsequent decade a variety of noisy-light analogues to more traditional ultrafast experiments were developed \cite{Kurokawa:1987aa,Kobayashi:1988aa,Nakatsuka:1988aa,Apanasevich:1988aa,Misawa:1989aa,Apanasevich:1989aa,Nakatsuka:1989aa,Hattori:1991aa,Vodchits:1991aa,Apanasevich:1992aa,Apanasevich:1992ab,Apanasevich:1993aa,Kobayashi:1994aa,Vodchits:1996aa,Kummrow:1996aa,Lau:1998aa,Kirkwood:1999aa,Kirkwood:1999ab,Kozich:1999aa,Ulness:1999aa,Ulness:1999ab,Zhang:2000aa,Kirkwood:2000aa,Zhang:2000ab,Zhang:2000ac,Kirkwood:2000aa,Kozich:2000aa,Kozich:2001aa,Rao:2001aa,Zhang:2001aa,Dawlaty:2001aa,Alghamdi:2001aa,Menezes:2001aa,Zhang:2002aa}. Although noisy light offers some advantages, it is not as generally useful as its short-pulse counterpart. One exception is coherent anti-Stokes Raman scattering spectroscopy (I$^{(2)}$CARS), where the noisy-light version has proven to be a very useful tool that is still being used to investigate a variety of systems \cite{Lee:1985aa,Dugan:1988aa,Dugan:1991aa,Dugan:1991ab,Lau:1994aa,Schaertel:1994aa,Schaertel:1995aa,Stimson:1996aa,Ulness:1997aa,Ulness:1997ab,Kozich:1999aa,Ulness:2003aa,Aung:2005aa,Berg:2006aa,Weisel:2007aa,Berg:2008aa}. Noisy sources have found occasional uses and advantages in other spectral regimes as well \cite{Ernst:1970aa,Knight:1982aa,Ernst:1987aa,Xu:2008aa,Meyer:2012aa}. 

The main feature and utility of noisy-light spectroscopy is that the time resolution is given by the coherence time of the light, not the temporal envelope. In principle, the noisy beam may be continuous wave (cw), although in practice it is often pulsed on the ordered of nanoseconds (still essentially cw relative to femtosecond and picosecond material dynamics). In a typical experiment such as I$^{(2)}$CARS, the noisy-light beam enters a Michelson interferometer to generate identical twin beams. One of the beams is delayed in time relative to the other by a controllable spatial delay in one arm of the interferometer. This can be generalized to multiple identical noisy beams with several controllable delays for more complicated techniques. 

Both short pulses and noisy light are spectrally broad, and in principle they could have identical optical spectral densities. But the differences between the excitation sources give rise to fundamentally different physical processes. For femtosecond pulses the frequency components must be \emph{phase locked} (not to be confused with phase matched). That is, all frequency components must have a specific phase relationship so that the interference generates a stable femtosecond laser pulse \cite{Cundiff:2002aa,Rulliere:2005aa}. Short pulses are ideally suited for direct time measurements; however, there are at least two disadvantages. First, femtosecond-pulse experimental setups are expensive and the experiments have strict stability requirements. In contrast, noisy-light experiments have more relaxed stability requirements and require fewer expensive pieces of equipment. There are several reasons for this, but the most important is that noisy light is unaffected by dispersion \cite{Ulness:2003aa}, unlike short pulses where extensive efforts are often required \cite{Fork:1987aa}. Second, finely resolved spectral information must be determined through analysis of the time information. The phase-locking requirement prevents direct probing of the sample spectrally, although gross spectral probing is possible since experimentally accessible femtosecond pulses do not have infinitely broad spectra. On the other hand, the spectrum of noisy light is completely \emph{phase unlocked}. That is, the phase of any one frequency component is completely independent of any other frequency component. In a sense each frequency component is behaving as if it came from an independent cw source. In other words, the noisy light source is built from an incoherent superposition of monochromatic cw light; this random superposition produces a spatially coherent laser beam whose electric field is a stochastic function of time. 

It is worthwhile to consider how noisy light would interact with a photosynthetic pigment-protein complex. Because of its phase-incoherent nature, noisy light is an important step closer to sunlight than are femtosecond pulses. In this work we use the well-established technique of factorized time correlation (FTC) diagram analysis \cite{Ulness:1996aa,Ulness:1997aa,Ulness:2003aa,Aung:2005aa}---along with direct calculation of the spectroscopic signals---to describe a noisy-light version of two-dimensional electronic spectroscopy (I$^{(4)}$ 2D ES, where I$^{(4)}$ indicates four incoherent beams). The theory is presented and discussed for noisy light interacting with a Bloch four-level system. This system is complex enough to capture the essential physics of the interaction yet sufficiently simple to yield results that can be understood at a level giving significant physical insight.

\section{Theory}
\subsection{Input Fields}
\begin{figure}
\centering
  \includegraphics[width=0.35\textwidth]{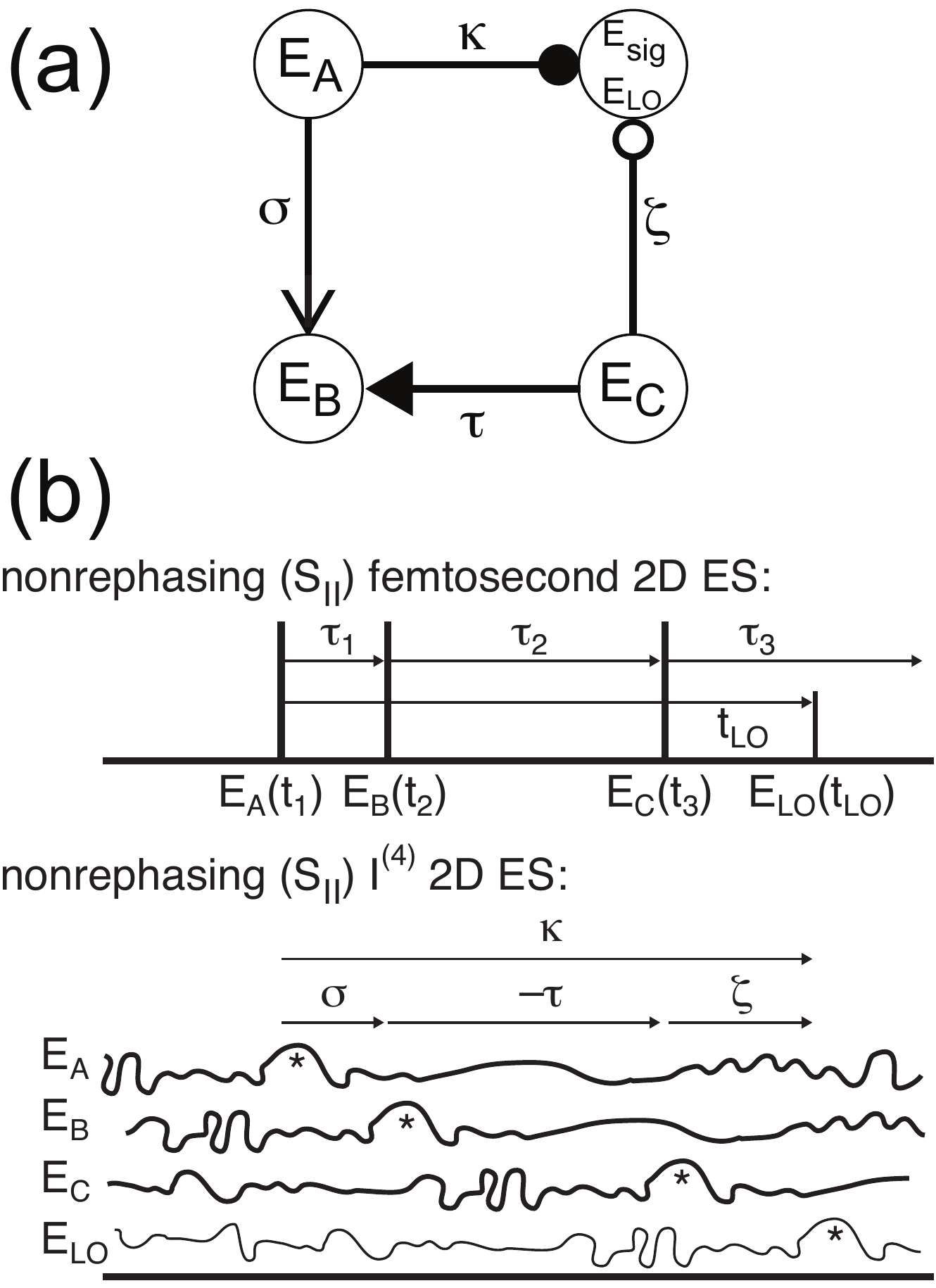} 
  \caption{Parameters that would be used for I$^{(4)}$ 2D ES. (a) The BOX beam geometry. Noisy-light excitation fields $E_A$, $E_B$, and $E_C$ generate a third-order nonlinear response in the sample, which emits a phase-matched beam in the direction of $E_{\mathrm{sig}}$. The local oscillator field, $E_{LO}$, is also in the direction of $E_{\mathrm{sig}}$. The four input noisy-light beams, $E_A$, $E_B$, $E_C$, and $E_{LO}$, are identical except for the relative experimentally controlled time-delay variables $\sigma$, $\tau$, $\zeta$, and $\kappa$.  (b) The relative pulse time orderings for femtosecond and I$^{(4)}$ 2D ES measurements, using the nonrephasing time ordering as an example. In femtosecond 2D ES, the experimental time-delay variables $\tau_1$, $\tau_2$, and $\tau_3$ directly relate to the interaction times of the femtosecond pulses, $t_1$, $t_2$, and $t_3$.  In I$^{(4)}$ 2D ES, the experimental time-delay variables have an ambiguous relation to the interaction times. The * indicates a common relative time point in all four noisy beams for ease of viewing.} 
  \label{fig:geometry}
\end{figure}

The two-dimensional noisy-light experiment analyzed here involves three identical excitation beams, $E_A$, $E_B$, and $E_C$, arranged in the standard BOX geometry as indicated in Fig. \ref{fig:geometry}(a). The fields produce a third-order nonlinear response from the system under study.  Beams $E_A$ and $E_C$ act in-phase while beam $E_B$ acts out-of-phase; in the nomenclature of femtosecond 2D ES, beams $E_A$ and $E_C$ are `nonconjugates' and beam $E_B$ is the `conjugate'. The resultant third-order signal is brought to quadrature at the detector with a fourth beam, the local oscillator, $E_{LO}$. That is, the signal is heterodyne detected. As we detail below, the action of the LO is somewhat different than in femtosecond measurements. Working with the complex analytic signal, the temporal characteristics of the electric fields of the four input beams can be written as 
\begin{subequations}
\label{eqn:fields}
\begin{align}
E_{A}(t)& = E_{0}p(t)e^{-i\overline{\omega }t}\,  \\
E_{B}(t)& = E_{0}p(t-\sigma )e^{-i\overline{\omega }(t-\sigma )}\,  \\
E_{C}(t)& = E_{0}p(t+\tau -\sigma )e^{-i\overline{\omega }(t+\tau -\sigma )}\,  \\
E_{LO}(t)& = E_{0}p(t-\kappa )e^{-i\overline{\omega }(t-\kappa )}\, ,
\end{align}
\end{subequations}
where $E_{0}$ is a constant representing the field strength, $p(t)$ is a complex stochastic function representing the random envelope of the noisy field and $\overline{\omega }$ is the central carrier wave frequency of the noisy light. The relative experimentally controlled temporal delays $\sigma$, $\tau$, $\zeta$, and $\kappa$ are shown in Fig. \ref{fig:geometry}. Time-delay variable $\kappa$ is redundant, $\kappa = \zeta + \sigma - \tau$, but it is introduced as a matter of convenience that will become clear below. We have also changed the sign of time-delay variable $\tau$ to be opposite of what might be expected from femtosecond measurements to match previous noisy-light work and as another convenience that will become clear.  We will also describe and show that although the light is on all the time, in the mathematical analysis we label specific interaction times and then integrate over them to account for all possible interaction times. The relation between interaction times and the experimentally controlled time-delay variables is simple for femtosecond 2D ES but complicated for I$^{(4)}$ 2D ES. We depict this in Fig. \ref{fig:geometry}(b), where experimental time-delay variables for the femtosecond measurement are given as $\tau_1$, $\tau_2$, and $\tau_3$ (in the literature there is an equivalent, alternative convention of $\tau$, $T$, and $t$, respectively).

\subsection{System}
\begin{figure}
\centering
  \includegraphics[width=0.2\textwidth]{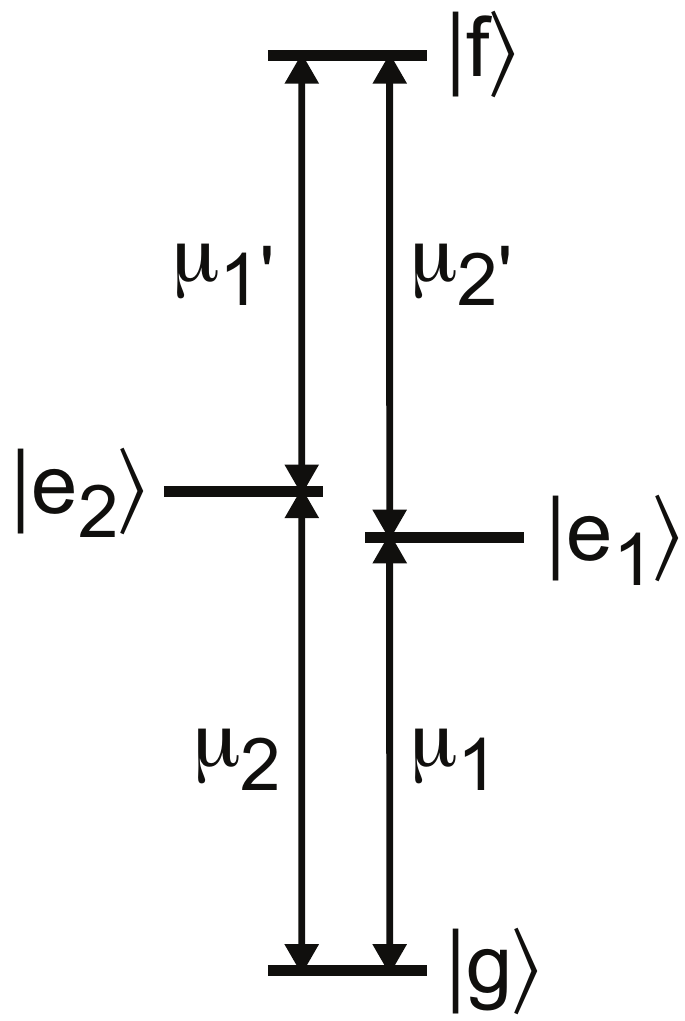} 
  \caption{The energy-level scheme represents a pair of coupled two-level systems. We include all possible third-order signals produced by the transitions ($\mu_1$, $\mu_{1'}$, $\mu_2$, and $\mu_{2'}$) among the four states: a common ground state, $\vert g \rangle$, two single-exciton states, $\vert e_{1} \rangle$ and $\vert e_{2} \rangle$, and a double-excitation state, $\vert f \rangle$.} 
  \label{fig:elevels}
\end{figure}

In this work we calculate signals expected in a noisy-light spectroscopy experiment of the Bloch four-level system representing a pair of coupled two-level systems as indicated in Fig. \ref{fig:elevels}. Coupling between the two-level systems leads to an excitonic basis having energy shifts of the excited states and renormalization of the transition-dipole moments. In the exciton basis, there is a common ground state, $\vert g \rangle$, two single-exciton states, $\vert e_{1} \rangle$ and $\vert e_{2} \rangle$, and a double-excitation state, $\vert f \rangle$, which can be reached via excited-state absorption from either of the single-exciton states.  For completeness, we mention that this energy-level scheme could also represent a pair of uncoupled two-level systems. However, here we are interested in the coupled case, which means $\mu_1 \ne \mu_{1'}$ and $\mu_2 \ne \mu_{2'}$.

\subsection{Material Response}
The material response is handled via a Bloch model \cite{Mukamel:1995aa} such that
\begin{eqnarray}
R^{(3)} & \propto  & (-1)^n \rho_0 \left(\frac{i}{\hbar}\right)^3 \mu^{(4)} \nonumber \\
& & \times e^{-i\Omega _{I}(t_{2}-t_{1})}e^{-i\Omega_{II}(t_{3}-t_{2})}e^{-i\Omega _{III}(t-t_{3})}, 
\end{eqnarray}
where $t_{1}$, $t_{2}$, and $t_{3}$ are the times at which the fields interact with the material. The $e^{-i\Omega_x (t_{i}-t_{j})}$ factors are the Liouville propagators during the intervals between field-matter interaction events, and $\Omega_x =\omega_x -i\gamma_x$ where $\omega_x$ is the Bohr frequency of the coherence during the time interval between field interaction events, and $\gamma_x$ is the phenomenological decay rate constant for that coherence. For population terms, $\Omega_{xx} = -i \Gamma_{xx}$, where $\Gamma_{xx}$ is the excited-state lifetime of state $x$. A factor of $(-1)^n$, where $n$ is the number of interactions on the ket side of the diagram, is also present to account for the sign difference of the excited-state--absorption pathways. The factor of $\mu^{(4)}$ represents the product of four transition dipole moments specific to the Liouville pathway.

\begin{figure*}
\centering
  \includegraphics[width=0.85\textwidth]{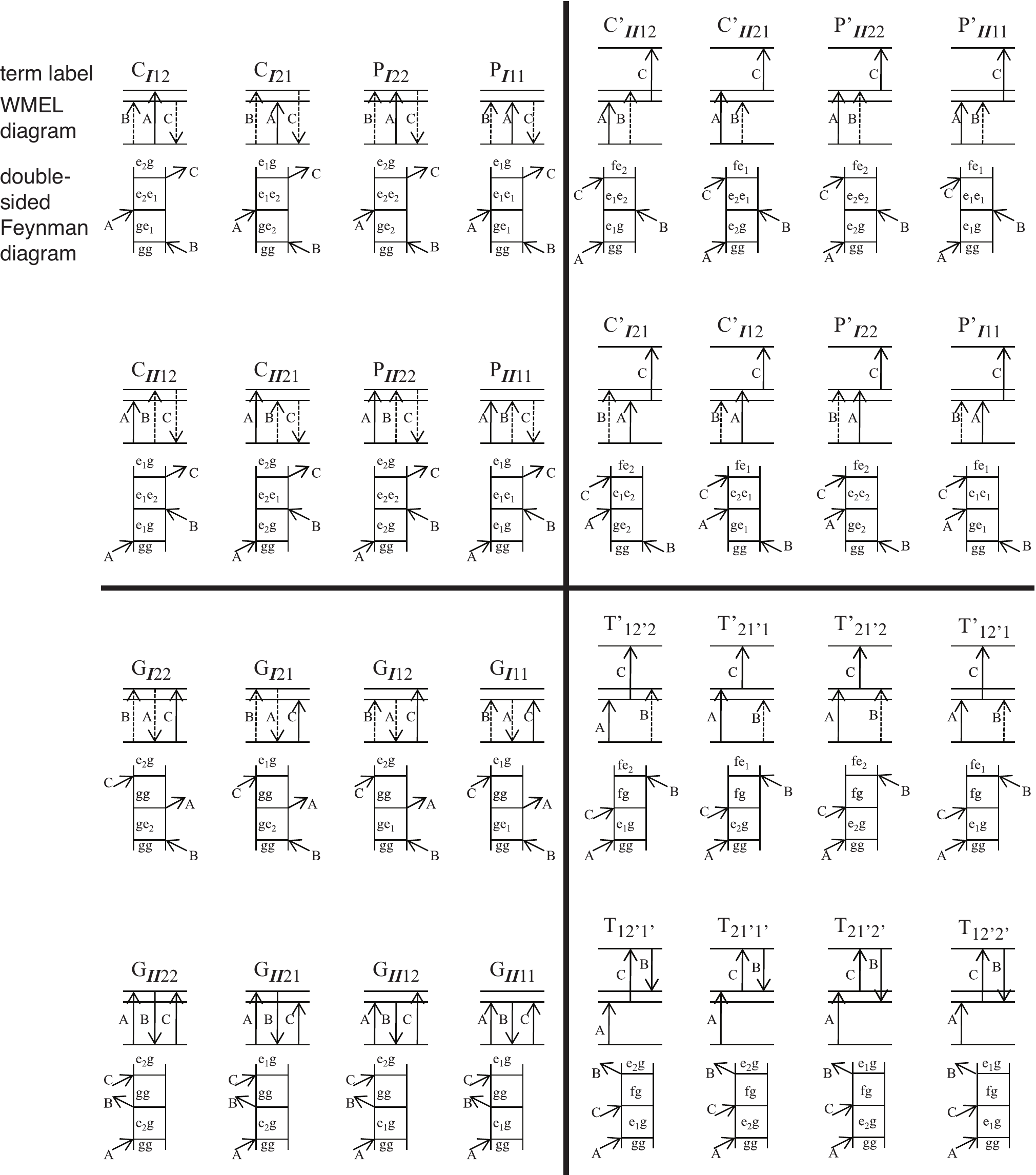} 
  \caption{The sets of WMEL and double-sided Feynman diagrams for third-order noisy-light spectroscopy of the four-level system. The upper left, lower left, upper right, and lower right quadrants contain the stimulated-emission, ground-state bleach, excited-state absorption, and the two-quantum pathways, respectively. Shown are the pathways in which field $E_A$ acts before field $E_C$, the $\alpha$ pathways. Not shown are the pathways in which field $E_C$ acts before field $E_A$, the $\beta$ pathways. The labelling scheme is described in the text.}
  \label{fig:feynmans}
\end{figure*}

For the four-level system depicted in Fig. \ref{fig:elevels}, there are 32 distinct Liouville paths that are triply resonant, 16 of which involve the higher energy state, $f$, and 16 which do not. Since noisy light is effectively cw, all orderings of fields $E_A$, $E_B$, and $E_C$ must be considered. Because fields $E_A$ and $E_C$ are in-phase with one another, the 32 possible Liouville paths double to 64 when field ordering is taken into account. Each of the 64 Liouville pathways can be represented by a wave-mixing energy level (WMEL) diagram \cite{Lee:1985aa} or by a double-sided Feynman diagram \cite{Mukamel:1995aa}. Both sets of diagrams are shown in Fig. \ref{fig:feynmans}. Rephasing ($S_{I}$), nonrephasing ($S_{II}$), and two-quantum ($S_{III}$) pathways all potentially contribute to the signal at all delay times in the I$^{(4)}$ 2D ES experiment, whereas in a femtosecond 2D ES experiment these sets of pathways are distinguishable by time ordering the fields. Because differentiation by time ordering is not particularly helpful, we have instead chosen to organize and label the groups of pathways according to the type of dynamics occurring during the second time period: $G$ for ground-state bleach, $P$ for population (stimulated emission), $C$ for coherent oscillation, and $T$ for two-quantum coherence. Excited-state absorption pathways carry an additional $'$ symbol. 

We then further differentiate the pathways using subscript notation. The numeral subscripts $I$ and $II$ indicate $S_{I}$ and $S_{II}$ pathways, respectively. Because all of the $T$ pathways are by definition $S_{III}$ pathways, they do not have a numeral subscript $III$ as that would be redundant.  The numbered subscripts such as $1$ and $2$ indicate the first two or three transition dipoles following the convention set forth in Fig. \ref{fig:elevels}. 

\subsection{Detected Signal}
The third-order polarization is calculated in the same manner as for femtosecond nonlinear spectroscopy \cite{Mukamel:1995aa} by integrating over the fields and the material response
\begin{equation}
E_{sig}^{(3)} \propto P^{(3)}\propto \int_{-\infty }^{t}dt_{3}\int_{-\infty}^{t_{3}}dt_{2}\int_{-\infty }^{t_{2}}dt_{1}E_{A}E_{B}^{\ast }E_{C}R^{(3)}.
\end{equation}
In general all orderings of fields $E_A$, $E_B$, and $E_C$ must be considered, although phase matching does restrict the possibilities somewhat. To maintain generality at this point the interaction time variables $t_1$, $t_2$, and $t_3$ are not associated with particular beams. To calculate a specific term in the polarization, the time ordering of the fields must be specified. The third-order polarization is responsible for generating the signal field. The field is then brought to the intensity level through quadrature with the local oscillator field and averaged, 
\begin{eqnarray}
I = \langle I(t) \rangle & = & \langle \vert E_{LO} + E_{sig}^{(3)} \vert ^2 \rangle \nonumber \\
 & = & \langle I_{LO} + I_{sig}^{(3)} + E_{sig}^{(3)} E^{\ast }_{LO} + E^{(3) \ast}_{sig} E_{LO}\rangle \nonumber \\
 & = & \langle I_{LO} \rangle + \langle I_{sig}^{(3)} \rangle + \langle E_{sig}^{(3)} E^{\ast }_{LO} \rangle + \langle c.c. \rangle ,
\end{eqnarray}
where $I_{LO} = E_{LO}E^{\ast}_{LO}$ and $I^{(3)}_{sig} = E_{sig}^{(3) \ast} E_{sig}^{(3)}$, and we have suppressed the dependence of these expressions on the experimentally controllable time-delay variables.  We consider that it is possible to remove term $\langle I^{(3)}_{sig} \rangle$ by appropriate balancing of beam intensities and to remove term $\langle I_{LO} \rangle$ through subtraction of a secondary measurement. Thus, suppressing the conjugate term, 
\begin{eqnarray}
I & \propto & \langle E_{sig}^{(3)} E_{LO}^{\ast }\rangle \propto \langle P^{(3)} E_{LO}^{\ast }\rangle \nonumber \\
I & \propto &\langle \int_{-\infty }^{t}dt_{3}\int_{-\infty}^{t_{3}}dt_{2}\int_{-\infty }^{t_{2}}dt_{1}E_{A}E_{B}^{\ast}E_{C}E_{LO}^{\ast }R^{(3)}\rangle \nonumber \\
& \propto &\int_{-\infty }^{t}dt_{3}\int_{-\infty}^{t_{3}}dt_{2}\int_{-\infty }^{t_{2}}dt_{1}\langle E_{A}E_{B}^{\ast}E_{C}E_{LO}^{\ast }\rangle R^{(3)}.  \label{eqn:fullInt}
\end{eqnarray}

Other than the stochastic envelope terms and the associated temporal averaging, the treatment thus far is similar to femtosecond methods; both are treatments of nonlinear optical signals in the semiclassical approximation. 

Appropriate treatment of the averaging of the noisy-light process results in a four-point time correlator. Because four-point time correlators are very difficult to handle mathematically, in noisy-light spectroscopy one assumes circular complex Gaussian statistics for the stochastic functions, $p(t)$ \cite{Goodman:1985aa}. This assumption---which has been used successfully in noisy-light spectroscopy for almost three decades---allows the four-point time correlator to be expressed in terms of two-point time correlators,
\begin{eqnarray}
\langle E_{A}E_{B}^{\ast }E_{C}E_{LO}^{\ast }\rangle & = & \langle E_{A}E_{B}^{\ast }\rangle \langle E_{C}E_{LO}^{\ast} \rangle \nonumber \\
& & + \langle E_{A}E_{LO}^{\ast }\rangle \langle E_{C}E_{B}^{\ast }\rangle.  
\label{eqn:fourpoint}
\end{eqnarray}
This approximation makes the mathematics more tractable and still captures the essential features of noisy-light spectroscopy. Zhang and coworkers have considered noisy-light fields with a variety of statistical characteristics and have considered the four-point correlator directly for special cases \cite{Zhang:2005aa,Zhang:2005ab,Gan:2009aa}. Normal treatment of higher-order correlators via the cumulant expansion leads to exceedingly complicated analysis, which, to the best of our understanding, does not impact the experimental outcome. Example calculations beginning with Eqns. \ref{eqn:fullInt} and \ref{eqn:fourpoint} are given in the Appendix. 

In short, the WMEL diagrams and the double-sided Feynman diagrams capture the light-matter interactions between the noisy fields and the Bloch four-level system that give rise to the third-order polarization that generates the emitted signal field. This signal field is brought to quadrature at the detector with a local-oscillator field, which is also a noisy beam. Time averaging results in a four-point time correlator which reduces to two terms each having a product of a pair of two-point time correlators. As we will see below, each pair of two-point time correlators can be depicted and understood using another diagrammatic technique.

\section{Results: Factorized Time Correlation Diagram Analysis}
The development of noisy-light spectroscopy has been hindered by the added mathematical complexity imposed by the quasi-cw and stochastic features of the light. The quasi-cw nature of the noisy light forces one to consider all possible time orderings of the field-matter interactions in the perturbative treatment of the nonlinear signal \cite{Ivanecky:1993aa}. The stochastic nature of the light requires explicit use of the bichromophoric model \cite{Dugan:1991aa} to handle the nontrivial averaging at the intensity level. The total field is the phase-matched sum of the signal fields launched from each of the individual chromophores in the sample. In the homodyne detection scheme, the mod-square of this sum is dominated by the cross terms, meaning those fields launched from two distinct chromophores. The total intensity is then very well represented by the sum of all pairwise, meaning two-chromophore, contributions. This bichromophoric model leads to the familiar $N^2$ dependence of the signal intensity in the coherent nonlinear spectroscopies, where $N$ is the number density of chromophores in the sample. For short-pulse spectroscopies, this point is not of practical relevance. The third-order polarization alone is sufficient for describing the nonlinear signal. Going to the intensity level is trivial---it is just the mod-square of the calculated polarization. For noisy-light spectroscopies, however, stochastic averaging of the noise at the intensity level requires that special attention be given to the signal intensity. The correlations among the various noisy-light field interactions on the two (otherwise independent) chromophores must be explicitly treated. These correlations are fundamental to the understanding of noisy-light spectroscopies. Explicit use of the bichromophoric model requires assigning distinct timelines ($t$ and $s$) to each of the two generic chromophores. The two timelines allow for the (in general) different histories of evolution for each chromophore. Subsequent stochastic averaging links the $t$ and $s$ time variables in a nontrivial way. The heterodyne detection scheme considered for this work replaces the second chromophore field in the bichromophoric model with the local-oscillator field. In principle one would not need explicit use of the bichromophoric model but doing so is desirable because it mirrors the homodyne scheme and, as will be shown below, is beneficial for subsequent analysis.

\begin{figure*}
\centering
  \includegraphics[width=0.63\textwidth]{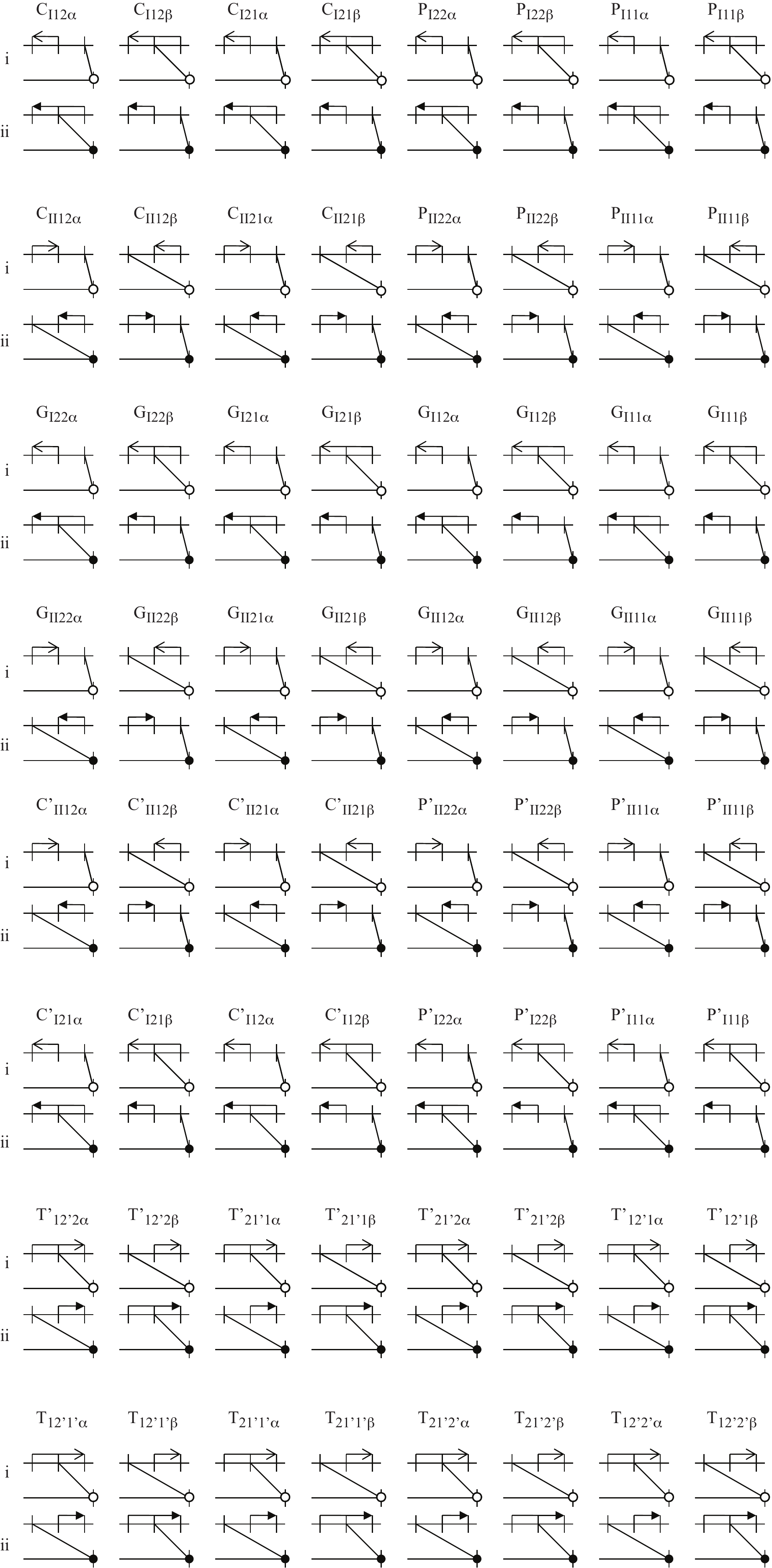}
  \caption{The set of 128 FTC diagrams. Although each FTC diagram represents a distinct mathematical term, many share the same or similar topological structure. The symbols $<$, $\blacktriangleleft$, $\circ$, and $\bullet$ at the ends of arrows represent time-delay variables $\sigma$, $\tau$, $\zeta$, and $\kappa$, respectively, and the direction of the arrow indicates the sign of the time variable.} 
  \label{fig:FTCs}
\end{figure*}

As it turns out, one can apply a diagrammatic representation of the mathematical expressions describing any given noisy-light signal \cite{Ulness:1996aa}. These diagrams are called factorized time correlation (FTC) diagrams, and the complete set of these diagrams represents the breakdown of the noisy-light signal into its elementary physical components. The set of FTC diagrams for the I$^{(4)}$ 2D ES measurement considered here are presented in Fig. \ref{fig:FTCs}. In general, the set of FTC diagrams is isomorphic with the set of intensity-level terms for a given noisy-light spectroscopy. Because the topological operations performed on the FTC diagrams represent mathematical operations performed on the analytic intensity-level terms, conceptual tools have been developed to connect the topological operations with the mathematical operations. The conceptual tools also allow one to glean considerable physical insight from the FTC diagrams \cite{Ulness:1996aa,Ulness:1997aa,Ulness:2003aa,Aung:2005aa}. 

\subsection{Conceptual Tools}
Three of the conceptual tools---accumulation, synchronization, and color locking---will be particularly important for FTC diagram analysis of the I$^{(4)}$ 2D ES signal. Although extensive discussions have been previously presented \cite{Ulness:1996aa,Ulness:2003aa}, these three concepts will be briefly reviewed here because of their utility in the remainder of this work. 

\subsubsection{Accumulation}
Across the ensemble of chromophores, the noisy fields may act on a given ensemble member at any time during the pulse duration. For those sequences of field actions that produce the noisy-light signal considered in this work, the ability of the fields to act at any time on a given chromophore must be summed over the ensemble of chromophores contributing to the production of the nonlinear signal. From a topological point of view this is represented on an FTC diagram as the freedom of a tick mark to `slide along a timeline'. Since the noisy beams are `always' present, any time intervention (represented by a tick mark on the FTC diagram), or correlated pair of interventions, is free to slide along the timeline provided the specific time ordering associated with a given FTC diagram is maintained. That is, during the sliding along the timeline, tick marks can not cross one another. To `slide along the timeline' indicates the potential for the field action to take place at any time over which the tick mark is permitted to slide. However, an individual tick mark is locked to a partner tick mark by the segment representing a pair correlator. The two tick marks, thus linked, correspond to a correlated event pair that must slide along the timeline together. This ability to `slide along the timeline' is called  \emph{accumulation}. Physically, accumulation represents the summation over the ensemble of chromophores in which the correlated noisy field interventions have the ability to act at any time relative to the other correlated pairs of actions that produce the signal. Mathematically, it represents integration over the time intervals between the light-matter interactions.

\subsubsection{Synchronization}
The second conceptual tool is concerned with the event coupling between the two timelines ($t$ and $s$). This coupling involves inter-timeline pair correlators. This implies tight \emph{synchronization} between the two chromophores of the events linked by a given correlator line. The `precision' or `strength' of this synchronization is inversely proportional to the coherence time of the noisy light. For simplicity the correlation functions are taken to be $\delta$-functions, thus synchronization is perfectly precise---the coherence time of the noisy light is taken to be zero. For this work synchronization occurs between the $t$ timeline where the polarization on the chromophore is being developed and the $s$ timeline representing the local oscillator field.

\subsubsection{Color Locking}
Noisy light has a broad spectrum similar to that of femtosecond pulses, with the important distinction that it is completely \emph{phase unlocked}. One says the noisy light is \emph{color locked}, because each color (or, equivalently, each frequency) is coherent only with itself---it is uncorrelated with any other color. Color locking is a consequence of the Wiener-Khintchine theorem \cite{McQuarrie:2000aa} which is expressed mathematically most conveniently by examining a pair correlator in frequency space: 
\begin{equation}
\langle \tilde{p}(\omega)\tilde{p}^{\ast }(\omega^{\prime })\rangle = \tilde{\Gamma}(\omega)\delta (\omega-\omega^{\prime })
\label{eqn:colorlock}
\end{equation}
where $\tilde{\Gamma}(\omega)$ is the optical spectral density of the noisy light. In the context of noisy-light spectroscopy, this implies that regardless of the spectral density of the sources, only identical frequencies may correlate to one another in a pairwise fashion. The consequence of color locking on FTC diagram analysis is extremely important: \emph{whichever frequency component happens to act from one field of a correlated pair (represented by one end of a line or arrow), the other field of the correlated pair (represented by the other end of the line or arrow) must act with the same frequency component}. This allows one to maintain the use of correlator terminology for a single color and refer to a $x$-$x$ color-locked pair correlator, where $x$ is a single frequency component of the noisy light. 

\subsection{Heterodyne Detection}
To date, FTC diagram analysis has been applied to several noisy-light spectroscopies \cite{Stimson:1997aa,Ulness:1998aa,Turner:2003aa,Aung:2005aa,Booth:2006aa} and also explored in a more general context \cite{Biebighauser:2002aa,Biebighauser:2003aa} to elucidate the underlying mathematical structure of the diagrams. In all previous work the noisy-light signal was homodyne detected. This is the first FTC diagram analysis of a heterodyne-detected signal. The transition to the heterodyne case is accommodated in a straightforward manner. For the heterodyne case, the $t$ timeline is drawn in the normal fashion as a horizontal line with three tick marks representing $t_{1}$, $t_{2}$, and $t_{3}$. (This is a condensation of the WMEL or double-sided Feynman diagram into a single line with tick marks.) The left hand side of the timeline represents $-\infty$ and the right hand side terminates with the quadrature event happening upon detection at time, $t$. The $s$ timeline represents that of the local oscillator. In the heterodyne scheme the distinction between the two timelines is trivial but nonetheless formally important. There is a single tick mark placed at the right terminal side of the horizontal line representing the local oscillator action at the detector during the quadrature event which brings the signal to the intensity level. This means $s=t$. 

In femtosecond 2D ES, the local-oscillator delay is not related to the signal-generation process other than as a \emph{temporal reference} that is removed via spectral interferometry in post-processing of the data \cite{Lepetit:1995aa}. Importantly, for simulations of femtosecond 2D ES, the local oscillator can be ignored altogether except mentioned that it allows one to recover the phase and amplitude of the emitted signal. In I$^{(4)}$ 2D ES, the action of the local oscillator is slightly different. The local oscillator serves as a \emph{temporal anchor} (setting $s=t$) because of synchronization (described above and in refs \cite{Ulness:2003aa} and \cite{Ulness:1996aa}). It is interesting, but somewhat of an aside, that for homodyning in noisy-light experiments, the detection event is not part of the $n$-point field correlator. Hence the concept of a temporal anchor is not present in previous noisy-light works. Instead one builds up the conjugate signal (three field actions) separately on the $s$ timeline. The $n$-point corrector (for the homodyne version of the measurement here it would a six-point correlator) and resultant pair correlators connect field events on and between the $t$ and $s$ timelines. But these pairs are free to accumulate (slide along the timeline) and thus, although synchronized, do not form a temporal anchor on the laboratory time frame.  Heterodyne detection thus has important implications both for the interpretation of the results and for experimental implementation as discussed below. 

\subsection{Analysis of the Topological Classes}
\begin{figure}
\centering
  \includegraphics[width=0.2\textwidth]{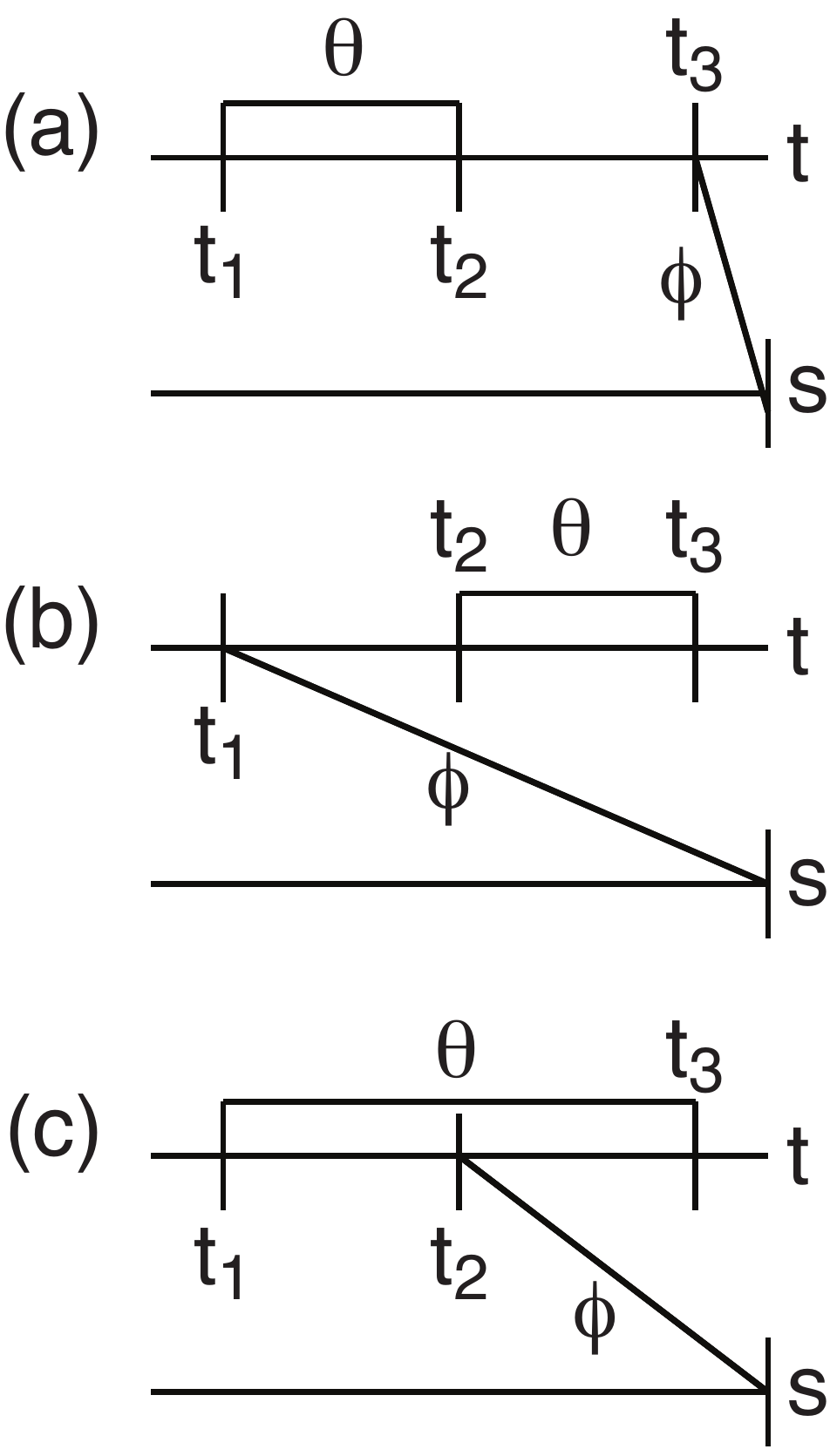} 
  \caption{FTC diagrams capture the nontrivial relationship between the experimentally controlled time-delay variables ($\theta$ and $\phi$) and the interaction-time variables ($t_i$). Interaction times $t_i$ on the $t$ timeline are indicated, as well as the time delays $\theta$ and $\phi$ as described in the text. The three topological classes of FTC diagrams for this experiment are (a) unrestricted, (b) singly restricted, and (c) doubly restricted. } 
  \label{fig:FTC_classes}
\end{figure}

Inspection of the set of 128 FTC diagrams in Fig. \ref{fig:FTCs} suggests a natural grouping based on topology. Three topological classes exist and are shown in Fig. \ref{fig:FTC_classes}. The line segments are labelled with the experimentally controllable delay variables: $\theta = \pm \sigma$ or $\pm \tau$, $\phi = \zeta$ or $\kappa$. It is understood that for leftward pointing arrows $\theta = - \sigma$ and $- \tau$ in all subsequent formulae. By way of example, a factor of the form $e^{-i \Omega \theta}$ would become $e^{+ i \Omega \tau}$ for leftward pointing solid arrow and $e^{-i \Omega \sigma}$ for a rightward pointing open arrow. Generally, rephasing (nonrephasing) terms contain $-\theta$ ($+\theta$);  all terms contain positive $\phi$.   It is interesting to note in Fig. \ref{fig:FTCs} that $\sigma$ and $\zeta$ only appear together or $\tau$ and $\kappa$ only appear together. There are 48 \emph{unrestricted} pathways, Fig. \ref{fig:FTC_classes}(a), 40 \emph{singly restricted} pathways, Fig. \ref{fig:FTC_classes}(b), and 40 \emph{doubly restricted} pathways, Fig. \ref{fig:FTC_classes}(c). 

One is able to perform FTC diagram analysis on each topological class to produce a single formula for generating expressions for all diagrams in that particular topological class. The set of 128 expressions can then be produced with relative ease. In essence we can perform FTC analysis on three diagrams and obtain the analytic result for all 128 terms. For readers less familiar with FTC diagram analysis, we reiterate that we have provided fully worked example calculations in the Appendix. FTC diagram analysis is preferred because, in addition to revealing the correct mathematical expression for each signal component, the analysis provides substantial physical insight and can help obviate mathematical errors. 

\subsubsection{Unrestricted Topology}
\label{sec:FTC_analysis_unrestricted}
\begin{table*}
\centering
\begin{tabular}{c | c | | c | c}
term & expression & term & expression  \\ 
\hline
$G_{I11 \alpha i}$  	&$\Theta[-\sigma] \Theta[\zeta] \frac{e^{+i\Omega_{g e_{1}}\sigma} e^{-i\Omega_{e_{1}g}\zeta}}{{\Omega_{gg}}}$ & 
$G_{II11 \alpha i}$  	&$\Theta[\sigma] \Theta[\zeta] \frac{e^{-i\Omega_{e_{1}g}(\sigma + \zeta)}}{\Omega_{gg}}$\\
$G_{I11 \beta ii}$   	&$\Theta[-\tau] \Theta[\kappa] \frac{e^{+i\Omega_{g e_{1}}\tau} e^{-i\Omega_{e_{1}g}\kappa}}{\Omega _{gg}}$ & 
$G_{II11 \beta ii}$   	&$\Theta[\tau] \Theta[\kappa]  \frac{e^{-i\Omega _{e_{1}g}(\tau +\kappa )}}{\Omega_{gg}}$\\
$G_{I12 \alpha i}$  	&$\Theta[-\sigma] \Theta[\zeta] \frac{e^{+i\Omega_{g e_{1}}\sigma} e^{-i\Omega_{e_{2}g}\zeta}}{\Omega _{gg}}$ & 
$G_{II12 \alpha i}$  	&$\Theta[\sigma] \Theta[\zeta] \frac{e^{-i\Omega _{e_{1}g}\sigma }e^{-i\Omega_{e_{2}g}\zeta}}{\Omega_{gg}}$\\
$G_{I12 \beta ii}$   	&$\Theta[-\tau] \Theta[\kappa] \frac{e^{+i\Omega_{g e_{1}}\tau }e^{-i\Omega _{e_{2}g}\kappa}}{\Omega _{gg}}$ & 
$G_{II12 \beta ii}$   	&$\Theta[\tau] \Theta[\kappa] \frac{e^{-i\Omega _{e_{1}g}\tau }e^{-i\Omega_{e_{2}g}\kappa}}{\Omega_{gg}}$\\
$G_{I21 \alpha i}$  	&$\Theta[-\sigma] \Theta[\zeta] \frac{e^{+i\Omega_{g e_{2}}\sigma}e^{-i\Omega_{e_{1}g}\zeta}}{\Omega_{gg}}$ & 
$G_{II21 \alpha i}$  	&$\Theta[\sigma] \Theta[\zeta] \frac{e^{-i\Omega _{e_{2}g}\sigma }e^{-i\Omega_{e_{1}g}\zeta}}{\Omega_{gg}}$\\
$G_{I21 \beta ii}$   	&$\Theta[-\tau] \Theta[\kappa] \frac{e^{+i\Omega _{g e_{2}}\tau }e^{-i\Omega _{e_{1}g}\kappa}}{\Omega _{gg}}$ & 
$G_{II21 \beta ii}$   	&$\Theta[\tau] \Theta[\kappa] \frac{e^{-i\Omega _{e_{2}g}\tau }e^{-i\Omega_{e_{1}g}\kappa}}{\Omega_{gg}}$\\
$G_{I22 \alpha i}$  	&$\Theta[-\sigma] \Theta[\zeta] \frac{e^{+i\Omega_{g e_{2}}\sigma} e^{-i\Omega_{e_{2}g}\zeta}}{\Omega _{gg}}$ & 
$G_{II22 \alpha i}$  	&$\Theta[\sigma] \Theta[\zeta] \frac{e^{-i\Omega _{e_{2}g}(\sigma + \zeta)}}{{\Omega_{gg}}}$\\
$G_{I22 \beta ii}$   	&$\Theta[-\tau] \Theta[\kappa] \frac{e^{+i\Omega_{g e_{2}}\tau} e^{-i\Omega_{e_{2}g}\kappa}}{\Omega _{gg}}$ & 
$G_{II22 \beta ii}$   	&$\Theta[\tau] \Theta[\kappa] \frac{e^{-i\Omega _{e_{2}g}(\tau + \kappa)}}{{\Omega_{gg}}}$\\

$P_{I11 \alpha i}$    	&$\Theta[-\sigma] \Theta[\zeta] \frac{e^{+i\Omega_{g e_{1}}\sigma} e^{-i\Omega_{e_{1}g}\zeta}}{\Omega _{e_{1}e_{1}}}$  &
$P'_{II 22 \alpha i}$ 	&$\Theta[\sigma] \Theta[\zeta] \frac{e^{-i\Omega _{e_{2}g} \sigma }e^{-i\Omega_{f e_{2}}\zeta}}{\Omega _{e_{2} e_{2}}}$\\
$P_{I11 \beta ii}$     	&$\Theta[-\tau] \Theta[\kappa] \frac{e^{+i\Omega_{g e_{1}}\tau} e^{-i\Omega_{e_{1}g}\kappa}}{\Omega _{e_{1}e_{1}}}$ & 
$P'_{II 22 \beta ii}$ 	&$\Theta[\tau] \Theta[\kappa] \frac{e^{-i\Omega _{e_{2}g} \tau }e^{-i\Omega_{f e_{2}}\kappa}}{\Omega _{e_{2} e_{2}}}$\\
$P_{I22 \alpha i}$   	&$\Theta[-\sigma] \Theta[\zeta] \frac{e^{+i\Omega_{g e_{2}}\sigma}e^{-i\Omega_{e_{2}g}\zeta}}{\Omega _{e_{2}e_{2}}}$ & 
$P'_{II 11 \alpha i}$ 	&$\Theta[\sigma] \Theta[\zeta] \frac{e^{-i\Omega _{e_{1}g} \sigma }e^{-i\Omega_{f e_{1}}\zeta}}{\Omega _{e_{1} e_{1}}}$\\
$P_{I22 \beta ii}$    	&$\Theta[-\tau] \Theta[\kappa] \frac{e^{+i\Omega_{g e_{2}}\tau} e^{-i\Omega_{e_{2}g}\kappa}}{\Omega _{e_{2}e_{2}}}$ & 
$P'_{II 11 \beta ii}$ 	&$\Theta[\tau] \Theta[\kappa] \frac{e^{-i\Omega _{e_{1} g} \tau }e^{-i\Omega_{f e_{1}}\kappa}}{\Omega _{e_{1} e_{1}}}$\\

$P'_{I 22 \alpha i}$ 	&$\Theta[-\sigma] \Theta[\zeta] \frac{e^{+i\Omega _{g e_{2}} \sigma }e^{-i\Omega_{f e_{2}}\zeta}}{\Omega _{e_{2} e_{2}}}$	&
$P_{II11 \alpha i}$   	&$\Theta[\sigma] \Theta[\zeta] \frac{e^{-i\Omega _{e_{1}g}(\sigma + \zeta )}}{\Omega_{e_{1}e_{1}}}$\\
$P'_{I 22 \beta ii}$ 	&$\Theta[-\tau] \Theta[\kappa] \frac{e^{+i\Omega _{g e_{2}} \tau }e^{-i\Omega_{f e_{2}}\kappa}}{-\Omega _{e_{2} e_{2}}}$	&
$P_{II11 \beta ii}$    	&$\Theta[\tau] \Theta[\kappa] \frac{e^{-i\Omega _{e_{1}g}(\tau + \kappa)}}{\Omega_{e_{1}e_{1}}}$\\
$P'_{I 11 \alpha i}$ 	&$\Theta[-\sigma] \Theta[\zeta] \frac{e^{+i\Omega _{g e_{1}} \sigma }e^{-i\Omega_{f e_{1}}\zeta}}{\Omega _{e_{1} e_{1}}}$	&
$P_{II22 \alpha i}$   	&$\Theta[\sigma] \Theta[\zeta] \frac{e^{-i\Omega _{e_{2}g}(\sigma + \zeta)}}{\Omega_{e_{2}e_{2}}}$\\
$P'_{I 11 \beta ii}$ 	&$\Theta[-\tau] \Theta[\kappa] \frac{e^{+i\Omega _{g e_{1}} \tau }e^{-i\Omega_{f e_{1}}\kappa}}{\Omega _{e_{1} e_{1}}}$	&
$P_{II22 \beta ii}$    	&$\Theta[\tau] \Theta[\kappa] \frac{e^{-i\Omega _{e_{2}g}(\tau + \kappa)}}{\Omega_{e_{2}e_{2}}}$\\
\end{tabular}
\caption{Expressions for the 32 nonzero FTC diagrams with unrestricted topology. Each expression also gains a factor of $\left(\frac{i}{\hbar}\right)^3 (-1)^n \rho _{0}\mu ^{(4)} I_0^2$. As detailed in the text, the 16 `C' type terms with unrestricted topology are zero due to color locking. Interestingly, no terms in this class originate from two-quantum pathways.} 
\label{tab:tab1}
\end{table*}

We first consider the unrestricted topological class. For this class, Fig. \ref{fig:FTC_classes}(a), the line segments representing pair correlators are topologically disjoint. The intra-timeline segment is free to `slide along the timeline' such that it can accumulate over the entire interval between $t_{2}$ and $t_{3}.$ The inter-timeline segment synchronizes the last field event and the local oscillator. The intra-timeline segment probes the response function on the interval between the first and second field actions. This gives rise to a $e^{-i\Omega_{I}\theta}$ factor. The inter-timeline segment probes the interval between the third field action and the quadrature event with the local oscillator, giving rise to a $e^{-i\Omega _{III}\phi }$ factor. The intra-timeline segment is free to accumulate fully over the entire response function between the second and third field actions. This contributes a factor $1/\Omega _{II}$ to the signal. Assembled together---along with a factor of $\left(\frac{i}{\hbar}\right)^3 (-1)^n \rho _{0}\mu ^{(4)} I_0^2$, where $\rho _{0}$ is the `dark' density operator factor, $n$ is the number of ket-side interactions, $\mu ^{(4)}$ is the particular sequence of four transition dipoles determined from the WMEL or double-sided Feynman diagram responsible for this term, and $I_0^2$ is the cumulative intensity of the incident beams and local oscillator---the analytic expressions associated with FTC diagrams in the unrestricted topological class are 
\begin{equation}
I_{UR}(\theta,\phi) = \Theta[ \theta] \Theta[\phi] \left( \frac{i}{\hbar}\right)^3 (-1)^n \rho _{0}\mu ^{(4)} I_0^2\frac{e^{-i\Omega _{I}\theta}e^{-i\Omega _{III}\phi }}{\Omega _{II}}.  
\label{eqn:unrestricted}
\end{equation}
To obtain the term for a specific FTC diagram one needs only to read $\Omega_{I}$, $\Omega_{III}$, $n$, and $\mu ^{(4)}$ from the WMEL diagram and $\theta$ and $\phi$ from the FTC diagram. The results are collected in Table \ref{tab:tab1}. 

Color locking plays an important role in this topological class. It forces the first and second field actions to be identical in frequency; therefore it is impossible to create coherent oscillations during the second time period for FTC diagrams in the unrestricted topological class. Thus all $C$ terms in the unrestricted class are zero, as would be all $T$ terms, however there are none. Only $P$ and $G$ terms persist. As we will show below, even though unrestricted terms cannot produce quantum beats from the $C$ type terms, some of the unrestricted terms can produce oscillations during $\tau$ due to polarization interference. 

\subsubsection{Singly Restricted Topology}
\begin{figure}
\centering
  \includegraphics[width=0.2\textwidth]{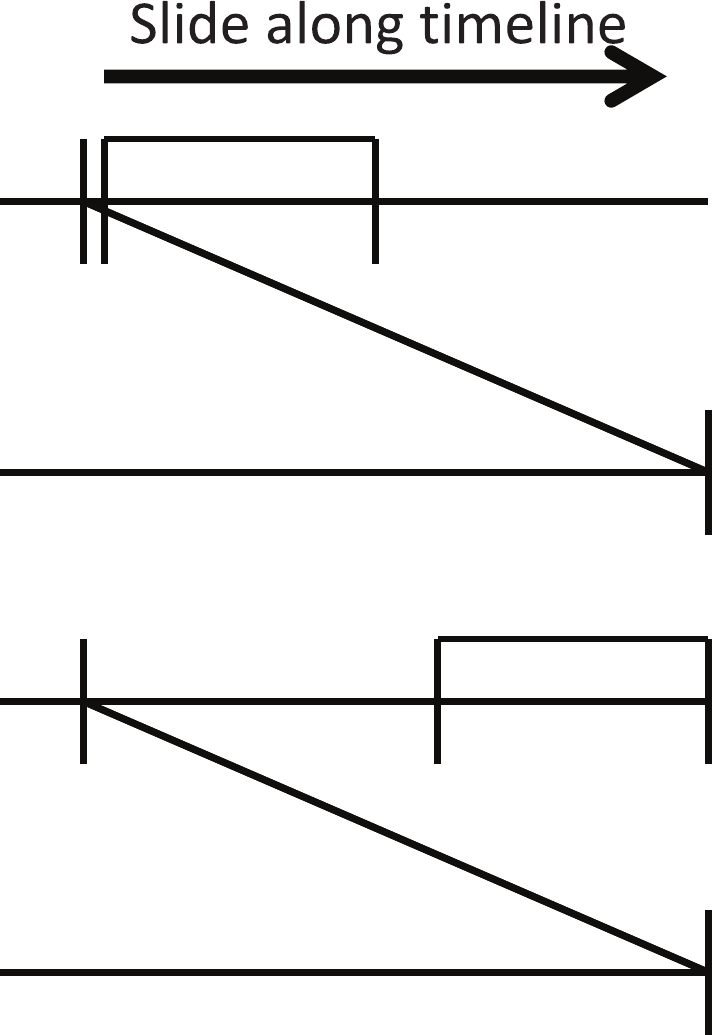} 
  \caption{Analysis of the singly restricted FTC topological class. Because of the experimentally fixed time delays, the correlated pair of events at $t_2$ and $t_3$ are not able to slide along the timeline in an unrestricted manner---there is a limited range over which they can slide. The directly correlated relationship between $s$ and $t_1$ and the directly correlated relationship between $t_2$ and $t_3$ result in an indirectly correlated relationship between $t_1$ and $t_3$ \cite{Ulness:2003aa}. Indirect correlation is also present in this topological class between events $t_3$ and $s$.}
  \label{fig:FTC_singly}
\end{figure}

\begin{table*}
\footnotesize
\centering
\begin{tabular}{c | c || c | c}
term & expression & term & expression \\
\hline
$C_{II 12 \alpha ii}$	&$\Theta[-\tau] \Theta[\kappa + \tau] (\tau + \kappa) e^{-i\Omega _{e_{1}g}(\tau + \kappa)}e^{+i\Omega_{e_{1} e_{2}}\tau}$ &
$C'_{II 12 \alpha ii}$	&$\Theta[-\tau] \Theta[\kappa + \tau] \frac{e^{+i\Omega _{e_{1}e_{2}}\tau} \left(e^{-i\Omega_{e_{1} g}(\kappa + \tau)} - e^{-i\Omega_{f e_{2}}(\kappa + \tau)} \right)}{{\Omega _{f e_{2}}-\Omega _{e_{1}g}}}$ \\
$C_{II 12 \beta i}$ 	&$\Theta[-\sigma] \Theta[\zeta + \sigma] (\sigma + \zeta) e^{-i\Omega _{e_{1}g}(\sigma + \zeta)}e^{+i\Omega_{e_{1} e_{2}}\sigma}$  &		
$C'_{II 12 \beta i}$	&$\Theta[-\sigma] \Theta[\zeta + \sigma]  \frac{e^{+i\Omega _{e_{1}e_{2}}\sigma} \left(e^{-i\Omega_{e_{1} g}(\zeta + \sigma)} - e^{-i\Omega_{f e_{2}}(\zeta + \sigma)} \right)}{{\Omega _{f e_{2}}-\Omega _{e_{1}g}}}$ \\
$C_{II 21 \alpha ii}$ 	&$\Theta[-\tau] \Theta[\kappa + \tau]  (\tau + \kappa) e^{-i\Omega _{e_{2}g}(\tau + \kappa)}e^{+i\Omega_{e_{2} e_{1}}\tau}$  &
$C'_{II 21 \alpha ii}$	&$\Theta[-\tau] \Theta[\kappa + \tau]  \frac{e^{+i\Omega _{e_{2}e_{1}}\tau} \left(e^{-i\Omega_{e_{2} g}(\kappa + \tau)} - e^{-i\Omega_{f e_{1}}(\kappa + \tau)} \right)}{{\Omega _{f e_{1}}-\Omega _{e_{2}g}}}$ \\
$C_{II 12 \beta i}$ 	&$\Theta[-\sigma] \Theta[\zeta + \sigma] (\sigma + \zeta) e^{-i\Omega _{e_{2}g}(\sigma + \zeta)}e^{+i\Omega_{e_{2} e_{1}}\sigma}$  &
$C'_{II 21 \beta i}$	&$\Theta[-\sigma] \Theta[\zeta + \sigma] \frac{e^{+i\Omega _{e_{2}e_{1}}\sigma} \left(e^{-i\Omega_{e_{2} g}(\zeta + \sigma)} - e^{-i\Omega_{f e_{1}}(\zeta + \sigma)} \right)}{{\Omega _{f e_{1}}-\Omega _{e_{2}g}}}$ \\
\hline

$P_{II 22 \alpha ii}$ 	&$\Theta[-\tau] \Theta[\kappa + \tau]  (\tau + \kappa) e^{-i\Omega _{e_{2}g}(\tau + \kappa)}e^{+i\Omega_{e_{2} e_{2}}\tau}$  & 
$P'_{II 22 \alpha ii}$	&$\Theta[-\tau] \Theta[\kappa + \tau]  \frac{e^{+i\Omega _{e_{2}e_{2}}\tau} \left(e^{-i\Omega_{e_{2} g}(\kappa + \tau)} - e^{-i\Omega_{f e_{2}}(\kappa + \tau)} \right)}{{\Omega _{f e_{2}}-\Omega _{e_{2}g}}}$ \\
$P_{II 22 \beta i}$ 	&$\Theta[-\sigma] \Theta[\zeta + \sigma] (\sigma + \zeta) e^{-i\Omega _{e_{2}g}(\sigma + \zeta)}e^{+i\Omega_{e_{2} e_{2}}\sigma}$  &
$P'_{II 22 \beta i}$	&$\Theta[-\sigma] \Theta[\zeta + \sigma]  \frac{e^{+i\Omega _{e_{2}e_{2}}\sigma} \left(e^{-i\Omega_{e_{2} g}(\zeta + \sigma)} - e^{-i\Omega_{f e_{2}}(\zeta + \sigma)} \right)}{{\Omega _{f e_{2}}-\Omega _{e_{2}g}}}$ \\
$P_{II 11 \alpha ii}$ 	&$\Theta[-\tau] \Theta[\kappa + \tau]  (\tau + \kappa) e^{-i\Omega _{e_{1}g}(\tau + \kappa)}e^{+i\Omega_{e_{1} e_{1}}\tau}$  &
$P'_{II 11 \alpha ii}$	&$\Theta[-\tau] \Theta[\kappa + \tau]  \frac{e^{+i\Omega _{e_{1}e_{1}}\tau} \left(e^{-i\Omega_{e_{1} g}(\kappa + \tau)} - e^{-i\Omega_{f e_{1}}(\kappa + \tau)} \right)}{{\Omega _{f e_{1}}-\Omega _{e_{1}g}}}$ \\
$P_{II 11 \beta i}$ 	&$\Theta[-\sigma] \Theta[\zeta + \sigma] (\sigma + \zeta) e^{-i\Omega _{e_{1}g}(\sigma + \zeta)}e^{+i\Omega_{e_{1} e_{1}}\sigma}$  &
$P'_{II 11 \beta i}$	&$\Theta[-\sigma] \Theta[\zeta + \sigma]  \frac{e^{+i\Omega _{e_{1}e_{1}}\sigma} \left(e^{-i\Omega_{e_{1} g}(\zeta + \sigma)} - e^{-i\Omega_{f e_{1}}(\zeta + \sigma)} \right)}{{\Omega _{f e_{1}}-\Omega _{e_{1}g}}}$  \\
\hline

$G_{II 22 \alpha ii}$	&$\Theta[-\tau] \Theta[\kappa + \tau] (\tau + \kappa) e^{-i\Omega _{e_{2}g}(\tau + \kappa)}e^{+i\Omega_{g g}\tau}$  &
$G_{II12 \alpha ii}$  	&$\Theta[-\tau] \Theta[\kappa + \tau] \frac{e^{+i\Omega _{gg}\tau} \left(e^{-i\Omega_{e_{1} g}(\kappa + \tau)} - e^{-i\Omega_{e_{2}g}(\kappa + \tau)} \right)}{{\Omega _{e_{2}g}-\Omega _{e_{1}g}}}$ \\
$G_{II 22 \beta i}$ 	&$\Theta[-\sigma] \Theta[\zeta + \sigma] (\sigma + \zeta) e^{-i\Omega _{e_{2}g}(\sigma + \zeta)}e^{+i\Omega_{g g}\sigma}$  &
$G_{II12 \beta i}$     	&$\Theta[-\sigma] \Theta[\zeta + \sigma] \frac{e^{+i\Omega _{gg}\sigma} \left(e^{-i\Omega_{e_{1} g}(\zeta + \sigma)} - e^{-i\Omega_{e_{2}g}(\zeta + \sigma)} \right)}{{\Omega _{e_{2}g}-\Omega _{e_{1}g}}}$ \\
$G_{II 11 \alpha ii}$ 	&$\Theta[-\tau] \Theta[\kappa + \tau] (\tau + \kappa) e^{-i\Omega _{e_{1}g}(\tau + \kappa)}e^{+i\Omega_{g g}\tau}$  &
$G_{II21 \alpha ii}$  	&$\Theta[-\tau] \Theta[\kappa + \tau] \frac{e^{+i\Omega _{gg}\tau} \left(e^{-i\Omega_{e_{2} g}(\kappa + \tau)} - e^{-i\Omega_{e_{1}g}(\kappa + \tau)} \right)}{{\Omega _{e_{1}g}-\Omega _{e_{2}g}}}$ \\
$G_{II 11 \beta i}$ 	&$\Theta[-\sigma] \Theta[\zeta + \sigma] (\sigma + \zeta) e^{-i\Omega _{e_{1}g}(\sigma + \zeta)}e^{+i\Omega_{g g}\sigma}$  &
$G_{II21 \beta i}$     	&$\Theta[-\sigma] \Theta[\zeta + \sigma] \frac{e^{+i\Omega _{gg}\sigma} \left(e^{-i\Omega_{e_{2} g}(\zeta + \sigma)} - e^{-i\Omega_{e_{1}g}(\zeta + \sigma)} \right)}{{\Omega _{e_{1}g}-\Omega _{e_{2}g}}}$ \\
\hline

$T_{21'1' \alpha ii}$ 	&$\Theta[\tau] \Theta[\kappa - \tau] (\kappa - \tau) e^{-i\Omega _{e_{2}g}(\kappa - \tau)}e^{-i\Omega_{f g}\tau}$  &
$T'_{12'2 \alpha ii}$	&$\Theta[\tau] \Theta[\kappa - \tau] \frac{e^{-i\Omega _{fg}\tau} \left(e^{-i\Omega_{e_{1} g}(\kappa - \tau)} - e^{-i\Omega_{f e_{2}}(\kappa - \tau)} \right)}{{\Omega _{f e_{2}}-\Omega _{e_{1}g}}}$ \\
$T_{21'1' \beta i}$ 	&$\Theta[\sigma] \Theta[\zeta - \sigma] (\zeta - \sigma) e^{-i\Omega _{e_{2}g}(\zeta - \sigma)}e^{-i\Omega_{f g}\sigma}$  &
$T'_{12'2 \beta i}$	&$\Theta[\sigma] \Theta[\zeta - \sigma] \frac{e^{-i\Omega _{fg}\sigma} \left(e^{-i\Omega_{e_{1} g}(\zeta - \sigma)} - e^{-i\Omega_{f e_{2}}(\zeta - \sigma)} \right)}{{\Omega _{f e_{2}}-\Omega _{e_{1}g}}}$ \\
$T_{12'2' \alpha ii}$	&$\Theta[\tau] \Theta[\kappa - \tau] (\kappa - \tau) e^{-i\Omega _{e_{1}g}(\kappa - \tau)}e^{-i\Omega_{f g}\tau}$  &
$T'_{21'1 \alpha ii}$ 	&$\Theta[\tau] \Theta[\kappa - \tau] \frac{e^{-i\Omega _{fg}\tau} \left(e^{-i\Omega_{e_{2} g}(\kappa - \tau)} - e^{-i\Omega_{f e_{1}}(\kappa - \tau)} \right)}{{\Omega _{f e_{1}}-\Omega _{e_{2}g}}}$ \\
$T_{12'2' \beta i}$	&$\Theta[\sigma] \Theta[\zeta - \sigma] (\zeta - \sigma) e^{-i\Omega _{e_{1}g}(\zeta - \sigma)}e^{-i\Omega_{f g}\sigma}$  &
$T'_{21'1 \beta i}$ 	&$\Theta[\sigma] \Theta[\zeta - \sigma] \frac{e^{-i\Omega _{fg}\sigma} \left(e^{-i\Omega_{e_{2} g}(\zeta - \sigma)} - e^{-i\Omega_{f e_{1}}(\zeta - \sigma)} \right)}{{\Omega _{f e_{1}}-\Omega _{e_{2}g}}}$ \\
$T_{12'1' \alpha ii}$ 	&$\Theta[\tau] \Theta[\kappa - \tau] \frac{e^{-i\Omega _{fg}\tau} \left(e^{-i\Omega_{e_{1} g}(\kappa - \tau)} - e^{-i\Omega_{e_{2}g}(\kappa - \tau)} \right)}{{\Omega _{e_{2}g}-\Omega _{e_{1}g}}}$  &
$T'_{21'2 \alpha ii}$ 	&$\Theta[\tau] \Theta[\kappa - \tau] \frac{e^{-i\Omega _{fg}\tau} \left(e^{-i\Omega_{e_{2} g}(\kappa - \tau)} - e^{-i\Omega_{f e_{2}}(\kappa - \tau)} \right)}{{\Omega _{f e_{2}}-\Omega _{e_{2}g}}}$ \\
$T_{12'1' \beta i}$ 	&$\Theta[\sigma] \Theta[\zeta - \sigma] \frac{e^{-i\Omega _{fg}\sigma} \left(e^{-i\Omega_{e_{1} g}(\zeta - \sigma)} - e^{-i\Omega_{e_{2}g}(\zeta - \sigma)} \right)}{{\Omega _{e_{2}g}-\Omega _{e_{1}g}}}$  &
$T'_{21'2 \beta i}$ 	&$\Theta[\sigma] \Theta[\zeta - \sigma] \frac{e^{-i\Omega _{fg}\sigma} \left(e^{-i\Omega_{e_{2} g}(\zeta - \sigma)} - e^{-i\Omega_{f e_{2}}(\zeta - \sigma)} \right)}{{\Omega _{f e_{2}}-\Omega _{e_{2}g}}}$ \\
$T_{21'2' \alpha ii}$	&$\Theta[\tau] \Theta[\kappa - \tau] \frac{e^{-i\Omega _{fg}\tau} \left(e^{-i\Omega_{e_{2} g}(\kappa - \tau)} - e^{-i\Omega_{e_{1}g}(\kappa - \tau)} \right)}{{\Omega _{e_{1}g}-\Omega _{e_{2}g}}}$  &		
$T'_{12'1 \alpha ii}$	&$\Theta[\tau] \Theta[\kappa - \tau] \frac{e^{-i\Omega _{fg}\tau} \left(e^{-i\Omega_{e_{1} g}(\kappa - \tau)} - e^{-i\Omega_{f e_{1}}(\kappa - \tau)} \right)}{{\Omega _{f e_{1}}-\Omega _{e_{1}g}}}$ \\
$T_{21'2' \beta i}$	&$\Theta[\sigma] \Theta[\zeta - \sigma] \frac{e^{-i\Omega _{fg}\sigma} \left(e^{-i\Omega_{e_{2} g}(\zeta - \sigma)} - e^{-i\Omega_{e_{1}g}(\zeta - \sigma)} \right)}{{\Omega _{e_{1}g}-\Omega _{e_{2}g}}}$  &		
$T'_{12'1 \beta i}$ 	&$\Theta[\sigma] \Theta[\zeta - \sigma] \frac{e^{-i\Omega _{fg}\sigma} \left(e^{-i\Omega_{e_{1} g}(\zeta - \sigma)} - e^{-i\Omega_{f e_{1}}(\zeta - \sigma)} \right)}{{\Omega _{f e_{1}}-\Omega _{e_{1}g}}}$ \\
\end{tabular}
\caption{Expressions for the 40 FTC diagrams with singly restricted topology. Each expression also gains a factor of $\left(\frac{i}{\hbar}\right)^3 (-1)^n \rho _{0}\mu ^{(4)} I_0^2$. Interestingly, no terms in this class originate from rephasing pathways.} 
\label{tab:tab2}
\end{table*}

We consider now the singly restricted class. For this class, Fig. \ref{fig:FTC_classes}(b), the line segments are still topologically disjoint. However, the inter-timeline segment synchronizes the first field action and the local oscillator. This has the effect of restricting the lower limit of accumulation of the intra-timeline segment. The inter-timeline segment determines the range over which the intra-timeline can accumulate. Further, these FTC diagrams represent nonzero terms only if $\phi > \vert \theta \vert$ because the topology cannot accommodate $\phi < \vert \theta \vert$ (the diagram cannot be drawn). When $\phi > \vert \theta \vert$, the intra-timeline segment directly probes the response function between the second and third field actions via direct correlation. Indirect correlation \cite{Ulness:2003aa,Aung:2005aa} is also present in these diagrams. The limited range of accumulation indirectly probes the response function between the first and second and between the third and quadrature intervals. These FTC diagrams represent terms that offer the cleanest probe of coherent oscillations during time period $\tau$ for `C' type terms.

The FTC diagram analysis for this class shows that accumulation occurs over both the interval between $t_{1}$ and $t_{2}$ and the interval between $t_{3}$ and the quadrature event at $s$. As the intra-timeline segment slides along the timeline, the accumulation over each of those intervals is opposite from the other. This gives rise to a $\frac{1}{\Omega _{III}-\Omega _{I}}$ factor. The interval between $t_{2}$ and $t_{3}$ is directly probed by $\theta $ so one obtains a $e^{-i\Omega _{II}\theta }$ factor. In addition to direct correlation, indirect correlation also appears as illustrated in Fig. \ref{fig:FTC_singly}. This gives rise to a $\left( e^{-i\Omega _{I}(\phi - \theta)} - e^{-i\Omega _{III}(\phi - \theta )}\right)$ factor. Taken together, this yields 
\begin{eqnarray}
I_{SR} & = & \Theta[\theta] \Theta[\phi - \theta ] \frac{(-1)^n \rho _{0}\mu ^{(4)} I_0^2}{\hbar ^{3}} \nonumber \\
& & \times \frac{e^{-i\Omega _{II}\theta}\left( e^{-i\Omega _{I}(\phi - \theta)}-e^{-i\Omega_{III}(\phi -  \theta)}\right) }{\Omega _{III}-\Omega _{I}}.
\label{eqn:singlyres}
\end{eqnarray} 

This topological class is pathological because of the apparent pole when $\Omega _{III} = \Omega _{I}.$ This is a removable pole, however, which is most easily seen by rearranging Eqn. \ref{eqn:singlyres} to 
\begin{eqnarray}
I_{SR} & = & \Theta[\theta] \Theta[\phi - \theta ] \frac{(-1)^n \rho _{0}\mu ^{(4)} I_0^2}{\hbar^{3}}\frac{e^{-i\Omega _{II}\theta}e^{-i\Omega _{I}(\phi - \theta )}} {\Omega_{III}-\Omega _{I}} \nonumber \\
& & \times \left( 1-e^{-i(\Omega_{III}-\Omega _{I})(\phi - \theta )}\right).
\end{eqnarray}
which upon Taylor series expansion becomes, when $\Omega _{III}\rightarrow \Omega _{I}$ and dropping higher-order terms,
\begin{eqnarray}
I_{SR} & = & \Theta[\theta] \Theta[\phi - \theta ] \frac{(-1)^n \rho _{0}\mu ^{(4)} I_0^2}{\hbar ^{3}}\frac{e^{-i\Omega_{II}\theta }e^{-i\Omega _{I}(\phi -  \theta)}  }{\Omega_{III}-\Omega _{I}}  \nonumber \\
& & \times \left(1-1+i(\Omega _{III}-\Omega _{I})(\phi - \theta )\right) \nonumber \\
 & = & \Theta[\theta] \Theta[\phi - \theta ] \frac{(-1)^n \rho _{0}\mu ^{(4)} I_0^2}{\hbar ^{3}}i(\phi -  \theta )\nonumber \\
 & & \times e^{-i\Omega _{I}(\phi - \theta )}e^{-i\Omega _{II}\theta }. \quad
\label{eqn:singlyres2}
\end{eqnarray}
The mathematical analysis for such terms reproduces this result exactly, without the need for a Taylor expansion, because the factor of $\Omega_{III} - \Omega_{I}$ appears in the argument of an exponential function before integration occurs. See the example in the Appendix for more details. The results for this topological class are collected in Table \ref{tab:tab2}. 

\subsubsection{Doubly Restricted Topology}
\begin{figure}
\centering
  \includegraphics[width=0.5\textwidth]{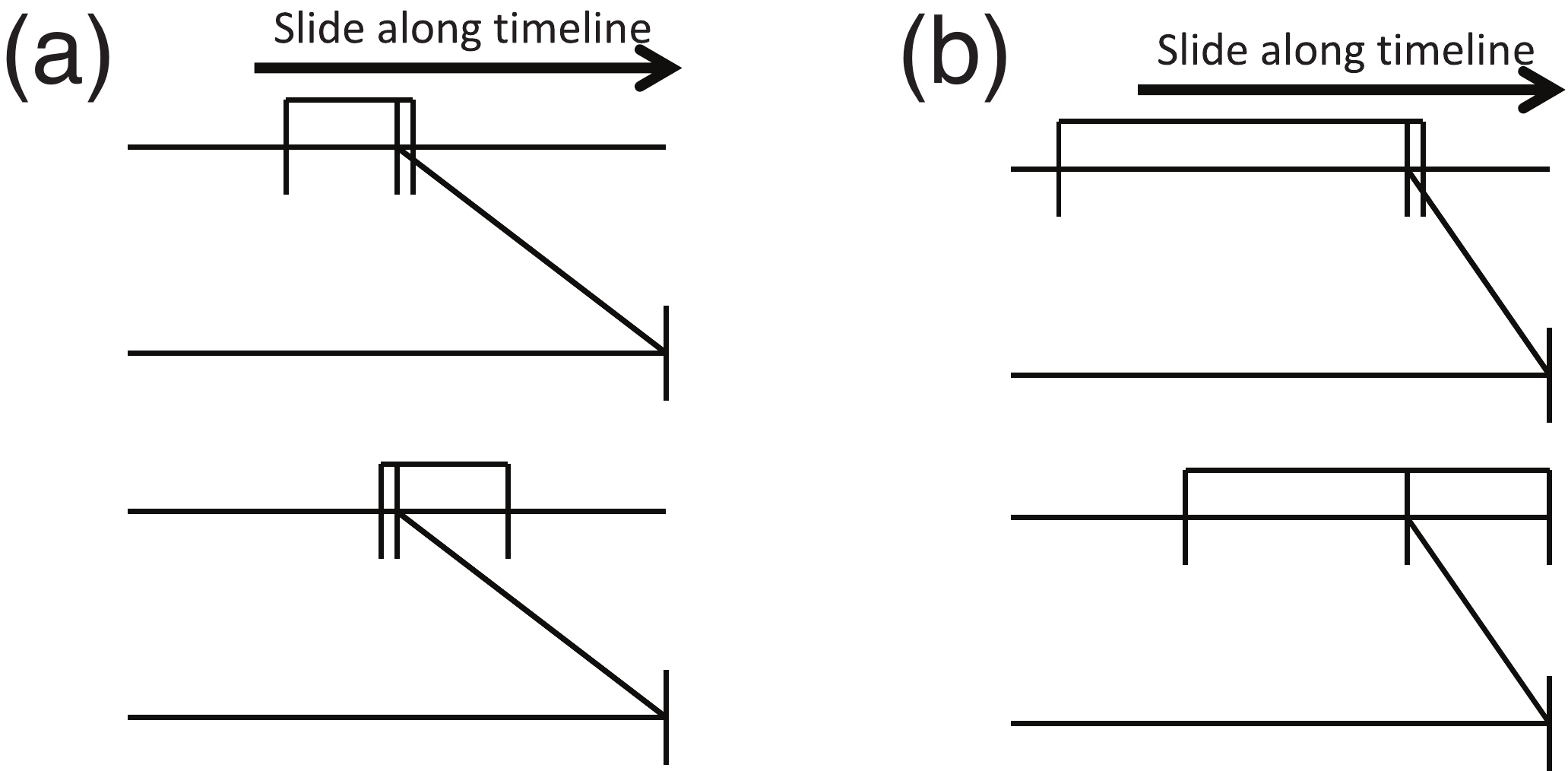} 
  \caption{Analysis of the doubly restricted FTC topological class. Direct correlations, manifest as line segments, exist between $t_1$ and $t_3$ and between $t_2$ and $s$.  Because the two segments are topologically linked, indirect correlation is complicated but nevertheless present. Most importantly, $t_1$ and $t_2$ are indirectly correlated, which relaxes the color-locking requirement. This enables coherent superpositions between the two single-exciton states to be developed on the `C' type terms.} 
  \label{fig:FTC_doubly}
\end{figure}

Finally, we consider the doubly restricted topological class. For this class, Fig. \ref{fig:FTC_classes}(c), the line segments are not topologically disjoint. The inter-timeline segment correlates the second field event with the local oscillator. Since the intra-timeline segment straddles the tick mark representing the second field event, the accumulation range is restricted from both the lower and upper limits. This topology is more complicated and must be piecewise defined about the point $\phi= \vert \theta \vert.$ If $\vert \theta \vert <\phi$ then the accumulation range of the intra-timeline segment is determined by its length, Fig. \ref{fig:FTC_doubly}(a). Conversely, if $\vert \theta \vert>\phi$, then the accumulation range is determined by the length of the inter-timeline segment, Fig. \ref{fig:FTC_doubly} (b). Another complication is that the intra-timeline segment does not probe a single time interval exclusively but rather it probes both the interval between the first and second field actions and the interval between the second and third field actions. These are indirect-correlation--based probes, so color locking does not apply \cite{Ulness:2003aa}. Thus these FTC diagrams can also probe coherent oscillations, albeit in a more complicated manner than those in the singly restricted class. 

\begin{table*}
\footnotesize
\centering
\begin{tabular}{c | c | c | c}
term & expression numerator ($\phi > \vert \theta \vert$, $\Theta[\phi - \theta]$)  & expression numerator ($\phi < \vert \theta \vert$,  $\Theta[-\phi + \theta]$)   & denominator\\
\hline
$C_{I12 \alpha ii}$	& 	$e^{+i\Omega _{e_{2} e_{1}}\tau}e^{-i\Omega_{e_{2}g}(\kappa + \tau)} - e^{+i\Omega _{ge_{1}} \tau} e^{-i\Omega _{e_{2}g} \kappa}$	&	$e^{+i\Omega _{g e_{1}}(\kappa + \tau)}e^{-i\Omega_{e_{2} e_{1}}\kappa} - e^{+i\Omega _{ge_{1}} \tau} e^{-i\Omega _{e_{2}g} \kappa}$	&	$\Omega _{ge_{1}}-\Omega _{e_{2}e_{1}}+\Omega _{e_{2}g}$\\
$C_{I12 \beta i}$	& 	$e^{+i\Omega _{e_{2} e_{1}}\sigma}e^{-i\Omega_{e_{2}g}(\zeta + \sigma)} - e^{+i\Omega _{ge_{1}} \sigma} e^{-i\Omega _{e_{2}g} \zeta}$	&	$e^{+i\Omega _{g e_{1}}(\zeta + \sigma)}e^{-i\Omega_{e_{2} e_{1}}\zeta} - e^{+i\Omega _{ge_{1}} \sigma} e^{-i\Omega _{e_{2}g} \zeta}$	&	$\Omega _{ge_{1}}-\Omega _{e_{2}e_{1}}+\Omega _{e_{2}g}$\\
$C_{I21 \alpha ii}$	& 	$e^{+i\Omega _{e_{1} e_{2}}\tau}e^{-i\Omega_{e_{1}g}(\kappa + \tau)} - e^{+i\Omega _{ge_{2}} \tau} e^{-i\Omega _{e_{1}g} \kappa}$	&	$e^{+i\Omega _{g e_{2}}(\kappa + \tau)}e^{-i\Omega_{e_{1} e_{2}}\kappa} - e^{+i\Omega _{ge_{2}} \tau} e^{-i\Omega _{e_{1}g} \kappa}$	&	$\Omega _{ge_{2}}-\Omega _{e_{1}e_{2}}+\Omega _{e_{1}g}$\\
$C_{I21 \beta i}$	& 	$e^{+i\Omega _{e_{1} e_{2}}\sigma}e^{-i\Omega_{e_{1}g}(\zeta + \sigma)} - e^{+i\Omega _{ge_{2}} \sigma} e^{-i\Omega _{e_{1}g} \zeta}$	&	$e^{+i\Omega _{g e_{2}}(\zeta + \sigma)}e^{-i\Omega_{e_{1} e_{2}}\zeta} - e^{+i\Omega _{ge_{2}} \sigma} e^{-i\Omega _{e_{1}g} \zeta}$	&	$\Omega _{ge_{2}}-\Omega _{e_{1}e_{2}}+\Omega _{e_{1}g}$\\
\hline

$P_{I22 \alpha ii}$	&	 $e^{+i\Omega _{e_{2} e_{2}}\tau}e^{-i\Omega_{e_{2}g}(\kappa + \tau)} - e^{+i\Omega _{ge_{2}} \tau} e^{-i\Omega _{e_{2}g} \kappa}$	&	$e^{+i\Omega _{g e_{2}}(\kappa + \tau)}e^{-i\Omega_{e_{2} e_{2}}\kappa} - e^{+i\Omega _{ge_{2}} \tau} e^{-i\Omega _{e_{2}g} \kappa}$	&	$\Omega _{g e_{2}} - \Omega _{e_{2}e_{2}} + \Omega _{e_{2} g}$\\
$P_{I22 \beta i}$	&	$e^{+i\Omega _{e_{2} e_{2}}\sigma}e^{-i\Omega_{e_{2}g}(\zeta + \sigma)} - e^{+i\Omega _{ge_{2}} \sigma} e^{-i\Omega _{e_{2}g} \zeta}$	&	$e^{+i\Omega _{g e_{2}}(\zeta + \sigma)}e^{-i\Omega_{e_{2} e_{2}}\zeta} - e^{+i\Omega _{ge_{2}} \sigma} e^{-i\Omega _{e_{2}g} \zeta}$	&	$\Omega _{g e_{2}} - \Omega _{e_{2}e_{2}} + \Omega _{e_{2} g}$\\
$P_{I11 \alpha ii}$	&	 $e^{+i\Omega _{e_{1} e_{1}}\tau}e^{-i\Omega_{e_{1}g}(\kappa + \tau)} - e^{+i\Omega _{ge_{1}} \tau} e^{-i\Omega _{e_{1}g} \kappa}$	&	$e^{+i\Omega _{g e_{1}}(\kappa + \tau)}e^{-i\Omega_{e_{1} e_{1}}\kappa} - e^{+i\Omega _{ge_{1}} \tau} e^{-i\Omega _{e_{1}g} \kappa}$	&	$\Omega _{g e_{1}} - \Omega _{e_{1}e_{1}} + \Omega _{e_{1} g}$\\
$P_{I11 \beta i}$	&	 $e^{+i\Omega _{e_{1} e_{1}}\sigma}e^{-i\Omega_{e_{1}g}(\zeta + \sigma)} - e^{+i\Omega _{ge_{1}} \sigma} e^{-i\Omega _{e_{1}g} \zeta}$	&	$e^{+i\Omega _{g e_{1}}(\zeta + \sigma)}e^{-i\Omega_{e_{1} e_{1}}\zeta} - e^{+i\Omega _{ge_{1}} \sigma} e^{-i\Omega _{e_{1}g} \zeta}$	&	$\Omega _{g e_{1}} - \Omega _{e_{1}e_{1}} + \Omega _{e_{1} g}$\\
\hline

$G_{I22 \alpha ii}$	&	 $e^{+i\Omega _{gg}\tau}e^{-i\Omega_{e_{2}g}(\kappa + \tau)} - e^{+i\Omega _{ge_{2}} \tau} e^{-i\Omega _{e_{2}g} \kappa}$	&	$e^{+i\Omega _{g e_{2}}(\kappa + \tau)}e^{-i\Omega_{gg}\kappa} - e^{+i\Omega _{ge_{2}} \tau} e^{-i\Omega _{e_{2}g} \kappa}$	&	$\Omega _{g e_{2}} - \Omega _{gg} + \Omega _{e_{2} g}$\\
$G_{I22 \beta i}$	&	 $e^{+i\Omega _{gg}\sigma}e^{-i\Omega_{e_{2}g}(\zeta + \sigma)} - e^{+i\Omega _{ge_{2}} \sigma} e^{-i\Omega _{e_{2}g} \zeta}$	&	$e^{+i\Omega _{g e_{2}}(\zeta + \sigma)}e^{-i\Omega_{gg}\zeta} - e^{+i\Omega _{ge_{2}} \sigma} e^{-i\Omega _{e_{2}g} \zeta}$	&	$\Omega _{g e_{2}} - \Omega _{gg} + \Omega _{e_{2} g}$\\
$G_{I21 \alpha ii}$	&	$e^{+i\Omega _{gg}\tau}e^{-i\Omega_{e_{2}g}(\kappa + \tau)} - e^{+i\Omega _{ge_{2}} \tau} e^{-i\Omega _{e_{1}g} \kappa}$	&	$e^{+i\Omega _{g e_{2}}(\kappa + \tau)}e^{-i\Omega_{gg}\kappa} - e^{+i\Omega _{ge_{2}} \tau} e^{-i\Omega _{e_{1}g} \kappa}$	&	$\Omega _{ge_{2}}-\Omega _{gg}+\Omega _{e_{1}g}$\\
$G_{I21 \beta i}$	&	$e^{+i\Omega _{gg}\sigma}e^{-i\Omega_{e_{1}g}(\zeta + \sigma)} - e^{+i\Omega _{ge_{2}} \sigma} e^{-i\Omega _{e_{1}g} \zeta}$	&	$e^{+i\Omega _{g e_{2}}(\zeta + \sigma)}e^{-i\Omega_{gg}\zeta} - e^{+i\Omega _{ge_{2}} \sigma} e^{-i\Omega _{e_{1}g} \zeta}$	&	$\Omega _{ge_{2}}-\Omega _{gg}+\Omega _{e_{1}g}$\\
$G_{I12 \alpha ii}$	&	$e^{+i\Omega _{gg}\tau}e^{-i\Omega_{e_{2}g}(\kappa + \tau)} - e^{+i\Omega _{ge_{1}} \tau} e^{-i\Omega _{e_{2}g} \kappa}$	&	$e^{+i\Omega _{g e_{1}}(\kappa + \tau)}e^{-i\Omega_{gg}\kappa} - e^{+i\Omega _{ge_{1}} \tau} e^{-i\Omega _{e_{2}g} \kappa}$	&	$\Omega _{ge_{1}}-\Omega _{gg}+\Omega _{e_{2}g}$\\
$G_{I12 \beta i}$	&	$e^{+i\Omega _{gg}\sigma}e^{-i\Omega_{e_{2}g}(\zeta + \sigma)} - e^{+i\Omega _{ge_{1}} \sigma} e^{-i\Omega _{e_{2}g} \zeta}$	&	$e^{+i\Omega _{g e_{1}}(\zeta + \sigma)}e^{-i\Omega_{gg}\zeta} - e^{+i\Omega _{ge_{1}} \sigma} e^{-i\Omega _{e_{2}g} \zeta}$	&	$\Omega _{ge_{1}}-\Omega _{gg}+\Omega _{e_{2}g}$\\
$G_{I11 \alpha ii}$	& 	 $e^{+i\Omega _{gg}\tau}e^{-i\Omega_{e_{1}g}(\kappa + \tau)} - e^{+i\Omega _{ge_{1}} \tau} e^{-i\Omega _{e_{1}g} \kappa}$	&	$e^{+i\Omega _{g e_{1}}(\kappa + \tau)}e^{-i\Omega_{gg}\kappa} - e^{+i\Omega _{ge_{1}} \tau} e^{-i\Omega _{e_{1}g} \kappa}$	&	$\Omega _{g e_{1}} - \Omega _{gg} + \Omega _{e_{1} g}$\\
$G_{I11 \beta i}$	& 	 $e^{+i\Omega _{gg}\sigma}e^{-i\Omega_{e_{1}g}(\zeta + \sigma)} - e^{+i\Omega _{ge_{1}} \sigma} e^{-i\Omega _{e_{1}g} \zeta}$	&	$e^{+i\Omega _{g e_{1}}(\zeta + \sigma)}e^{-i\Omega_{gg}\zeta} - e^{+i\Omega _{ge_{1}} \sigma} e^{-i\Omega _{e_{1}g} \zeta}$	&	$\Omega_{g e_{1}} - \Omega_{gg} + \Omega _{e_{1} g}$\\
\hline

$C'_{I 21 \alpha ii}$	&	$e^{+i\Omega_{e_{1} e_{2}}\tau}e^{-i\Omega_{f e_{2}}(\kappa + \tau)} - e^{+i\Omega_{ge_{2}} \tau} e^{-i\Omega_{f e_{2}} \kappa}$	&	$e^{+i\Omega_{g e_{2}}(\kappa + \tau)}e^{-i\Omega_{e_{1} e_{2}}\kappa} - e^{+i\Omega_{ge_{2}} \tau} e^{-i\Omega_{f e_{2}} \kappa}$	&	$\Omega_{ge_{2}}-\Omega_{e_{1} e_{2}}+\Omega_{f e_{2}}$\\
$C'_{I 21 \beta i}$ 	&	$e^{+i\Omega_{e_{1} e_{2}}\sigma}e^{-i\Omega_{f e_{2}}(\zeta + \sigma)} - e^{+i\Omega_{ge_{2}} \sigma} e^{-i\Omega_{f e_{2}} \zeta}$	&	$e^{+i\Omega_{g e_{2}}(\zeta + \sigma)}e^{-i\Omega_{e_{1} e_{2}}\zeta} - e^{+i\Omega_{ge_{2}} \sigma} e^{-i\Omega_{f e_{2}} \zeta}$	&	$\Omega_{ge_{2}}-\Omega_{e_{1} e_{2}}+\Omega_{f e_{2}}$\\
$C'_{I 12 \alpha ii}$	&	$e^{+i\Omega_{e_{2} e_{1}}\tau}e^{-i\Omega_{f e_{1}}(\kappa + \tau)} - e^{+i\Omega_{ge_{1}} \tau} e^{-i\Omega_{f e_{1}} \kappa}$	&	$e^{+i\Omega_{g e_{1}}(\kappa + \tau)}e^{-i\Omega_{e_{2} e_{1}}\kappa} - e^{+i\Omega_{ge_{1}} \tau} e^{-i\Omega_{f e_{1}} \kappa}$	&	$\Omega_{ge_{1}}-\Omega_{e_{2} e_{1}}+\Omega_{f e_{1}}$\\
$C'_{I 12 \beta i}$	&	$e^{+i\Omega_{e_{2} e_{1}}\sigma}e^{-i\Omega_{f e_{1}}(\zeta + \sigma)} - e^{+i\Omega_{ge_{1}} \sigma} e^{-i\Omega_{f e_{1}} \zeta}$	&	$e^{+i\Omega_{g e_{1}}(\zeta + \sigma)}e^{-i\Omega_{e_{2} e_{1}}\zeta} - e^{+i\Omega_{ge_{1}} \sigma} e^{-i\Omega_{f e_{1}} \zeta}$	&	$\Omega_{ge_{1}}-\Omega_{e_{2} e_{1}}+\Omega_{f e_{1}}$\\
\hline

$P'_{I 22 \alpha ii}$   &	$e^{+i\Omega_{e_{2} e_{2}}\tau}e^{-i\Omega_{f e_{2}}(\kappa + \tau)} - e^{+i\Omega_{ge_{2}} \tau} e^{-i\Omega_{f e_{2}} \kappa}$	&	$e^{+i\Omega_{g e_{2}}(\kappa + \tau)}e^{-i\Omega_{e_{2} e_{2}}\kappa} - e^{+i\Omega_{ge_{2}} \tau} e^{-i\Omega_{f e_{2}} \kappa}$	&	$\Omega_{ge_{2}}-\Omega_{e_{2} e_{2}}+\Omega_{f e_{2}}$\\
$P'_{I 22 \beta i}$ 	&	$e^{+i\Omega_{e_{2} e_{2}}\sigma}e^{-i\Omega_{f e_{2}}(\zeta + \sigma)} - e^{+i\Omega_{ge_{2}} \sigma} e^{-i\Omega_{f e_{2}} \zeta}$	&	$e^{+i\Omega_{g e_{2}}(\zeta + \sigma)}e^{-i\Omega_{e_{2} e_{2}}\zeta} - e^{+i\Omega_{ge_{2}} \sigma} e^{-i\Omega_{f e_{2}} \zeta}$	&	$\Omega_{ge_{2}}-\Omega_{e_{2} e_{2}}+\Omega_{f e_{2}}$\\
$P'_{I 11 \alpha ii}$ 	&	$e^{+i\Omega_{e_{1} e_{1}}\tau}e^{-i\Omega_{f e_{1}}(\kappa + \tau)} - e^{+i\Omega_{ge_{1}} \tau} e^{-i\Omega_{f e_{1}} \kappa}$	&	$e^{+i\Omega_{g e_{1}}(\kappa + \tau)}e^{-i\Omega_{e_{1} e_{1}}\kappa} - e^{+i\Omega_{ge_{1}} \tau} e^{-i\Omega_{f e_{1}} \kappa}$	&	$\Omega_{ge_{1}}-\Omega_{e_{1} e_{1}}+\Omega_{f e_{1}}$\\
$P'_{I 11 \beta i}$ 	&	$e^{+i\Omega_{e_{1} e_{1}}\sigma}e^{-i\Omega_{f e_{1}}(\zeta + \sigma)} - e^{+i\Omega_{ge_{2}} \sigma} e^{-i\Omega_{f e_{1}} \zeta}$	&	$e^{+i\Omega_{g e_{1}}(\zeta + \sigma)}e^{-i\Omega_{e_{1} e_{1}}\zeta} - e^{+i\Omega_{ge_{1}} \sigma} e^{-i\Omega_{f e_{1}} \zeta}$	&	$\Omega_{ge_{1}}-\Omega_{e_{1} e_{1}}+\Omega_{f e_{1}}$\\
\hline

$T'_{12'2 \alpha i}$ 	&	$e^{-i\Omega_{fg}\sigma}e^{-i\Omega_{f e_{2}}(\zeta - \sigma)} - e^{-i\Omega_{e_{1} g} \sigma} e^{-i\Omega_{f e_{2}} \zeta}$	&	$e^{-i\Omega_{e_{1} g}(\sigma - \zeta)}e^{-i\Omega_{fg}\zeta} - e^{-i\Omega_{e_{1} g} \sigma} e^{-i\Omega_{f e_{2}} \zeta}$	&	$\Omega_{e_{1} g} - \Omega_{fg} + \Omega_{f e_{2}}$\\
$T'_{12'2 \beta ii}$ 	&	$e^{-i\Omega_{fg}\tau}e^{-i\Omega_{f e_{2}}(\kappa - \tau)} - e^{-i\Omega_{e_{1} g} \tau} e^{-i\Omega_{f e_{2}} \kappa}$	&	$e^{-i\Omega_{e_{1} g}(\tau - \kappa)}e^{-i\Omega_{fg}\kappa} - e^{-i\Omega_{e_{1} g} \tau} e^{-i\Omega_{f e_{2}} \kappa}$	&	$\Omega_{e_{1} g} - \Omega_{fg} + \Omega_{f e_{2}}$\\
$T'_{21'1 \alpha i}$ 	&	$e^{-i\Omega_{fg}\sigma}e^{-i\Omega_{f e_{1}}(\zeta - \sigma)} - e^{-i\Omega_{e_{2} g} \sigma} e^{-i\Omega_{f e_{1}} \zeta}$	&	$e^{-i\Omega_{e_{2} g}(\sigma - \zeta)}e^{-i\Omega_{fg}\zeta} - e^{-i\Omega_{e_{2} g} \sigma} e^{-i\Omega_{f e_{1}} \zeta}$	&	$\Omega_{e_{2} g} - \Omega_{fg} + \Omega_{f e_{1}}$\\
$T'_{21'1 \beta ii}$ 	&	$e^{-i\Omega_{fg}\tau}e^{-i\Omega_{f e_{1}}(\kappa - \tau)} - e^{-i\Omega_{e_{2} g} \tau} e^{-i\Omega_{f e_{1}} \kappa}$	&	$e^{-i\Omega_{e_{2} g}(\tau - \kappa)}e^{-i\Omega_{fg}\kappa} - e^{-i\Omega_{e_{2} g} \tau} e^{-i\Omega_{f e_{1}} \kappa}$	&	$\Omega_{e_{2} g} - \Omega_{fg} + \Omega_{f e_{1}}$\\
$T'_{21'2 \alpha i}$ 	&	$e^{-i\Omega_{fg}\sigma}e^{-i\Omega_{f e_{2}}(\zeta - \sigma)} - e^{-i\Omega_{e_{2} g} \sigma} e^{-i\Omega_{f e_{2}} \zeta}$	&	$e^{-i\Omega_{e_{2} g}(\sigma - \zeta)}e^{-i\Omega_{fg}\zeta} - e^{-i\Omega_{e_{2} g} \sigma} e^{-i\Omega_{f e_{2}} \zeta}$	&	$\Omega_{e_{2} g} - \Omega_{fg} + \Omega_{f e_{2}}$\\
$T'_{21'2 \beta ii}$ 	&	$e^{-i\Omega_{fg}\tau}e^{-i\Omega_{f e_{2}}(\kappa - \tau)} - e^{-i\Omega_{e_{2} g} \tau} e^{-i\Omega_{f e_{2}} \kappa}$	&	$e^{-i\Omega_{e_{2} g}(\tau - \kappa)}e^{-i\Omega_{fg}\kappa} - e^{-i\Omega_{e_{2} g} \tau} e^{-i\Omega_{f e_{2}} \kappa}$	&	$\Omega_{e_{2} g} - \Omega_{fg} + \Omega_{f e_{2}}$\\
$T'_{12'1 \alpha i}$ 	&	$e^{-i\Omega_{fg}\sigma}e^{-i\Omega_{f e_{1}}(\zeta - \sigma)} - e^{-i\Omega_{e_{1} g} \sigma} e^{-i\Omega_{f e_{1}} \zeta}$	&	$e^{-i\Omega_{e_{1} g}(\sigma - \zeta)}e^{-i\Omega_{fg}\zeta} - e^{-i\Omega_{e_{1} g} \sigma} e^{-i\Omega_{f e_{1}} \zeta}$	&	$\Omega_{e_{1} g} - \Omega_{fg} + \Omega_{f e_{1}}$\\
$T'_{12'1 \beta ii}$ 	&	$e^{-i\Omega_{fg}\tau}e^{-i\Omega_{f e_{1}}(\kappa - \tau)} - e^{-i\Omega_{e_{1} g} \tau} e^{-i\Omega_{f e_{1}} \kappa}$	&	$e^{-i\Omega_{e_{1} g}(\tau - \kappa)}e^{-i\Omega_{fg}\kappa} - e^{-i\Omega_{e_{1} g} \tau} e^{-i\Omega_{f e_{1}} \kappa}$	&	$\Omega_{e_{1} g} - \Omega_{fg} + \Omega_{f e_{1}}$\\
\hline

$T_{12'1' \alpha i}$ 	&	$e^{-i\Omega_{fg}\sigma}e^{-i\Omega_{e_{2} g}(\zeta - \sigma)} - e^{-i\Omega_{e_{1} g} \sigma} e^{-i\Omega_{e_{2} g} \zeta}$	&	$e^{-i\Omega_{e_{1} g}(\sigma - \zeta)}e^{-i\Omega_{fg}\zeta} - e^{-i\Omega_{e_{1} g} \sigma} e^{-i\Omega_{e_{2} g} \zeta}$	&	$\Omega_{e_{1} g} - \Omega_{fg} + \Omega_{e_{2} g}$\\
$T_{12'1' \beta ii}$ 	&	$e^{-i\Omega_{fg}\tau}e^{-i\Omega_{e_{2} g}(\kappa - \tau)} - e^{-i\Omega_{e_{1} g} \tau} e^{-i\Omega_{e_{2} g} \kappa}$	&	$e^{-i\Omega_{e_{1} g}(\tau - \kappa)}e^{-i\Omega_{fg}\kappa} - e^{-i\Omega_{e_{1} g} \tau} e^{-i\Omega_{e_{2} g} \kappa}$	&	$\Omega_{e_{1} g} - \Omega_{fg} + \Omega_{e_{2} g}$\\
$T_{21'1' \alpha i}$ 	&	$e^{-i\Omega_{fg}\sigma}e^{-i\Omega_{e_{1} g}(\zeta - \sigma)} - e^{-i\Omega_{e_{2} g} \sigma} e^{-i\Omega_{e_{1} g} \zeta}$	&	$e^{-i\Omega_{e_{2} g}(\sigma - \zeta)}e^{-i\Omega_{fg}\zeta} - e^{-i\Omega_{e_{2} g} \sigma} e^{-i\Omega_{e_{1} g} \zeta}$	&	$\Omega_{e_{2} g} - \Omega_{fg} + \Omega_{e_{1} g}$\\
$T_{21'1' \beta ii}$ 	&	$e^{-i\Omega_{fg}\tau}e^{-i\Omega_{e_{1} g}(\kappa - \tau)} - e^{-i\Omega_{e_{2} g} \tau} e^{-i\Omega_{e_{1} g} \kappa}$	&	$e^{-i\Omega_{e_{2} g}(\tau - \kappa)}e^{-i\Omega_{fg}\kappa} - e^{-i\Omega_{e_{2} g} \tau} e^{-i\Omega_{e_{1} g} \kappa}$	&	$\Omega_{e_{2} g} - \Omega_{fg} + \Omega_{e_{1} g}$\\
$T_{21'2' \alpha i}$ 	&	$e^{-i\Omega_{fg}\sigma}e^{-i\Omega_{e_{2} g}(\zeta - \sigma)} - e^{-i\Omega_{e_{2} g} (\sigma + \zeta)}$	&	$e^{-i\Omega_{e_{2} g}(\sigma - \zeta)}e^{-i\Omega_{fg}\zeta} - e^{-i\Omega_{e_{2} g} (\sigma + \zeta)}$	&	$2\Omega_{e_{2} g} - \Omega_{fg}$\\
$T_{21'2' \beta ii}$ 	&	$e^{-i\Omega_{fg}\tau}e^{-i\Omega_{e_{2} g}(\kappa - \tau)} - e^{-i\Omega_{e_{2} g} (\tau + \kappa)}$	&	$e^{-i\Omega_{e_{2} g}(\tau - \kappa)}e^{-i\Omega_{fg}\kappa} - e^{-i\Omega_{e_{2} g} (\tau + \kappa)}$	&	$2\Omega_{e_{2} g} - \Omega_{fg}$\\
$T_{12'2' \alpha i}$ 	&	$e^{-i\Omega_{fg}\sigma}e^{-i\Omega_{e_{1} g}(\zeta - \sigma)} - e^{-i\Omega_{e_{1} g} (\sigma + \zeta)}$	&	$e^{-i\Omega_{e_{1} g}(\sigma - \zeta)}e^{-i\Omega_{fg}\zeta} - e^{-i\Omega_{e_{1} g} (\sigma + \zeta)}$	&	$2\Omega_{e_{1} g} - \Omega_{fg}$\\
$T_{12'2' \beta ii}$ 	&	$e^{-i\Omega_{fg}\tau}e^{-i\Omega_{e_{1} g}(\kappa - \tau)} - e^{-i\Omega_{e_{1} g} (\tau + \kappa)}$	&	$e^{-i\Omega_{e_{1} g}(\tau - \kappa)}e^{-i\Omega_{fg}\kappa} - e^{-i\Omega_{e_{1} g} (\tau + \kappa)}$	&	$2\Omega_{e_{1} g} - \Omega_{fg}$\\
\end{tabular}
\caption{Expressions for the 40 FTC diagrams with doubly restricted topology. Each expression also gains a factor of $\Theta[\phi] \Theta[\theta] \left(\frac{i}{\hbar}\right)^3 (-1)^n \rho _{0}\mu ^{(4)} I_0^2$. Interestingly, no terms in this class originate from (one-quantum) nonrephasing pathways.}
\label{tab:tab3}
\end{table*}

FTC diagram analysis for both the $\vert \theta \vert>\phi$ and the $\vert \theta \vert <\phi$ cases shows accumulation over all three intervals which yields a factor of $\frac{1}{\Omega _{I}-\Omega _{II}+\Omega _{III}}.$ The indirect correlation probes of the response function differ for cases $\vert \theta\vert >\phi $ and $\vert \theta \vert<\phi$ as illustrated in Fig. \ref{fig:FTC_doubly}. For $\vert \theta \vert>\phi$, this gives rise to a factor of 
\begin{equation}
e^{-i\Omega _{I}(\theta -\phi )}e^{-i\Omega_{II}\phi }-e^{-i\Omega _{I}\theta }e^{-i\Omega _{III}\phi }, \nonumber
\end{equation}
and for $\vert \theta \vert<\phi$, this gives a factor of 
\begin{equation}
e^{-i\Omega _{II}\theta }e^{-i\Omega _{III}(\phi - \theta )}-e^{-i\Omega _{I}\theta }e^{-i\Omega _{III}\phi }. \nonumber
\end{equation} 
Together, these factors yield
\begin{eqnarray}
I_{DR}  & = & \Theta[\theta] \Theta[\phi] \frac{(-1)^n \rho _{0}\mu ^{(4)} I_0^2}{\hbar ^{3}(\Omega _{I}-\Omega _{II}+\Omega _{III})}  \\
& & \times \bigg\{ \Theta[-\phi + \theta] \left( e^{-i\Omega_{I}(\theta -\phi )}e^{-i\Omega _{II}\phi }-e^{-i\Omega _{I}\theta}e^{-i\Omega _{III}\phi }\right) \nonumber \\
& & + \Theta[\phi - \theta] \left( e^{-i\Omega_{II}\theta }e^{-i\Omega _{III}(\phi - \theta )}-e^{-i\Omega_{I} \theta }e^{-i\Omega _{III}\phi }\right) \bigg\}. \nonumber 
\label{eqn:doubly}
\end{eqnarray}
We used Heaviside functions to write the separate results as one large expression. See the example in the Appendix for further details. The results for this topological class are collected in Table \ref{tab:tab3}.

\newpage
\section{Discussion}
The expressions contained in Tables \ref{tab:tab1}, \ref{tab:tab2}, and \ref{tab:tab3} are the main result of this work. The total expected I$^{(4)}$ 2D ES signal is the sum of these 128 expressions. In this section we present simulations for representative individual terms and describe how the signals compare to both the femtosecond-pulse 2D spectrum \cite{Hybl:2001aa,Brixner:2004aa,Gundogdu:2007aa,Engel:2007aa,Prokhorenko:2009aa,Womick:2009aa,Collini:2010aa,Panitchayangkoon:2010aa,Turner:2011ab,Christensson:2011aa,Turner:2012aa,Turner:2012ab,Dostal:2012aa,Wong:2012aa,Harel:2012aa,Fidler:2012aa,Lewis:2012aa,Turner:2012ac} and the less-common frequency-domain 2D spectrum (a collection of transient-absorption measurements using a narrowband but tuneable pump), which has been performed in the infrared \cite{Hamm:1998aa,Bredenbeck:2004aa,Cahoon:2008aa} and ultraviolet \cite{Aubock:2012aa}, but to the best of our knowledge not in the visible. We will relate our results to both the standard femtosecond 2D ES measurement---the sum of rephasing and nonrephasing components performed at a series of $\tau_2$ times---and the two-quantum femtosecond 2D ES measurement \cite{Stone:2009rt,Stone:2009aa,Nemeth:2009aa,Nemeth:2010aa,Turner:2010aa,Dai:2011aa,Abramavicius:2012aa}. 

In this work, we use the Bloch four-level system to represent a pair of coupled two-level systems characterized by the parameters (in the excitonic basis) given in Table \ref{tab:tab4}. We set $\frac{\rho_{0} I_0^2}{\hbar^3} = 1$, and the normalized transition dipole moments were $\mu_{1} = 1$, $\mu_{2} = 1.2$, $\mu_{1'} = 0.4$, and $\mu_{2'} = 0.5$ We computed the signals in the time domain using the given expressions; numerical Fourier transformation led to the presented I$^{(4)}$ 2D ES.  We describe several representative terms from each topological class that contain characteristics common to many terms. 

\begin{table}
\centering
\begin{tabular}{c | c | c }
 & $\nu$ (THz) & $\gamma$ (THz) \\
\hline
$\Omega_{e_1 g}$ 		& 525 THz 	& 5 THz \\ 
$\Omega_{e_2 g}$ 		& 575 THz 	& 5.5 THz \\ 
$\Omega_{e_2 e_1}$ 	& 50 THz 		& 9 THz \\ 
$\Omega_{f g}$ 		& 1100 THz 	& 9.5 THz \\ 
$\Omega_{f e_2}$ 		& 525 THz 	& 9.5 THz \\  
$\Omega_{f e_1}$ 		& 575 THz 	& 9.5 THz 
\end{tabular}
\caption{Simulation parameters. Each frequency is defined as $\Omega_{xy} = 2 \pi (\nu_{xy} - i \gamma_{xy})$ and its conjugate frequency defined as $\Omega_{yx} = 2 \pi (-\nu_{xy} - i \gamma_{xy})$.}
\label{tab:tab4}
\end{table}

It is important to keep in mind that in principle all 128 terms can contribute at all possible conditions of experimental time delays. Thus we carefully consider the conditions under which each term is zero or nonzero. In multiple-beam spectroscopy measurements, positive and negative delay times can be easily confused. Our convention is depicted in Fig. \ref{fig:geometry}(b) in which the field envelopes can be either advanced or retarded by, for example, retroreflectors on translation stages. The definition of positive $\sigma$ is such that path length is removed from field $E_A$ so that a relative moment of its envelope interacts with the sample before that same relative moment from field $E_B$ interacts with the sample.  Time delay $\tau$ is similarly defined on the interval between $E_B$ and $E_C$, except with the opposite convention; negative $\tau$ means that a relative moment of the envelope of field $E_B$ interacts with the sample before that same relative moment from field $E_C$. Finally, in the experimental implementation of the I$^{(4)}$ 2D ES measurement described here, time-delay variable $\zeta$ must be scanned. This will probably be most easily accomplished by adding path length to the local oscillator beam; at each positive $\zeta$ value where a relative moment of the envelope of field $E_C$ interacts with the sample before that same relative moment from field $E_{LO}$, the intensity of the signal would be measured with, for example, a photodiode. Scanning $\zeta$ is a necessary consequence that in noisy-light spectroscopy the envelope of the emitted signal has no deterministic time dependence. 

\subsection{Unrestricted Examples}
\begin{figure}
\centering
  \includegraphics[width=0.4\textwidth]{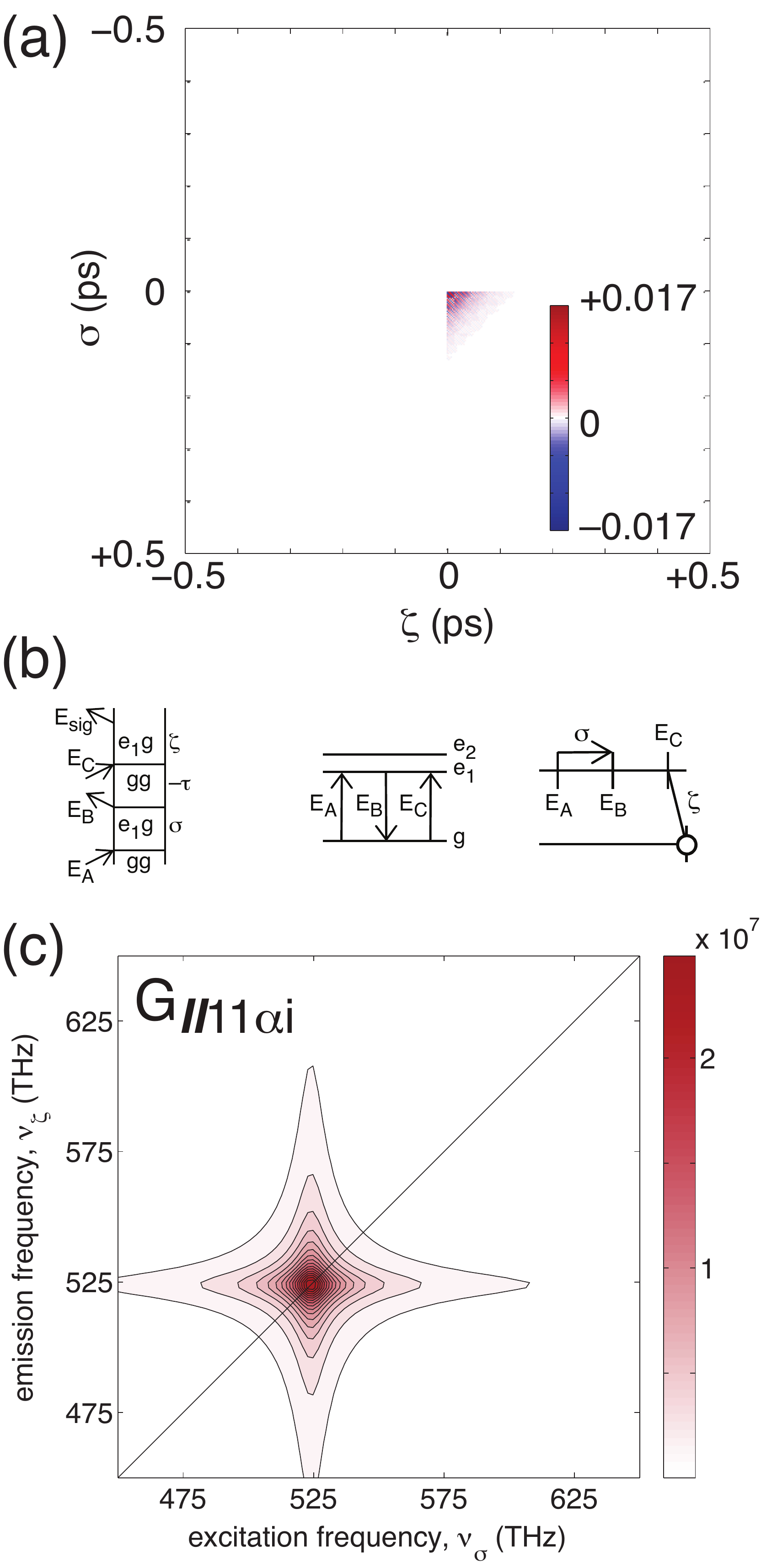} 
  \caption{The result for term $G_{II 1 1 \alpha i}$. (a) The real part of the signal in the time domain. (b) The double-sided Feynman diagram, the WMEL diagram, and the FTC diagram for this term. (c) Absolute value I$^{(4)}$ 2D ES spectrum.  This term is independent of time-delay variable $\tau$, meaning its amplitude will not oscillate coherently nor will it decay with the excited-state lifetime.} 
  \label{fig:spectrum_GII11ai}
\end{figure}

In this topological class we study terms $G_{II 11 \alpha i}$, $G_{I 11 \alpha i}$, $G_{II 11 \beta ii}$, $G_{I 21 \alpha i}$, and $G_{I 21 \beta ii}$. We begin with term $G_{II 11 \alpha i}$, a term that originates from a nonrephasing Liouville pathway. We can write its expression as
\begin{equation}
I (\sigma, \tau, \zeta) = \Theta[\sigma] \Theta[\zeta] \frac{\rho_{0} \vert \mu_1 \vert^4  I_{0}^{2}}{i \hbar^3 \Omega _{gg}} e^{-i\Omega_{e_{1} g}\sigma} e^{-i\Omega_{e_{1}g}\zeta},
\label{eqn:gII11ai_time_domain}
\end{equation}
to explicitly show the time-delay variables relevant to the measurement for this term. The Heaviside functions restrict this term to be nonzero only for positive values of time-delay variables $\sigma$ and $\zeta$, which means that this nonrephasing term is active under experimental time delays that correspond to a typical nonrephasing measurement. We present the time-domain signal in Fig. \ref{fig:spectrum_GII11ai}(a). 

Two-dimensional spectra are usually, but not always \cite{Pakoulev:2009aa,Mathew:2009aa,Mathew:2009ab,Mathew:2009ac,Mathew:2010aa,Pakoulev:2010aa}, plotted as a function of frequency for the first and third time-delay variables. For the $G_{II 11 \alpha i}$ expression, Fourier transformation of $\sigma$ and $\zeta$ yields 
\begin{eqnarray}
I (\omega_{\sigma},\tau,\omega_{\zeta}) & = & \frac{\rho_{0} \vert \mu_1 \vert^4 I_{0}^{2}}{\hbar^3 \Gamma_{g g}}  \frac{1}{\gamma_{e_{1} g} + i (\omega_\sigma - \omega_{e_{1} g})} \nonumber \\
& &  \times \frac{1}{\gamma_{e_{1} g} + i (\omega_\zeta - \omega_{e_{1} g})},
\label{eqn:gII11ai_2des}
\end{eqnarray}
where we inserted $\Omega_{gg} = -i \Gamma_{gg}$ and $\Omega_{e_{1} g} = \omega_{e_{1} g} -i \gamma_{e_{1} g}$. We present the absolute value (meaning amplitude) 2D spectrum for this term in Fig. \ref{fig:spectrum_GII11ai}(c) on a linear color scale.  There are several similarities---and two major differences---between the contribution from this term to femtosecond 2D ES and I$^{(4)}$ 2D ES measurements. 

In both the femtosecond and noisy-light measurements, the frequencies during the first and third time periods evolve phase of the same sign under exponential decay, which leads the peak for this term to appear in the $(+,+)$ quadrant of Fig. \ref{fig:spectrum_GII11ai}(c). In addition, this term leads to a diagonal peak at the lower-energy exciton frequency in I$^{(4)}$ 2D ES just as the corresponding term would in femtosecond 2D ES.  A significant difference between this component of the I$^{(4)}$ 2D ES signal and the corresponding component of the femtosecond 2D ES signal is that this term does not depend on the waiting time, $\tau$, not even as a population decay. Instead, the population lifetime appears in the $\frac{1}{\Gamma_{gg}}$ pre-factor, which leads to the very large amplitude of this peak. As will become clear, this is a common feature in I$^{(4)}$ 2D ES.   Because the signal is $\tau$-independent, this term is active even under conditions which seemingly violate causality: the probe sequence (fields $E_C$ and $E_{LO}$) can `happen' before the pump sequence (fields $E_A$ and $E_B$) as long as the relative delay internal to each sequence is correct. This is a natural consequence of the fact that pair correlators---not individual field action times---govern dynamics in noisy-light spectroscopy measurements. 

The second main difference, as mentioned above, involves signal detection. In a direct experimental implementation of I$^{(4)}$ 2D ES as described here, it will be necessary to scan the delay between fields $E_C$ and $E_{LO}$ to extract in a point-by-point fashion the time dependence of the signal during the third time period. Implementing spectral interferometry in I$^{(4)}$ 2D ES will be challenging because of the lack of causality and the asymmetry of the signal as a function of time-delay $\zeta$, both of which are used in the spectral interferometry analysis in femtosecond 2D ES. 

\begin{figure}
\centering
  \includegraphics[width=0.4\textwidth]{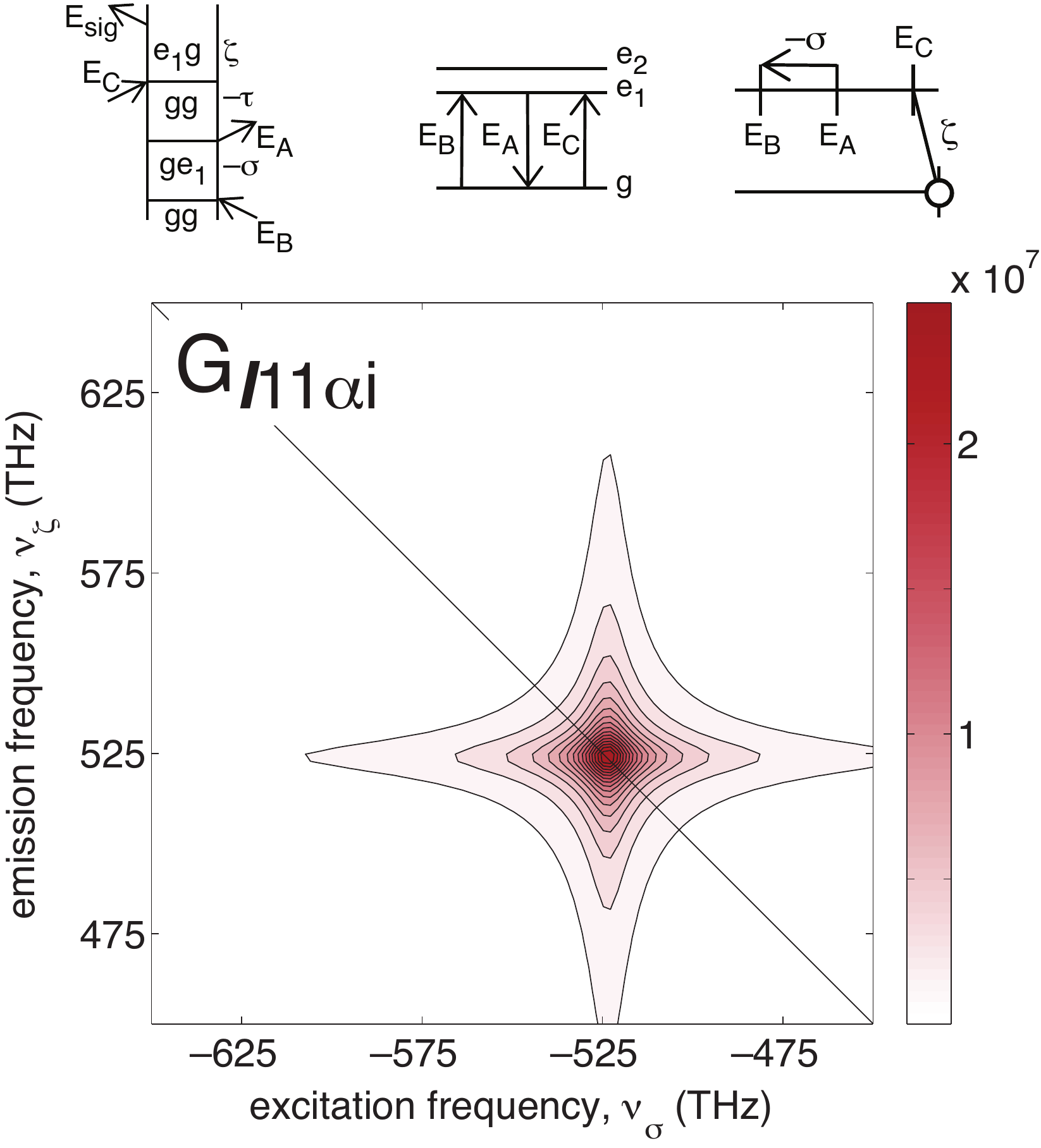} 
  \caption{The result for term $G_{I 1 1 \alpha i}$. (top) The double-sided Feynman diagram, the WMEL diagram, and the FTC diagram for this term. (bottom) Absolute value I$^{(4)}$ 2D ES spectrum. This term is also $\tau$ independent.} 
  \label{fig:spectrum_GI11ai}
\end{figure}

The second unrestricted topology term we consider is $G_{I 11 \alpha i}$, whose expression is
\begin{equation}
I (\sigma, \tau, \zeta) = \Theta[-\sigma] \Theta[\zeta] \frac{\rho_{0} \vert \mu_1 \vert^4  I_{0}^{2}}{i \hbar^3 \Omega _{gg}} e^{+i\Omega_{g e_{1}}\sigma} e^{-i\Omega_{e_{1}g}\zeta},
\end{equation}
and whose 2D spectrum is shown in Fig. \ref{fig:spectrum_GI11ai}. This term, which originates from a rephasing pathway, behaves in a manner similar to the femtosecond 2D ES measurement: it gains negative phase ($\omega_{g e}$ rather than $\omega_{e g}$) as a function of time-delay variable $- \sigma$. This results in the peak appearing in the $(-,+)$ quadrant, which also happens in femtosecond 2D ES although most practitioners flip the sign of the excitation axis to positive frequency. This peak is also large in amplitude due to the $\frac{1}{\Gamma_{gg}}$ factor and contributes to the lower-energy diagonal peak. Like the previous term, its amplitude is not $\tau$ dependent. 

The third unrestricted topology term we consider, $G_{II 11 \beta ii}$, exposes another difference between the femtosecond and I$^{(4)}$ 2D ES experiments. This term has the same response function as $G_{II 11 \alpha i}$, however, it is a $\beta$ pathway, which accounts for the possibility that field $E_C$ can interact before field $E_A$. In I$^{(4)}$ 2D ES the contribution from this pathway must be considered because of the quasi-cw nature of the fields.  The expression for this term is 
\begin{equation}
I = \Theta[\tau] \Theta[\kappa] \frac{\rho_{0} \vert \mu_1 \vert^4  I_{0}^{2}}{i \hbar^3 \Omega _{gg}} e^{-i\Omega_{e_{1} g} \tau} e^{-i\Omega_{e_{1}g} \kappa},
\end{equation}
where for the moment we suppress the time-delay variables in the argument of the function. The time-delay variables natural to the expression are $\tau$ and $\kappa$, but we must convert them into the variables we chose to represent the total signal: $\sigma$, $\tau$, and $\zeta$.  Using the identity $\tau + \kappa = \sigma + \zeta$, this change of variables results in 
\begin{equation}
I (\sigma, \tau, \zeta) = \Theta[\tau] \Theta[\zeta + \sigma - \tau] \frac{\rho_{0} \vert \mu_1 \vert^4  I_{0}^{2}}{i \hbar^3 \Omega _{gg}} e^{-i\Omega_{e_{1} g}\sigma} e^{-i\Omega_{e_{1}g}\zeta}.
\end{equation}
This expression is almost identical to Eqn. \ref{eqn:gII11ai_time_domain}, however, the seemingly slight change in the arguments of the Heaviside functions causes this term to behave in an unexpected manner. Even though the term originates from a nonrephasing pathway, nonzero signal can accrue in the rephasing pulse time ordering---when $\sigma$ is negative---if $\tau$ and $\zeta$ are appropriately selected. 

\begin{figure}
\centering
  \includegraphics[width=0.5\textwidth]{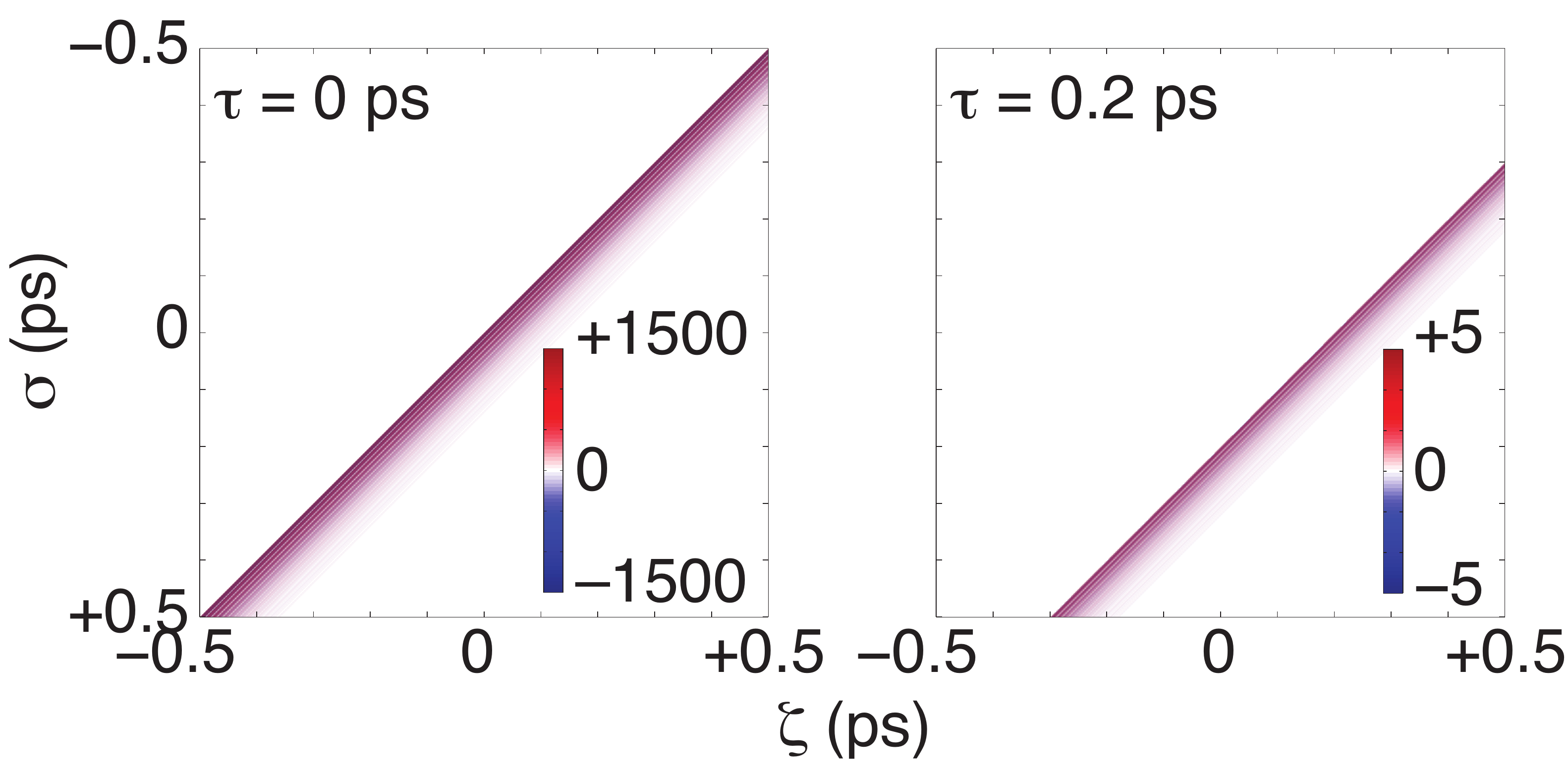} 
  \caption{The result for term $G_{II 1 1 \beta ii}$ in the time domain (real part of the signal) when $\tau =0$ (left) and $\tau = +0.2 $ ps (right). This term is zero when $\tau <0$.  The signal at $\tau = 0.2$ ps is significantly weaker than the signal at $\tau = 0$.} 
  \label{fig:spectrum_GII11bii_time}
\end{figure}

\begin{figure}
\centering
  \includegraphics[width=0.5\textwidth]{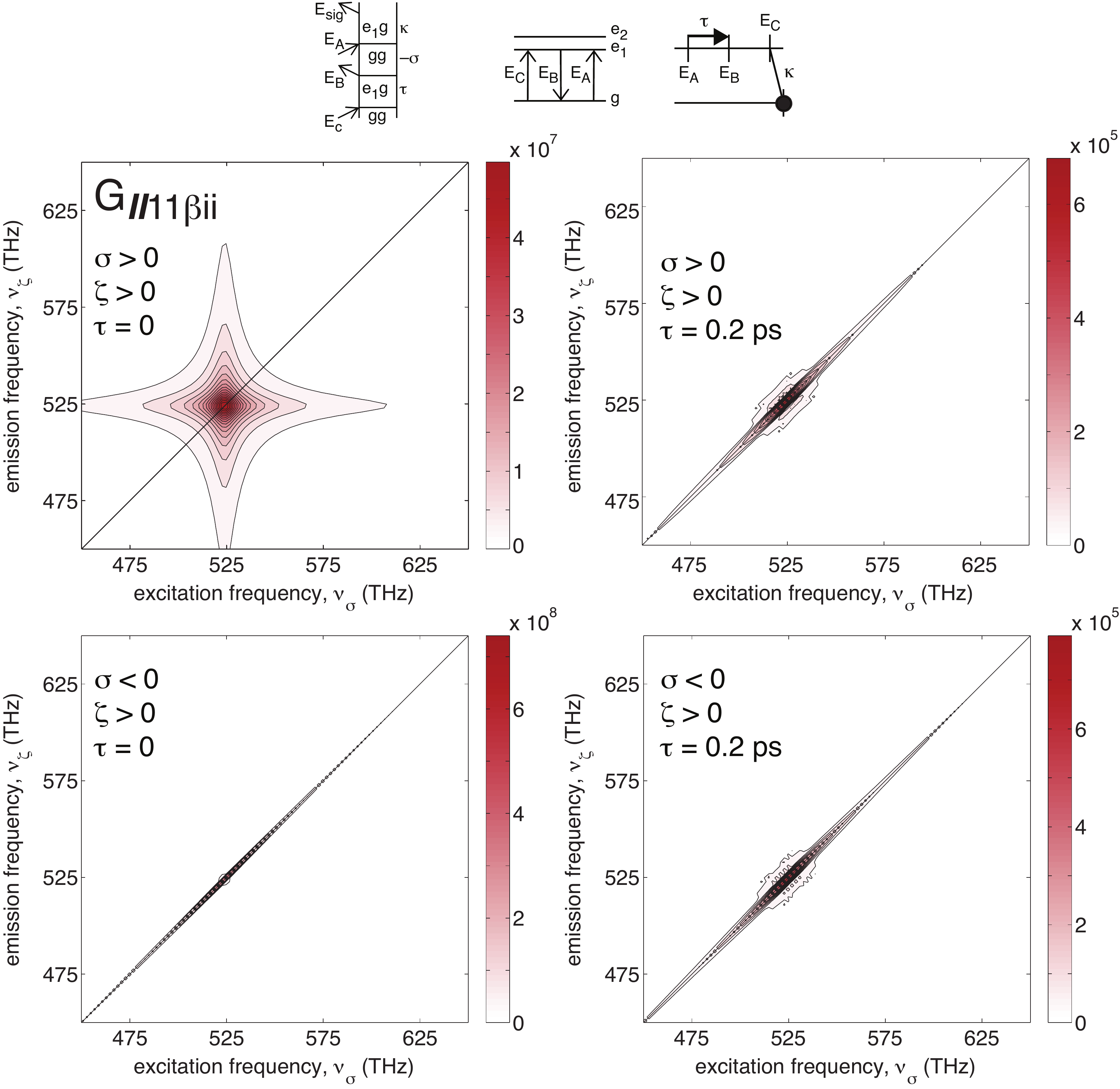} 
  \caption{The result for term $G_{II 1 1 \beta ii}$. (top) The double-sided Feynman diagram, the WMEL diagram, and the FTC diagram for this term. (middle) Absolute value I$^{(4)}$ 2D ES spectrum under nonrephasing conditions for two values of $\tau$. (bottom) Absolute value I$^{(4)}$ 2D ES spectrum under rephasing conditions for two values of $\tau$.} 
  \label{fig:spectrum_GII11bii}
\end{figure}

The result is easiest to understand in the time domain, see the two plots in Fig. \ref{fig:spectrum_GII11bii_time}. Even though this term can be nonzero under rephasing conditions, it still accrues phase at positive frequency, therefore Fourier transformation will produce a peak in the $(+,+)$ quadrant. We display the expected contribution to the 2D spectrum for this term under both rephasing and nonrephasing measurements, when $\tau = 0$ and when $\tau = 0.2$ ps, in Fig.   \ref{fig:spectrum_GII11bii}. The nonrephasing spectrum for $\tau = 0$ is expected; the signal that appears under the rephasing conditions, however, no has a very small anti-diagonal line width, and the maximum signal is almost an order of magnitude greater than the previously.  The undesirable nonzero signal under rephasing conditions can be eliminated in an experiment by proper Fourier filtering of the signal quadrants. When $\tau > 0 $, as in the 0.2 ps plots, the signal is also elongated about the diagonal but it is significantly weaker than the signal at $\tau = 0$ under both rephasing and nonrephasing conditions. 

\begin{figure}
\centering
  \includegraphics[width=0.4\textwidth]{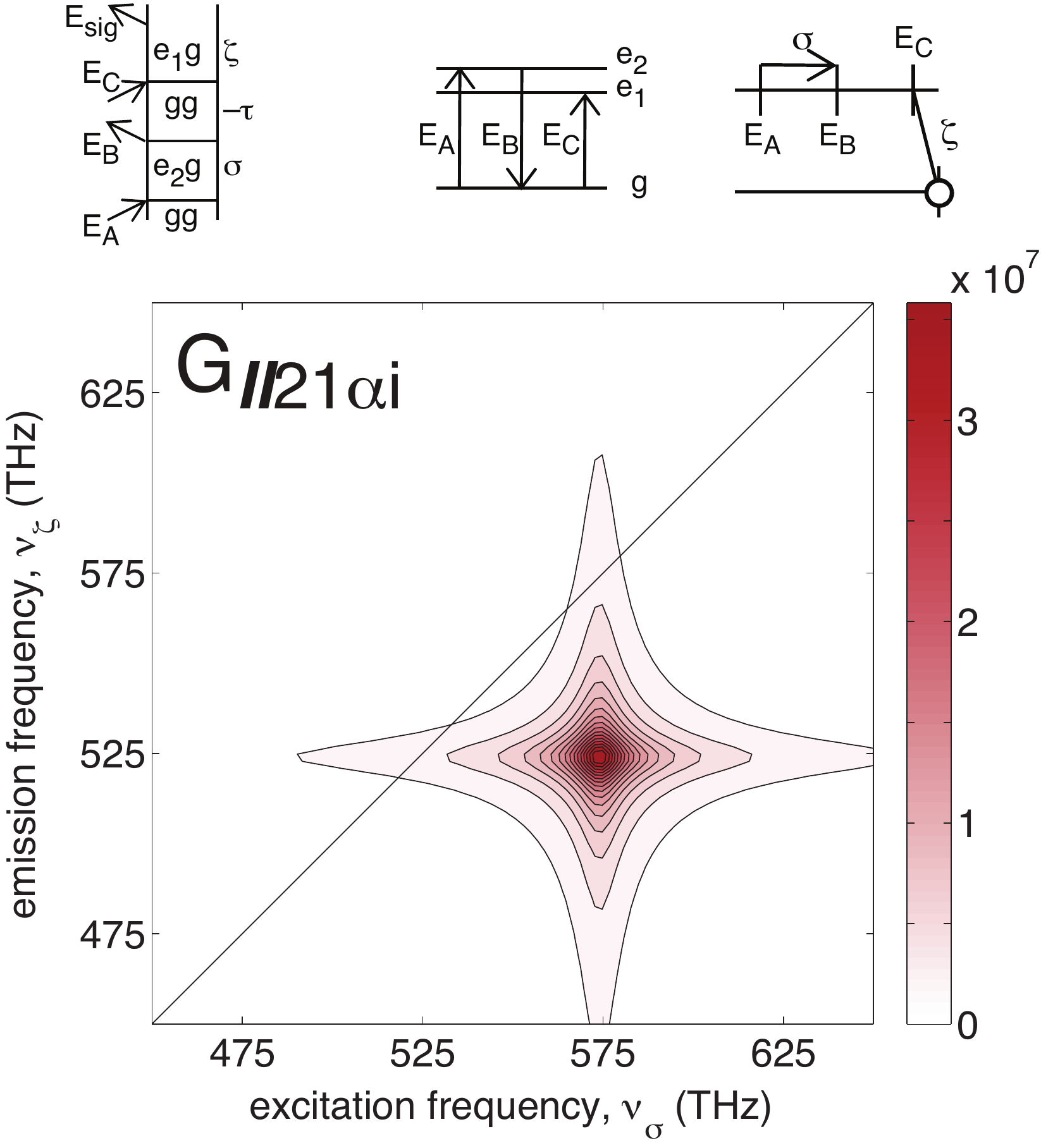} 
  \caption{The result for term $G_{II 2 1 \alpha i}$. (top) The double-sided Feynman diagram, the WMEL diagram, and the FTC diagram for this term. (bottom) Absolute value I$^{(4)}$ 2D ES spectrum. This term contributes with large amplitude to the cross peak and is independent of time-delay variable $\tau$.} 
  \label{fig:spectrum_GII21ai}
\end{figure}

The fourth term we present is $G_{II 21 \alpha i}$, which is different from the two above because it involves transitions to both of the single-exciton states. An initial expectation from the Liouville pathway suggests that this term should lead to a cross peak. Indeed, the expression
\begin{equation}
I (\sigma, \tau, \zeta) = \Theta[\sigma] \Theta[\zeta] \frac{\rho_{0} \vert \mu_1 \vert^2 \vert \mu_2 \vert^2  I_{0}^{2}}{i \hbar^3 \Omega_{gg}} e^{-i\Omega_{e_{2} g}\sigma}e^{-i\Omega_{e_{1}g}\zeta}
\end{equation} 
has different oscillation frequencies during $\sigma$ and $\zeta$. The 2D spectrum is displayed in Fig. \ref{fig:spectrum_GII21ai}; this term does give rise to a cross peak under nonrephasing conditions just as expected from the Liouville pathway and from femtosecond 2D ES.  Similar to most of the other terms in this topological class, there is no dependence on time-interval $\tau$, and again the excited-state lifetime is related to the peak amplitude. 

The fifth and final term we consider in this class is $G_{I 21 \beta ii}$. This is one of the most unusual unrestricted terms. Expressed in the variables of the experiment, this term becomes
\begin{eqnarray}
I (\sigma, \tau, \zeta) & = &  \Theta[-\tau] \Theta[\zeta + \sigma - \tau] \frac{\rho_{0} \vert \mu_1 \vert^2 \vert \mu_2 \vert^2  I_{0}^{2}}{\hbar^3 \Gamma_{gg}} e^{-i\Omega_{e_{1} g}\sigma} \nonumber \\
& & \times e^{-i\Omega_{e_{1}g}\zeta} e^{-i (\Omega_{e_{1} g} - \Omega_{g e_{2}}) \tau}.
\end{eqnarray} 
The Heaviside functions for this term need to be considered carefully.  This term is nonzero when $\kappa > 0$, meaning field $E_A$ before field $E_{LO}$, and when $\tau < 0$, meaning field $E_B$ before field $E_C$. In other words, there is not a strict requirement on the delays internal to the pump and probe sequences as long as the pump sequence `happens' before the probe sequence. 

\begin{figure}
\centering
  \includegraphics[width=0.5\textwidth]{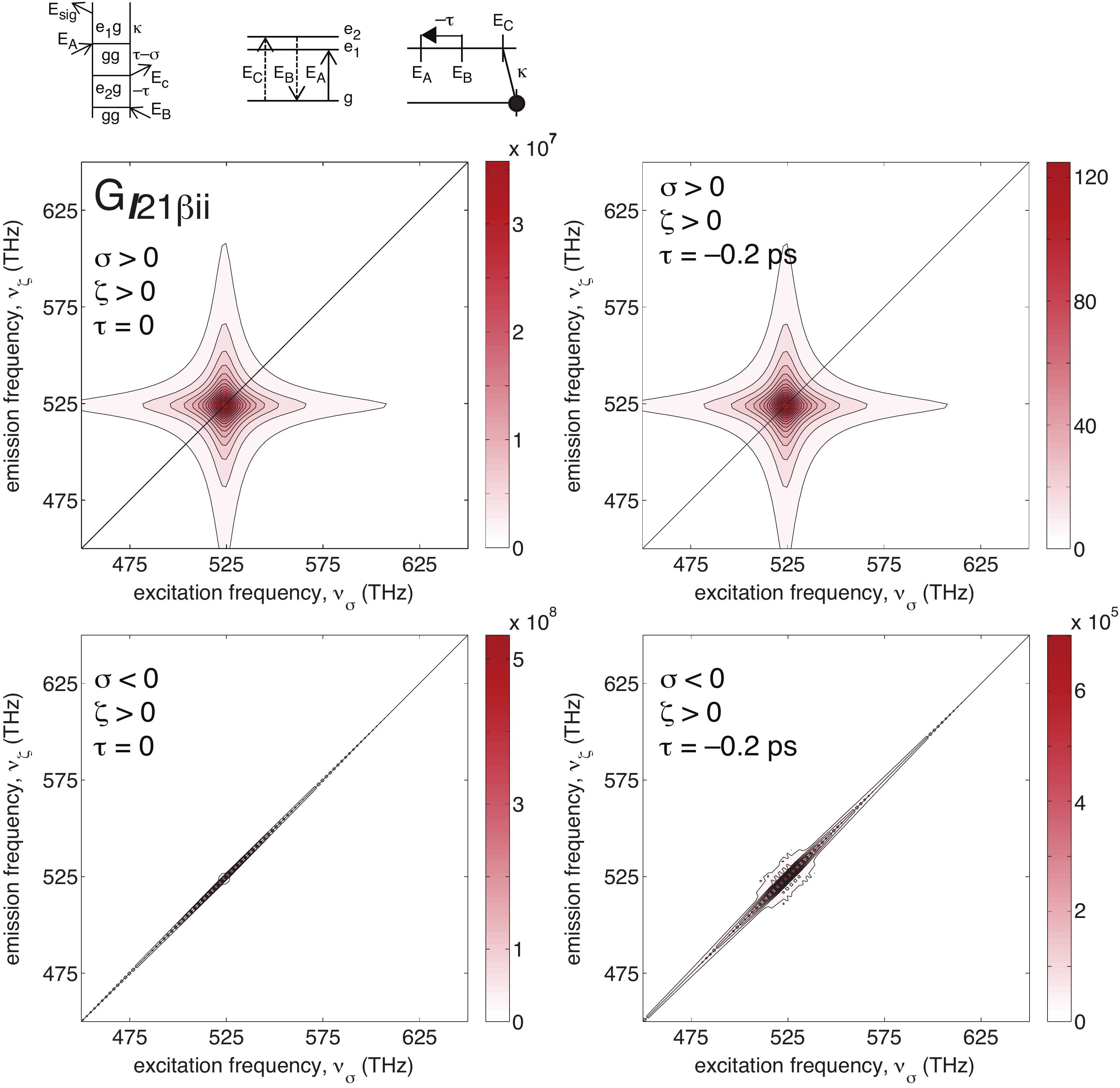} 
  \caption{The result for term $G_{I 2 1 \beta ii}$. (top) The double-sided Feynman diagram, the WMEL diagram, and the FTC diagram for this term. (bottom) Absolute value I$^{(4)}$ 2D ES spectrum. This term oscillates during $\tau$ at the difference frequency of the two single-exciton states due to polarization interference, and the oscillation decays at the difference of excitonic dephasing times, $\gamma_{e_1 g} - \gamma_{g e_{2}}$.} 
  \label{fig:spectrum_GI21bii}
\end{figure}

The 2D spectra for this term under four different timing conditions are displayed in Fig. \ref{fig:spectrum_GI21bii}.  The Liouville pathway suggests that, under rephasing time-delay conditions, this term should give rise to a cross peak that should decay slowly with no oscillatory behavior; in  I$^{(4)}$ 2D ES this term leads to a diagonal peak with an oscillating amplitude during time period $\tau$ which dephases (very slowly) at a rate given by $\gamma_{e_1 g} - \gamma_{g e_2}$. Under certain dephasing conditions the signal could even grow in amplitude as a function of $\tau$. As a reminder, here we have chosen $\gamma_{e_1 g} = \gamma_{g e_2}$, thus the oscillations neither grow nor dephase during $\tau$. 

At first glance, the oscillations during $\tau$ appear to be the quantum beating observed in many femtosecond 2D ES measurements. The oscillations are not true quantum beats; they are a manifestation of polarization interference, a topic considered in the femtosecond literature throughout the 1990s regarding how a pair of uncoupled transitions can lead to coherent oscillations in certain one-dimensional four wave-mixing measurements such as self-diffraction \cite{Koch:1992aa,Erland:1993aa,Lyssenko:1993aa,Leisching:1994aa,Weckendrup:1995aa}. It may be worthwhile to revisit and investigate this distinction between the femtosecond and noisy-light methods in future work. 

\begin{figure}
\centering 
  \includegraphics[width=0.5\textwidth]{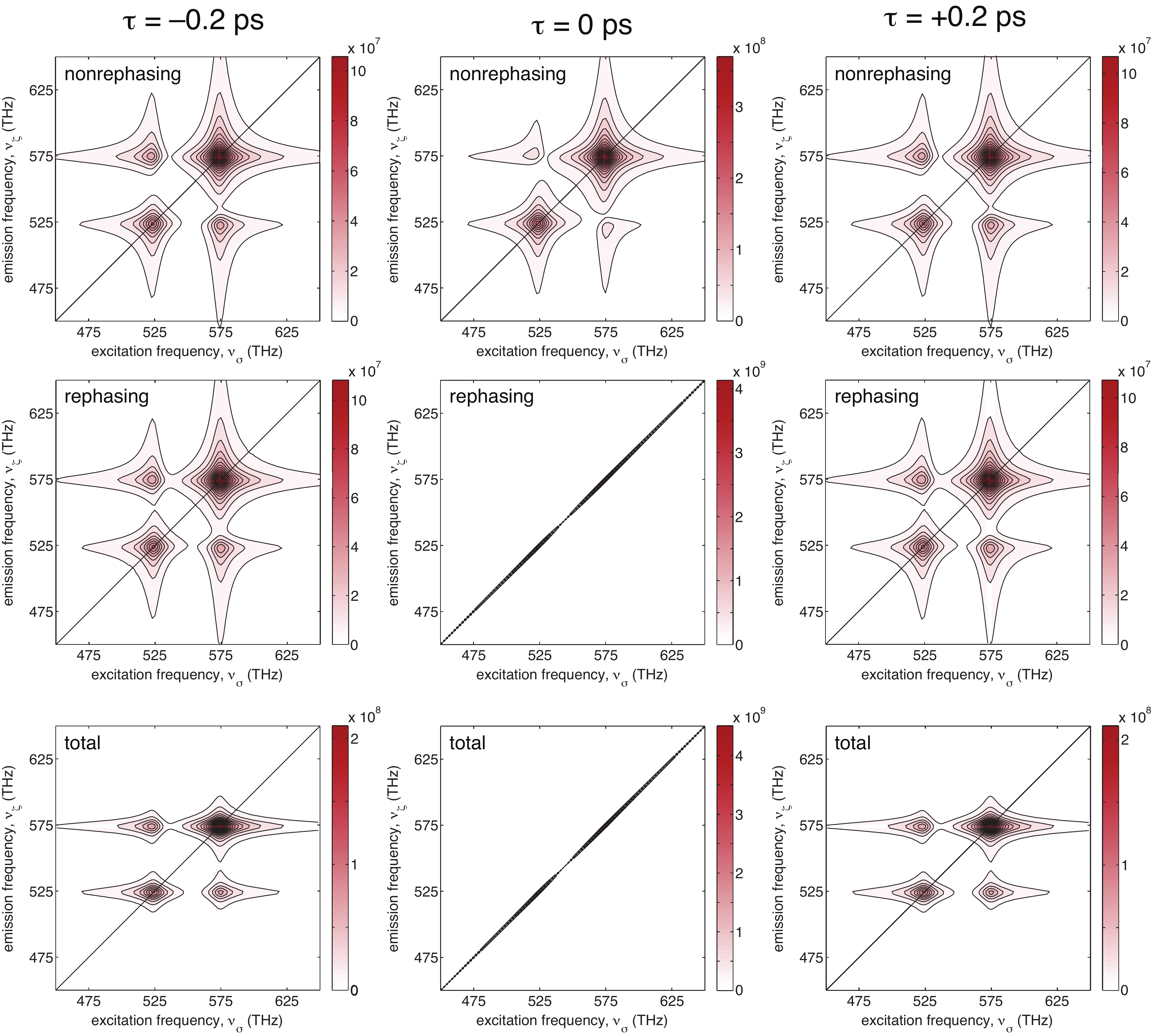} 
  \caption{The total result for the unrestricted topological class for three values of $\tau$. Each spectrum is the absolute value of the sum of all the terms which contribute under the specified time ordering. (top) The nonrephasing contributions. (middle) The rephasing contributions flipped into the $(+,+)$ quadrant. (bottom) The total I$^{(4)}$ 2D ES.}  
  \label{fig:spectrum_unrestricted} 
\end{figure}

There are 32 nonzero terms in this class, and their contributions can be categorized by the four types of Heaviside restrictions, $\Theta[ \pm \sigma] \Theta[ \zeta]$ and $\Theta[ \pm \tau] \Theta[\kappa]$.  In Fig. \ref{fig:spectrum_unrestricted} we show the total results under six different pulse-timing conditions, $\zeta$ from $0$ to $0.5$ ps, $\pm \sigma$ from $0$ to $0.5$ ps, for $\tau \in \{ 0, \pm 0.2 \}$ ps, and we show the total spectrum (sum of rephasing and nonrephasing) for each of the three $\tau$ values. We chose these values based on expected experimental conditions. Each spectrum is the absolute value of the sum---not the sum of the absolute value---of all the terms which contribute under the specified time ordering. 

Some general features for the unrestricted topology terms are as follows. The terms can lead to diagonal and cross peaks. A few terms lead to peaks that oscillate due to polarization interference during time-delay interval $\tau$, but most do not oscillate or even decay as a function of $\tau$.  No true coherent quantum beats are observed for terms of this topology because color locking eliminated the terms which could have done so. There are no two-quantum pathways in this topological class that contribute to the I$^{(4)}$ 2D ES signal.

\subsection{Singly Restricted Examples}
\begin{figure}
\centering
  \includegraphics[width=0.4\textwidth]{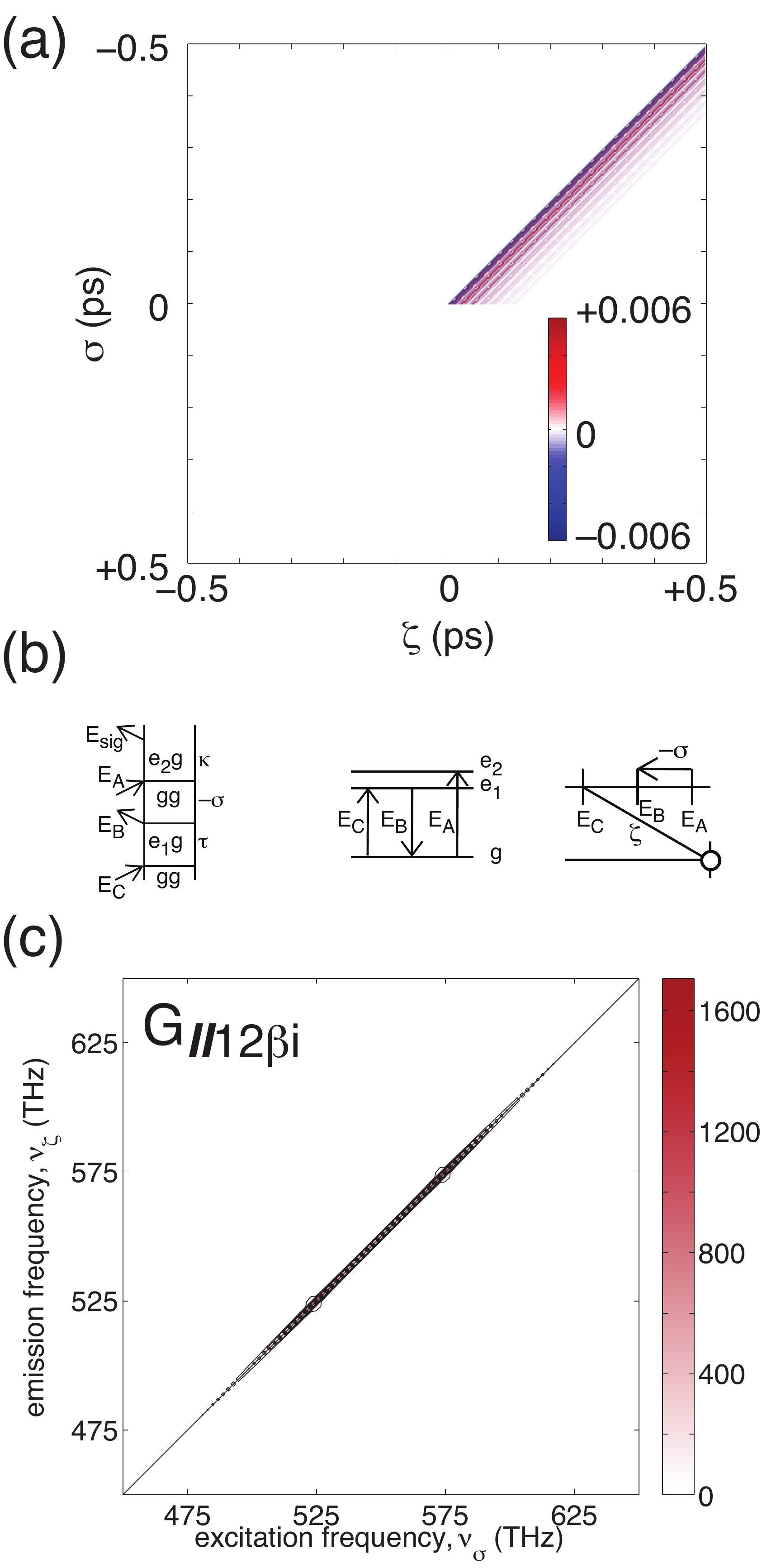} 
  \caption{The result for term $G_{II 12 \beta i}$. (a) The real part of the signal in the time domain. (b) The double-sided Feynman diagram, the WMEL diagram, and the FTC diagram for this term. (c) Absolute value I$^{(4)}$ 2D ES spectrum. This term contributes only along the diagonal at both exciton frequencies, and it is independent of time-delay variable $\tau$.} 
  \label{fig:spectrum_GII12bi}
\end{figure}

In this topological class we study terms $G_{II 12 \beta i}$, $G_{II 22 \beta i}$, $G_{II 22 \alpha ii}$, $C_{II 12 \beta i}$, $C_{II 12 \alpha ii}$, and $T_{21'1' \beta i}$. We begin with $G_{II 12 \beta i}$, 
\begin{eqnarray}
I(\sigma, \tau, \zeta) & = & \Theta[-\sigma] \Theta[\zeta + \sigma]  \\
& & \times \frac{e^{+i\Omega _{gg}\sigma} \left(e^{-i\Omega_{e_{1} g}(\zeta + \sigma)} - e^{-i\Omega_{e_{2}g}(\zeta + \sigma)} \right)}{{i(\Omega _{e_{2}g}-\Omega _{e_{1}g})}}. \nonumber
\end{eqnarray}
The product of Heaviside functions, $\Theta[-\sigma] \Theta[\zeta + \sigma]$, means that even though this term originates from a nonrephasing pathway, it is active only during the rephasing pulse time ordering, see Fig. \ref{fig:spectrum_GII12bi}(a). The restriction on $\zeta$ originates from its coupling with $\sigma$ in the second Heaviside function, where $\sigma$ is restricted by the first Heaviside function. Although the maximum amplitude in the time domain is quite small, Fourier transformation results in a peak in the 2D spectrum in Fig. \ref{fig:spectrum_GII12bi}(c) with a respectable amplitude because the oscillations persist for much longer than the optical dephasing time along $\zeta = - \sigma$. The peak will gain amplitude as either or both of the variables are scanned to longer times. The peak has a very small anti-diagonal linewidth. Close inspection reveals that there are two peaks, one along the diagonal at each exciton frequency. This term has no $\tau$ dependence. 

We next study term $G_{II 22 \beta i}$, which is very similar to the previous term except it was of the pathological type, which lead to the expression 
\begin{equation}
I(\sigma, \tau, \zeta) = \Theta[-\sigma] \Theta[\zeta + \sigma] (\sigma + \zeta) e^{-i\Omega _{e_{2}g}(\sigma + \zeta)}e^{+i\Omega_{g g}\sigma}. 
\end{equation}

\begin{figure}
\centering
  \includegraphics[width=0.4\textwidth]{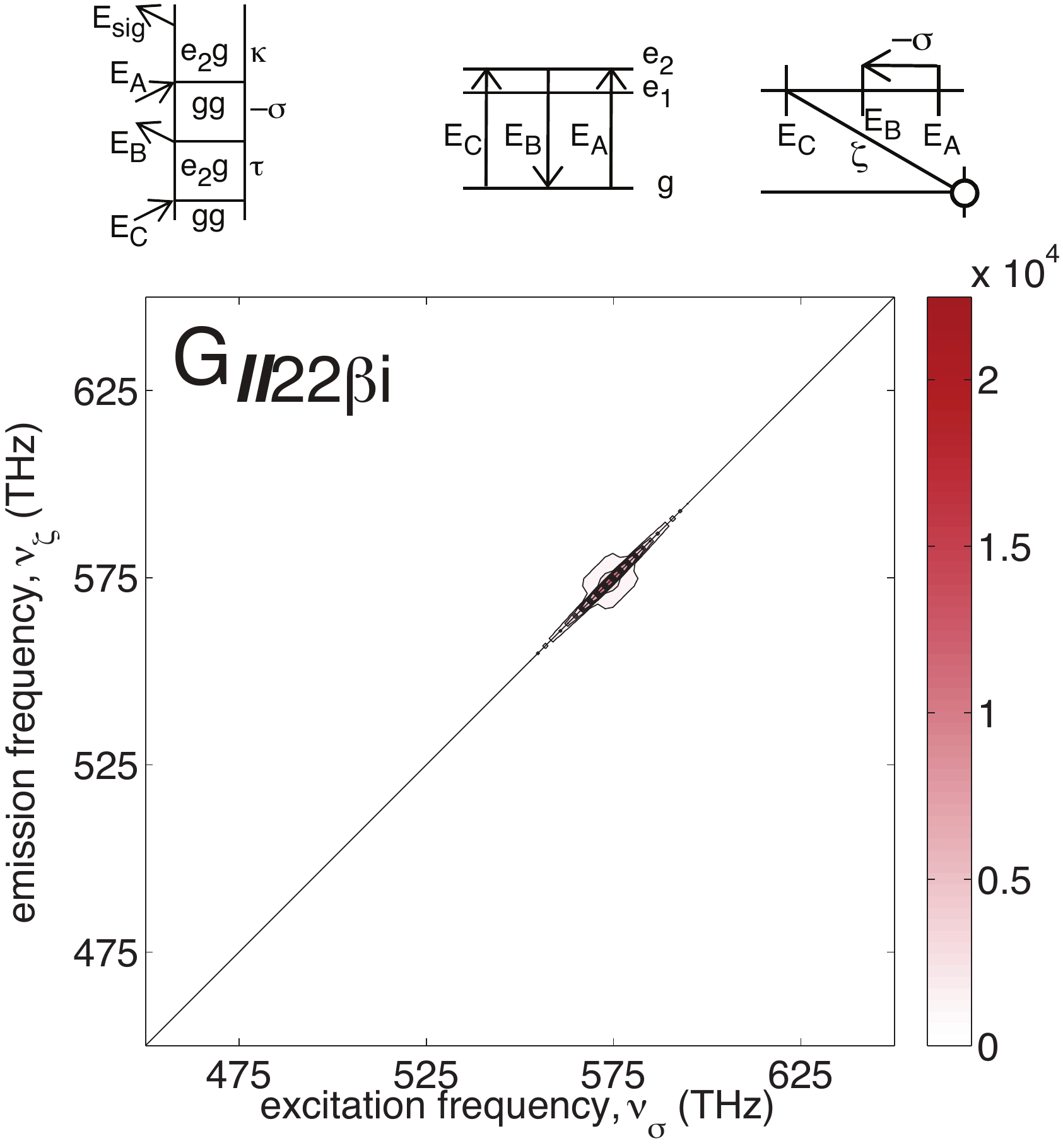} 
  \caption{The result for term $G_{II 22 \beta i}$. (top) The double-sided Feynman diagram, the WMEL diagram, and the FTC diagram for this term. (bottom) Absolute value I$^{(4)}$ 2D ES spectrum. This term contributes only along the diagonal near the higher-energy exciton state and is independent of time-delay variable $\tau$.} 
  \label{fig:spectrum_GII22bi}
\end{figure}

The resulting 2D spectrum is shown in Fig. \ref{fig:spectrum_GII22bi}.  This is one of the highest-amplitude terms in this topological class. This term contributes only along the diagonal near the higher-energy exciton state and is independent of time-delay variable $\tau$.  This peak will also gain amplitude as the delays $\sigma$ and $\zeta$ are scanned to longer times. 

\begin{figure}
\centering
  \includegraphics[width=0.5\textwidth]{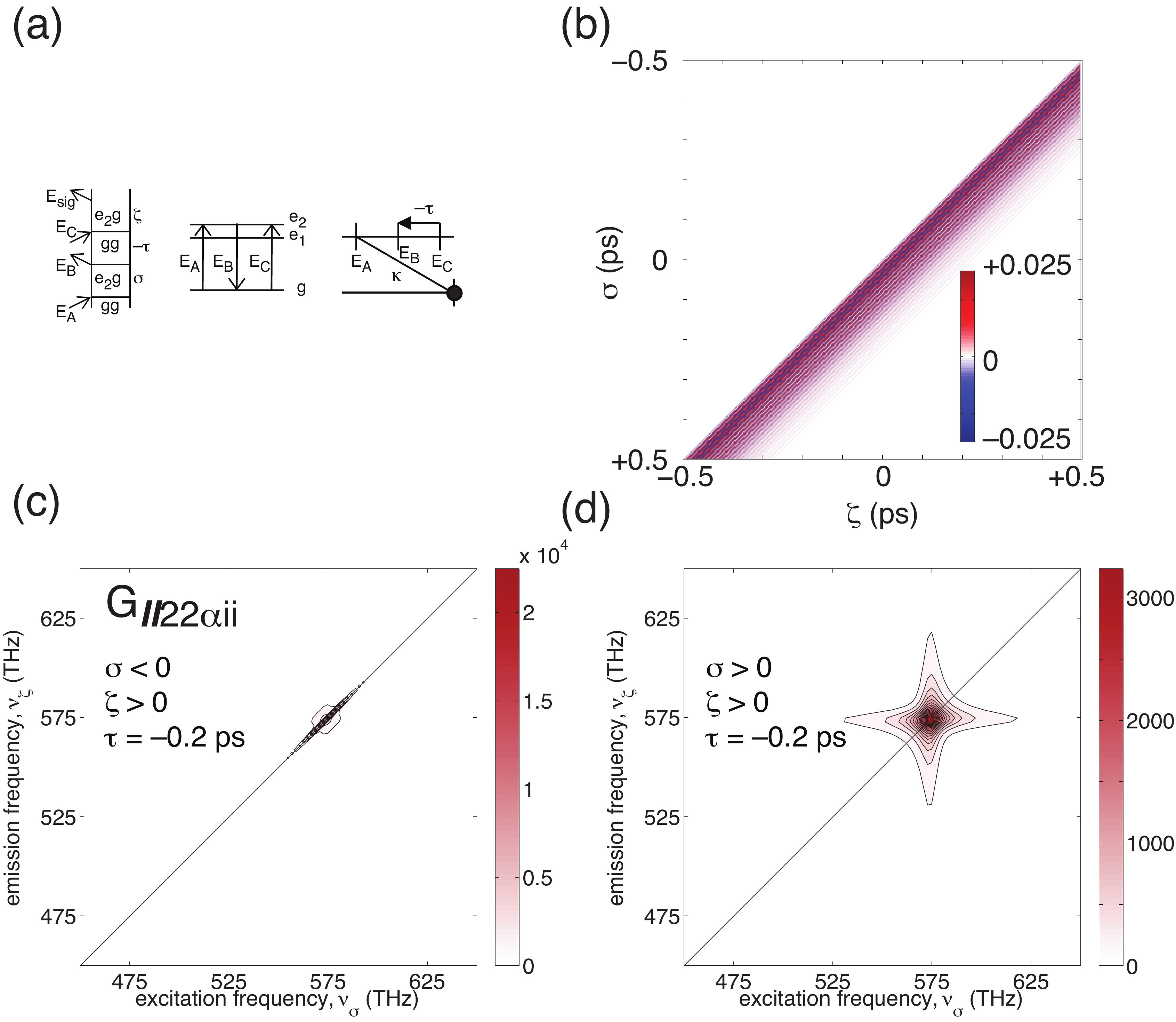} 
  \caption{The result for term $G_{II 22 \alpha ii}$. (a) The double-sided Feynman diagram, the WMEL diagram, and the FTC diagram for this term. (b) Time-domain signal. (c) Absolute value I$^{(4)}$ 2D ES nonrephasing spectrum. (d) Absolute value I$^{(4)}$ 2D ES rephasing spectrum. This term contributes only along the diagonal near the higher-energy exciton state and is independent of time-delay variable $\tau$.} 
  \label{fig:spectrum_GII22aii}
\end{figure}

The third term we consider is $G_{II 22 \alpha ii}$, where elimination of time-delay variable $\kappa$ leads to
\begin{equation}
I(\sigma,\tau,\zeta) = \Theta[-\tau] \Theta[\zeta + \sigma] (\zeta + \sigma) e^{-i\Omega _{e_{2}g}(\zeta + \sigma)}e^{+i\Omega_{g g}\tau}.
\end{equation}
Here the restraint on $\sigma$ is lifted, and it is replaced with the less-stringent restraint on $\tau$.  This means that the signal persists throughout the surface of the $\sigma$ and $\zeta$ variables, see Fig. \ref{fig:spectrum_GII22aii}(b). Like the previous terms, the amplitude is small but the signal can be acquired out to large values of $\zeta$ and $\sigma$. Here we display the resulting rephasing and nonrephasing spectra in Fig. \ref{fig:spectrum_GII22aii}(c) and (d), respectively. The shape of the peak in the nonrephasing spectrum is very similar---albeit at reduced amplitude---to many of the peaks that originate from terms in the unrestricted topological class.

\begin{figure}
\centering
  \includegraphics[width=0.4\textwidth]{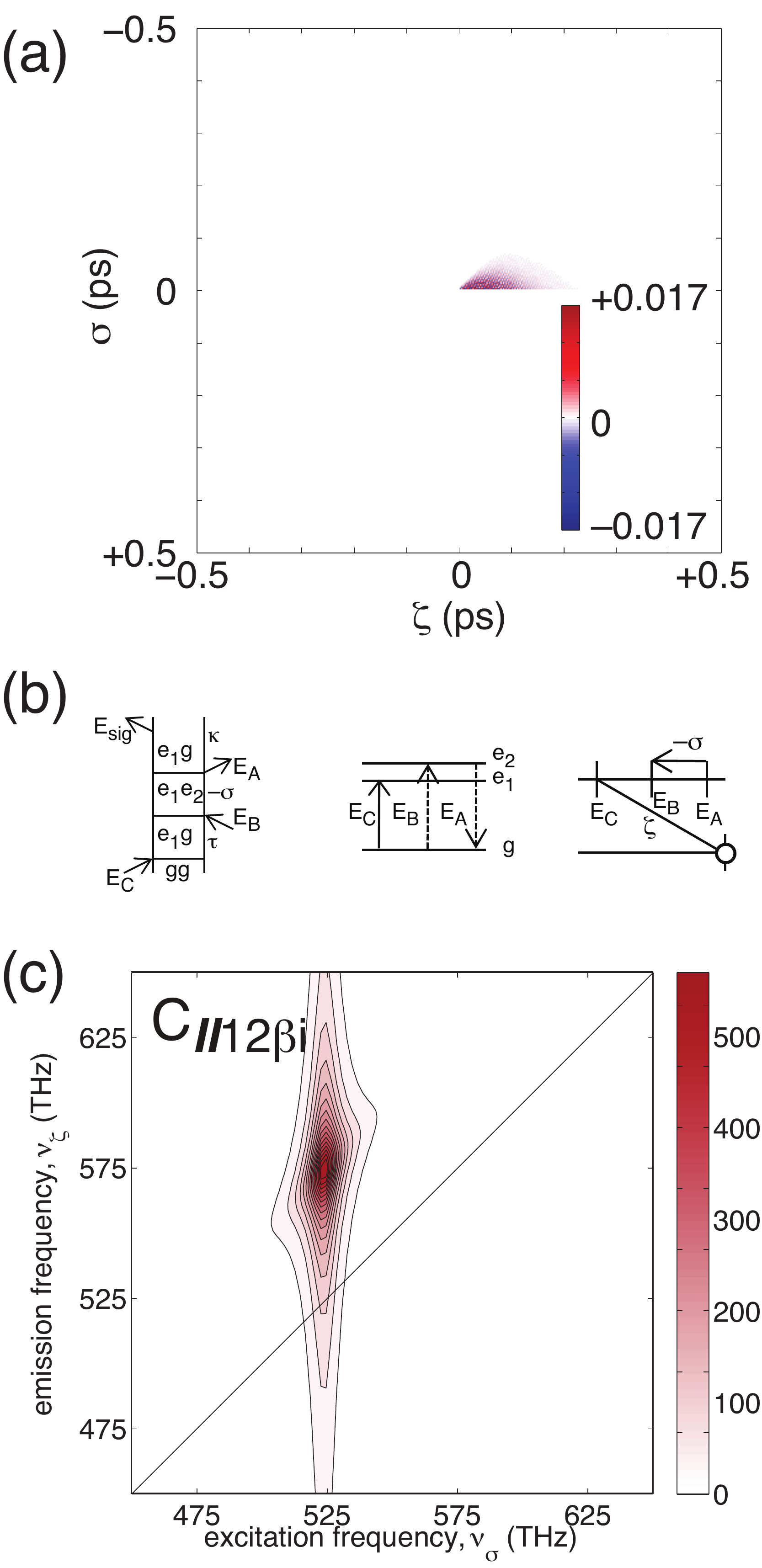} 
  \caption{The result for term $C_{I 12 \beta i}$. (a) The real part of the signal in the time domain.  (b) The double-sided Feynman diagram, the WMEL diagram, and the FTC diagram for this term. (c) Absolute value I$^{(4)}$ 2D ES spectrum. This term contributes only along the diagonal near the higher-energy exciton state and is independent of time-delay variable $\tau$.} 
  \label{fig:spectrum_CII12bi}
\end{figure}

The next term we consider is the first nonzero term that originated from a $C$ type response function. The time-domain data and the 2D spectrum for $C_{II 12 \beta i}$,
\begin{equation}
I(\sigma, \tau, \zeta) = \Theta[-\sigma] \Theta[\zeta + \sigma] (\sigma + \zeta) e^{-i\Omega _{e_{1}g}(\sigma + \zeta)}e^{+i\Omega_{e_1 e_2}\sigma}, 
\end{equation}
are shown in Fig. \ref{fig:spectrum_CII12bi}(a) and (c), respectively. 

The quantum beats of interest are convolved into $\sigma$, which leads to a shift of the peak into the off-diagonal position rather than coherent oscillations as a function of time-delay variable $\tau$. The signal amplitude is quite small, and since this term does not persist at long scan times, it will remain at a reduced amplitude in the frequency domain. There is no dependence on time-delay variable $\tau$ even though this a coherence pathway term. 

\begin{figure}
\centering
  \includegraphics[width=0.5\textwidth]{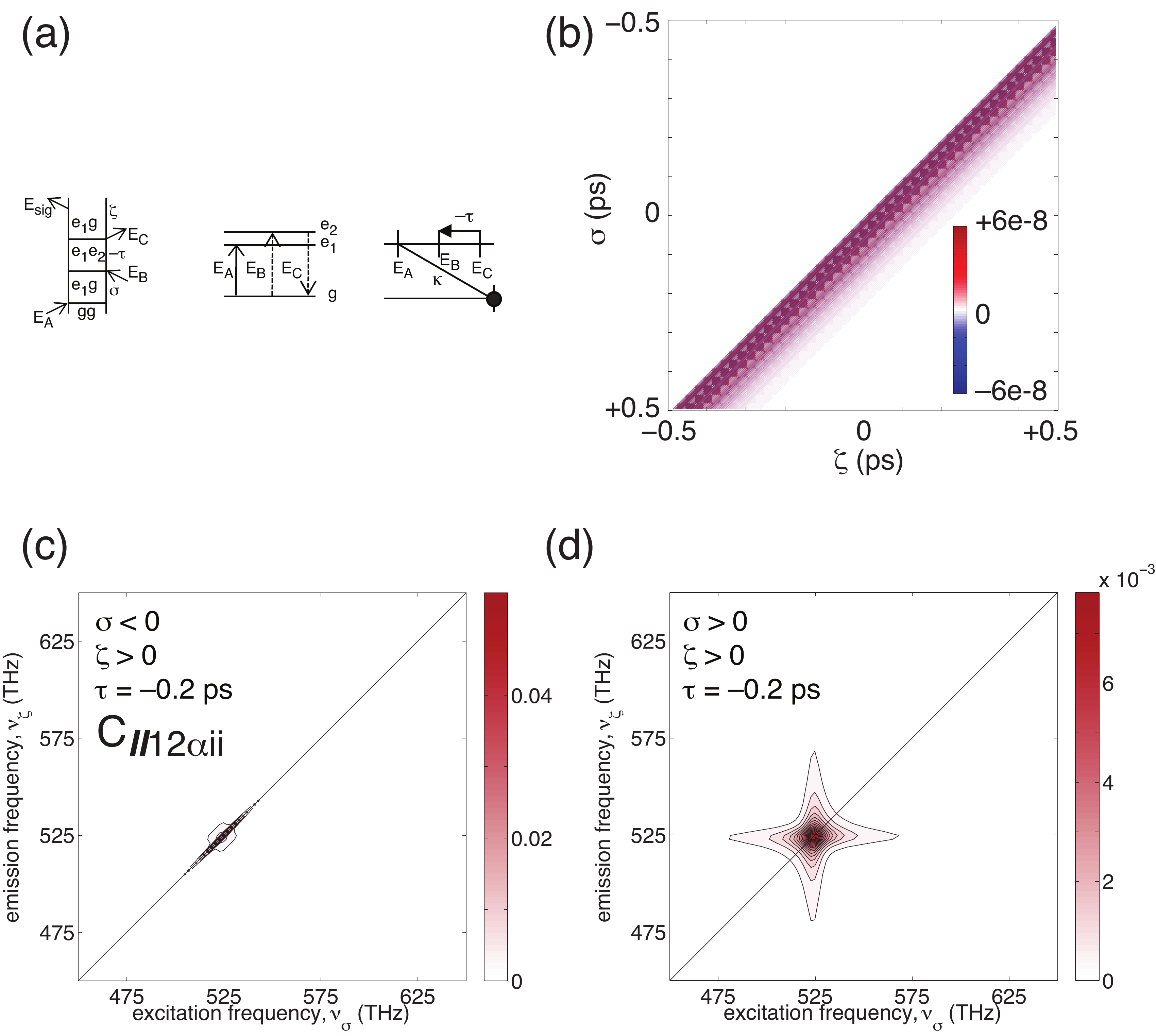} 
  \caption{The result for term $C_{II 12 \alpha ii}$. (top) The double-sided Feynman diagram, the WMEL diagram, and the FTC diagram for this term. (bottom) Absolute value I$^{(4)}$ 2D ES spectrum. These peaks oscillate as a function of time-delay variable $\tau$.} 
  \label{fig:spectrum_CII12aii}
\end{figure}

The fifth term of note in this topological class is $C_{II 12 \alpha ii}$, another $C$ type term, where elimination of $\kappa$ leads to
\begin{equation}
I(\sigma, \tau, \zeta) = \Theta[-\tau] \Theta[\zeta + \sigma] (\sigma + \zeta) e^{-i\Omega _{e_{1}g}(\sigma + \zeta)}e^{+i\Omega_{e_1 e_2}\tau}. 
\end{equation}
This is one of the most interesting terms of the experiment since it has true quantum beats during time period $\tau$.  This nonrephasing pathway, under time-delay conditions equivalent to a nonrephasing scan, appears almost identical to femtosecond measurement: it contributes to the diagonal peak in Fig. \ref{fig:spectrum_CII12aii}(d) and it oscillates as a function of $\tau$. In the noisy-light measurement, however, this term also contributes under the standard rephasing time-delay conditions and when $\zeta < 0$, both of which are not expected from femtosecond measurements, see Fig. \ref{fig:spectrum_CII12aii}(b). We show the case of standard rephasing conditions in Fig. \ref{fig:spectrum_CII12aii}(c). Unfortunately the contribution from this term is extremely weak; note the maximum amplitude in Fig. \ref{fig:spectrum_CII12aii}(d) is on the order of 10$^{-3}$ compared to many of the unrestricted terms that have amplitudes on the order of the 10$^7$. 

The final term of note in this class is $T_{21'1' \beta i}$, a term which originates from a two-quantum pathway,
\begin{equation}
I(\sigma, \tau, \zeta) = \Theta[\sigma] \Theta[\zeta - \sigma] (\sigma - \zeta) e^{-i\Omega _{e_{2}g}(\sigma - \zeta)}e^{-i\Omega_{fg}\sigma}. 
\end{equation}

\begin{figure}
\centering
  \includegraphics[width=0.3\textwidth]{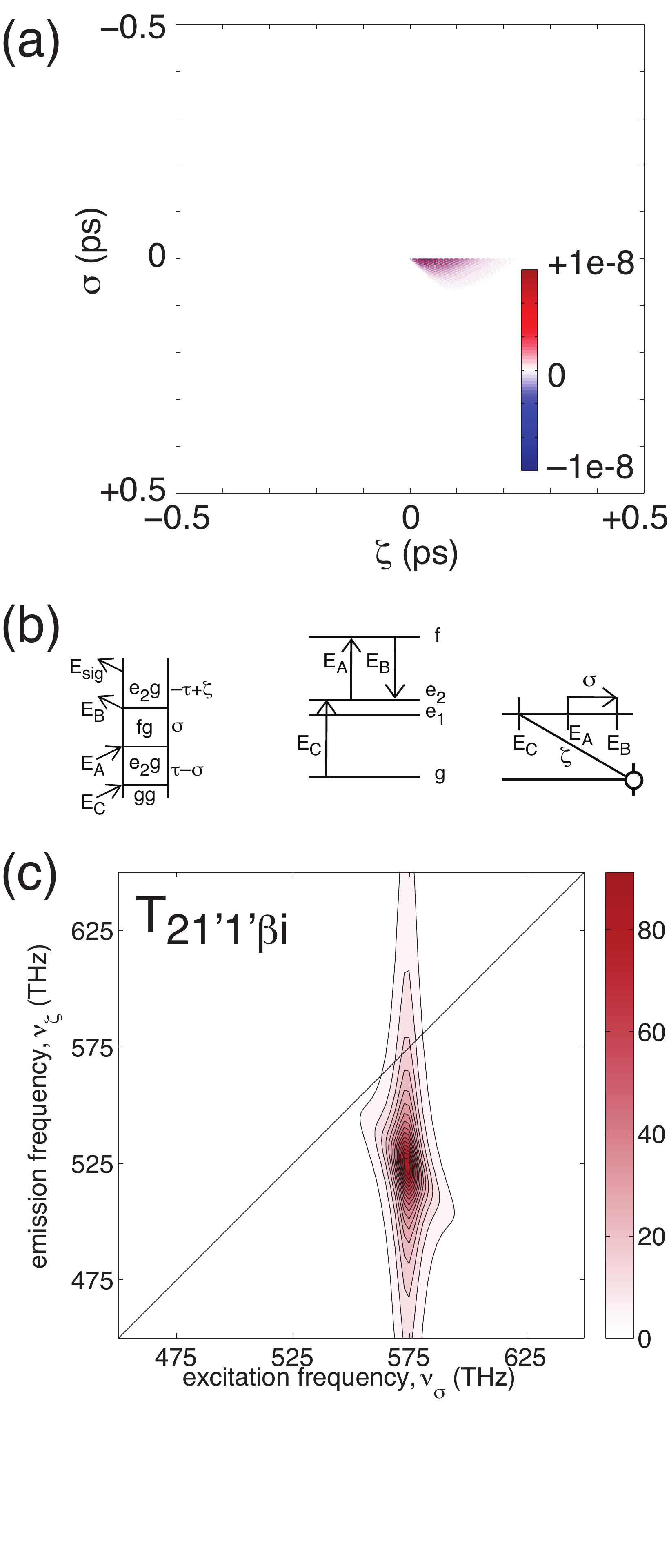} 
  \caption{The result for term $T_{21'1' \beta i}$. (a) The real part of the signal in the time domain. (b) The double-sided Feynman diagram, the WMEL diagram, and the FTC diagram for this term.  (c) Absolute value I$^{(4)}$ 2D ES nonrephasing spectrum.} 
  \label{fig:spectrum_T21p1pbi}
\end{figure}

The two-quantum frequency that makes this term interesting in femtosecond 2D ES is convolved with a one-quantum frequency during $\sigma$. This leads the peak to appear at the cross-peak position, but not at the two-quantum frequency. However, this term and the ones similar to it have an unusual, twisted peak shape, see Fig. \ref{fig:spectrum_T21p1pbi}(c). Similar to so many others, this term does not depend on the value of $\tau$.

\begin{figure} 
\centering 
  \includegraphics[width=0.5\textwidth]{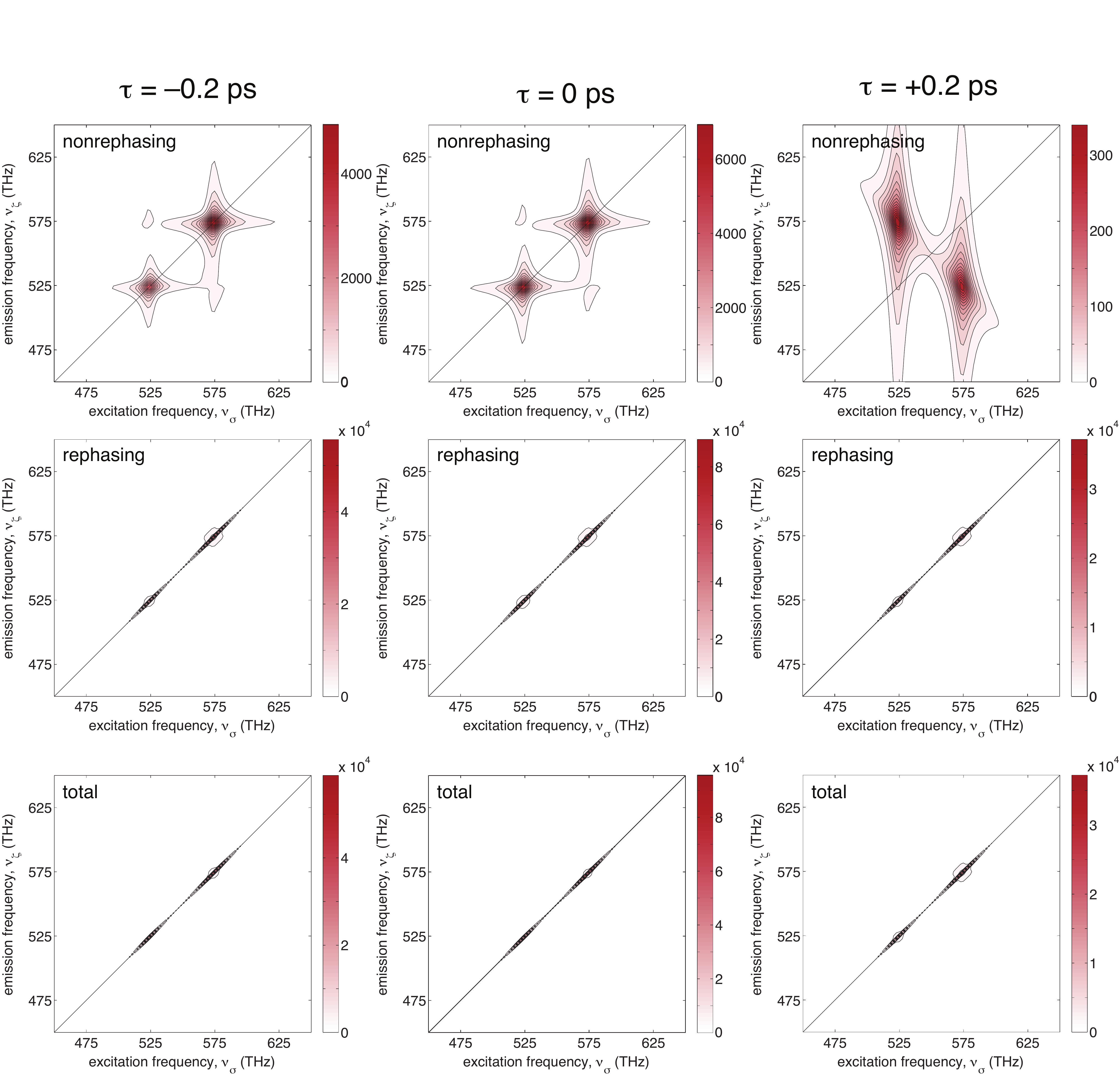}  
  \caption{The total result for the singly restricted topological class for three values of $\tau$. (top) The nonrephasing contributions. (middle) The rephasing contributions flipped into the $(+,+)$ quadrant. (bottom) The total I$^{(4)}$ 2D ES.} 
  \label{fig:spectrum_singly_restricted} 
\end{figure}

There are 40 nonzero terms in this class, and their contributions can be categorized by the four types of Heaviside restrictions, $\Theta[ - \sigma] \Theta[\zeta + \sigma]$, $\Theta[ \sigma] \Theta[\zeta - \sigma]$, $\Theta[ - \tau] \Theta[\kappa + \tau]$, and $\Theta[ \tau] \Theta[\kappa - \tau]$.  In Fig. \ref{fig:spectrum_singly_restricted} we show the total results under the same six different pulse-timing conditions used in the unrestricted topological class. The rephasing terms are about an order of magnitude stronger than the nonrephasing terms and thus dominate the total response. We have chosen to not present any results under which $\zeta$ is negative. 

Generally, terms in this topological class are much weaker than terms in the unrestricted class, and they can lead to diagonal and cross peaks. A few terms lead to peaks that oscillate due to true quantum beats, but they are quite weak.  Most terms do not oscillate or decay as a function of $\tau$.  There are no rephasing pathways in this topological class that contribute to the I$^{(4)}$ 2D ES signal. We have not shown the spectra that result from the signal that occurs at negative values of time-delay variable $\zeta$.

\subsection{Doubly Restricted Examples}
\begin{figure}
\centering
  \includegraphics[width=0.4\textwidth]{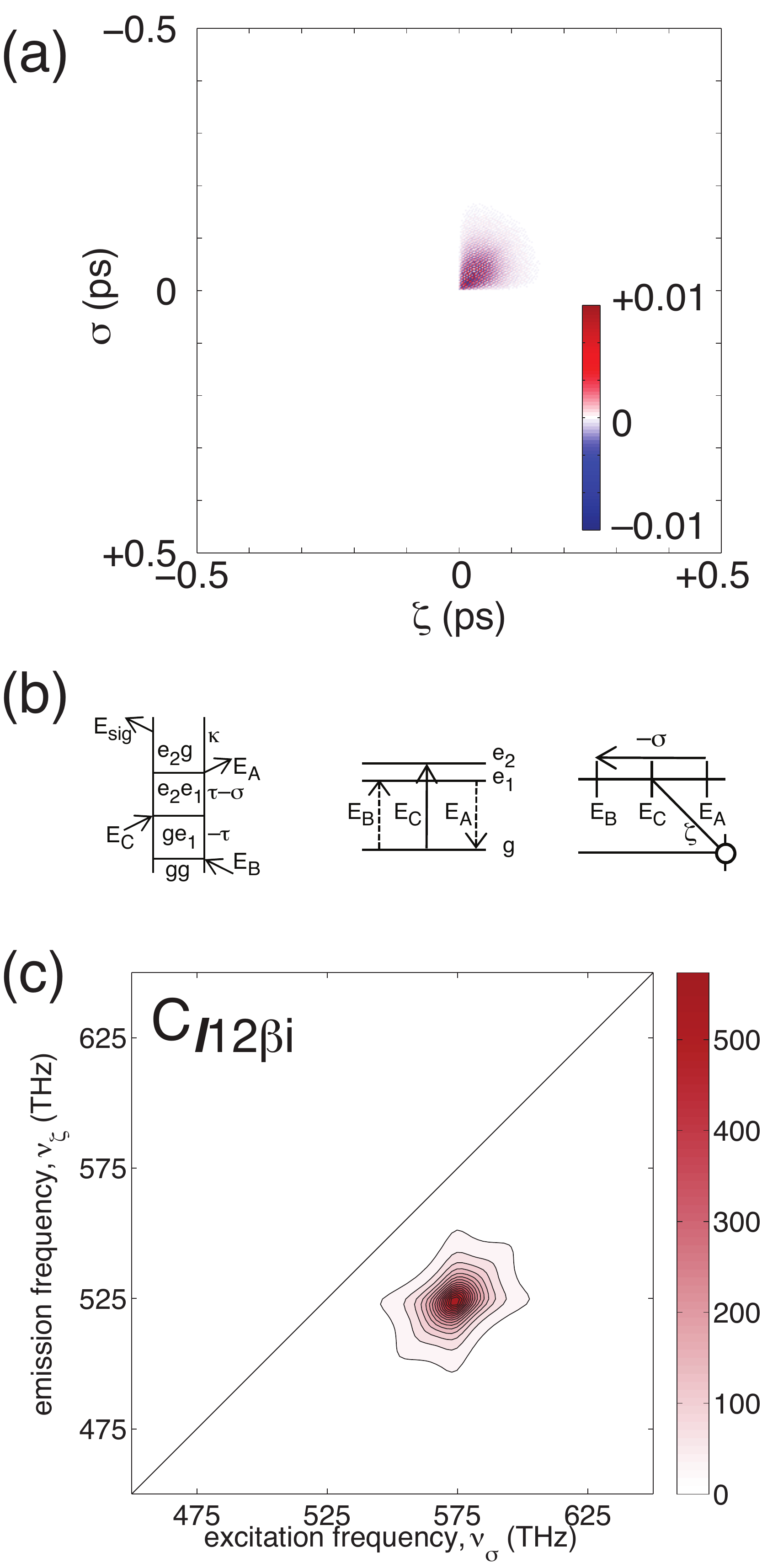} 
  \caption{The result for term $C_{I 12 \beta i}$. (a) The real part of the signal in the time domain.  (b) The double-sided Feynman diagram, the WMEL diagram, and the FTC diagram for this term. (c) Absolute value I$^{(4)}$ 2D ES spectrum.} 
  \label{fig:spectrum_CI12bi}
\end{figure}

The first term we consider in the doubly restricted topological class is the coherence term $C_{I 12 \beta i}$. Its expression is 
\begin{eqnarray}
I & = &  \frac{ \Theta[\zeta] \Theta[-\sigma] \rho _{0}\vert \mu _{1} \vert^{2} \vert \mu _{2} \vert^{2} I_0^2}{\hbar ^{3}(\Omega_{e_{2} g} - \Omega_{e_{2}e_{1}} + \Omega_{g e_{1}})} \Bigg\{ \Theta[\sigma + \zeta]  \nonumber \\
& & \times \bigg[e^{+i \Omega_{e_{2}e_{1}} \sigma} e^{-i \Omega_{e_{2} g} (\zeta + \sigma)} - e^{+i \Omega_{ge_{1}} \sigma} e^{-i \Omega_{e_{2} g} \zeta}\bigg] \nonumber \\
& & + \Theta[-\sigma - \zeta] \nonumber \\
& & \times \bigg[e^{-i \Omega_{e_2 e_1} \zeta} e^{+i \Omega_{ge_{1}} (\zeta + \sigma)}  -  e^{+i \Omega_{ge_{1}} \sigma} e^{-i \Omega_{e_{2} g} \zeta} \bigg] \Bigg\}, \qquad
\end{eqnarray} 
and the time-domain signal and rephasing 2D spectrum for this term are presented in Fig. \ref{fig:spectrum_CI12bi}(a) and (c), respectively. This term is only active under typical rephasing conditions when $\zeta > 0$ and $\sigma < 0$. The peak is located in the off-diagonal position, but as the expression shows, this term does not oscillate as a function of time-delay variable $\tau$. The shape of the peak is quite unusual due to the symmetry in the time-domain signal about the line $\zeta = \sigma$. The amplitude of this term is very small, and increasing the size of the temporal scan range will not increase its amplitude. 

\begin{figure}
\centering
  \includegraphics[width=0.4\textwidth]{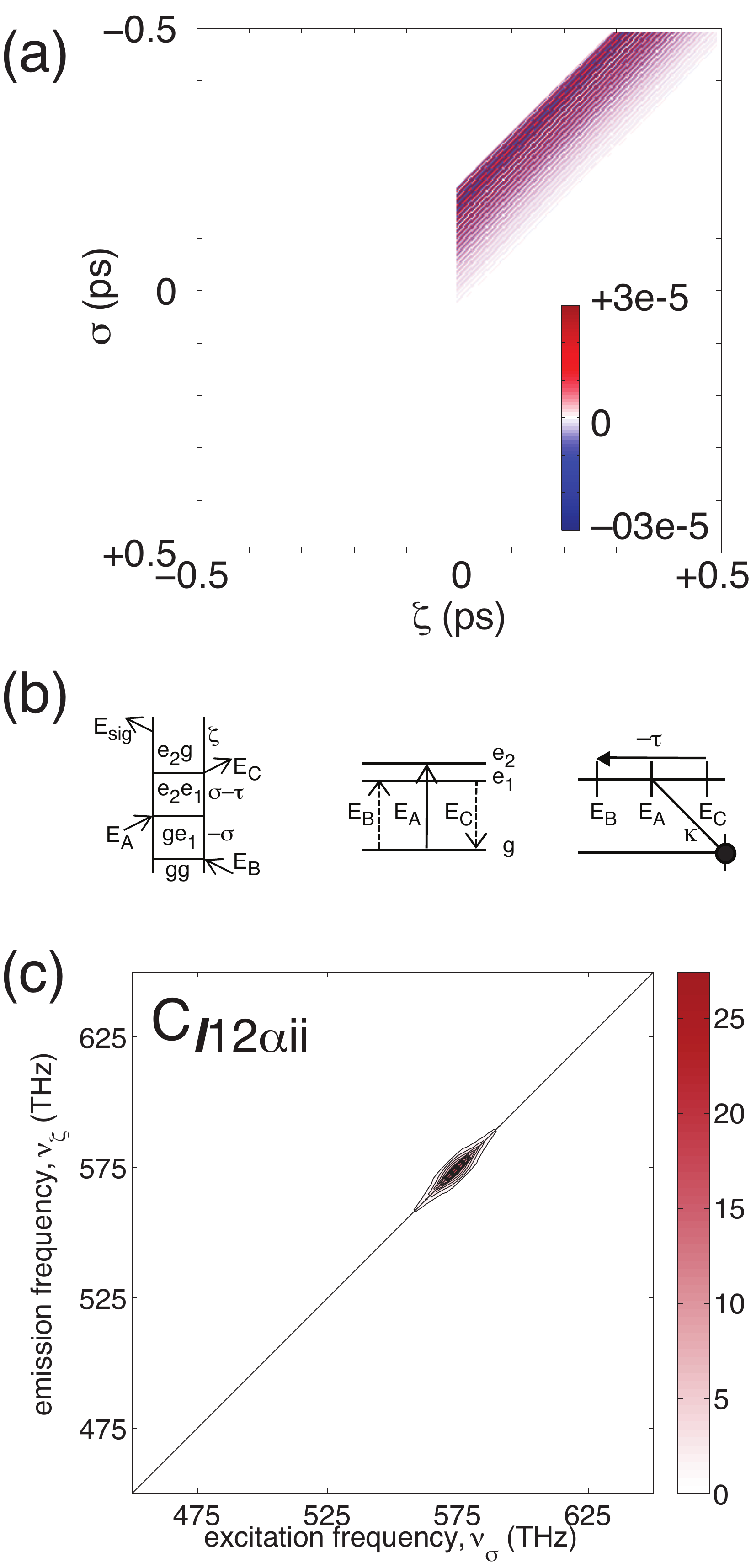} 
  \caption{The result for term $C_{I 12 \alpha ii}$. (a) The real part of the signal in the time domain.  (b) The double-sided Feynman diagram, the WMEL diagram, and the FTC diagram for this term. (c) Absolute value I$^{(4)}$ 2D ES spectrum.} 
  \label{fig:spectrum_CI12aii}
\end{figure}

The expression for the second term of note in this class is $C_{I 12 \alpha ii}$, whose expression is  
\begin{eqnarray}
I & = &  \frac{ \Theta[\kappa] \Theta[-\tau] \rho _{0}\vert \mu _{1} \vert^{2} \vert \mu _{2} \vert^{2} I_0^2}{\hbar ^{3}(\Omega_{e_{2} g} - \Omega_{e_{2}e_{1}} + \Omega_{g e_{1}})} \Bigg\{ \Theta[\tau + \kappa]  \nonumber \\
& & \times \bigg[e^{+i \Omega_{e_{2}e_{1}} \tau} e^{-i \Omega_{e_{2} g} (\kappa + \tau)} - e^{+i \Omega_{ge_{1}} \tau} e^{-i \Omega_{e_{2} g} \kappa}\bigg] \nonumber \\
& & + \Theta[-\tau - \kappa] \nonumber \\
& & \times \bigg[e^{-i \Omega_{e_2 e_1} \kappa} e^{+i \Omega_{ge_{1}} (\tau + \kappa)}  -  e^{+i \Omega_{ge_{1}} \tau} e^{-i \Omega_{e_{2} g} \kappa} \bigg] \Bigg\}, \qquad
\end{eqnarray} 
where, for brevity, in this case we have chosen not to eliminate time-delay variable $\kappa$ from the expression. Nevertheless, the simulation was performed in the space of $\sigma$, $\tau$, and $\zeta$. The signal and resulting 2D spectrum are shown in Fig. \ref{fig:spectrum_CI12aii}. Although this term does contain quantum-beat oscillations in the factor of $e^{-\Omega_{e_1 e_2} \tau}$ in the first part of the expression, the signal is very weak. Its contribution to the 2D spectrum in principle could be increased by scanning time variables $\sigma$ and $\zeta$ to larger values. The resultant peak has a very small anti-diagonal line width reminiscent of several terms in the singly restricted topological class. 

We turn to term $G_{I 12 \beta i}$ next
\begin{eqnarray}
I & = &  \frac{ \Theta[\zeta] \Theta[-\sigma] \rho _{0}\vert \mu _{1} \vert^{2} \vert \mu _{2} \vert^{2} I_0^2}{\hbar ^{3}(\Omega_{e_{2} g} - \Omega_{gg} + \Omega_{g e_{1}})} \Bigg\{ \Theta[\sigma + \zeta]  \nonumber \\
& & \times \bigg[e^{+i \Omega_{gg} \sigma} e^{-i \Omega_{e_{2} g} (\zeta + \sigma)} - e^{+i \Omega_{ge_{1}} \sigma} e^{-i \Omega_{e_{2} g} \zeta}\bigg] \nonumber \\
& & + \Theta[-\sigma - \zeta] \nonumber \\
& & \times \bigg[e^{-i \Omega_{gg} \zeta} e^{+i \Omega_{ge_{1}} (\zeta + \sigma)}  -  e^{+i \Omega_{ge_{1}} \sigma} e^{-i \Omega_{e_{2} g} \zeta} \bigg] \Bigg\}. \qquad
\end{eqnarray} 
This term is an example of one of the stronger terms in this class, primarily because the signal persists about the line $\zeta = -\sigma$ to large values of either variable. Moreover, the peak amplitude in the 2D spectrum is large because the signal is compressed about the diagonal. There is the a second portion of the signal---small and rapidly decaying---that leads to the weak cross peak. 

\begin{figure}
\centering
  \includegraphics[width=0.4\textwidth]{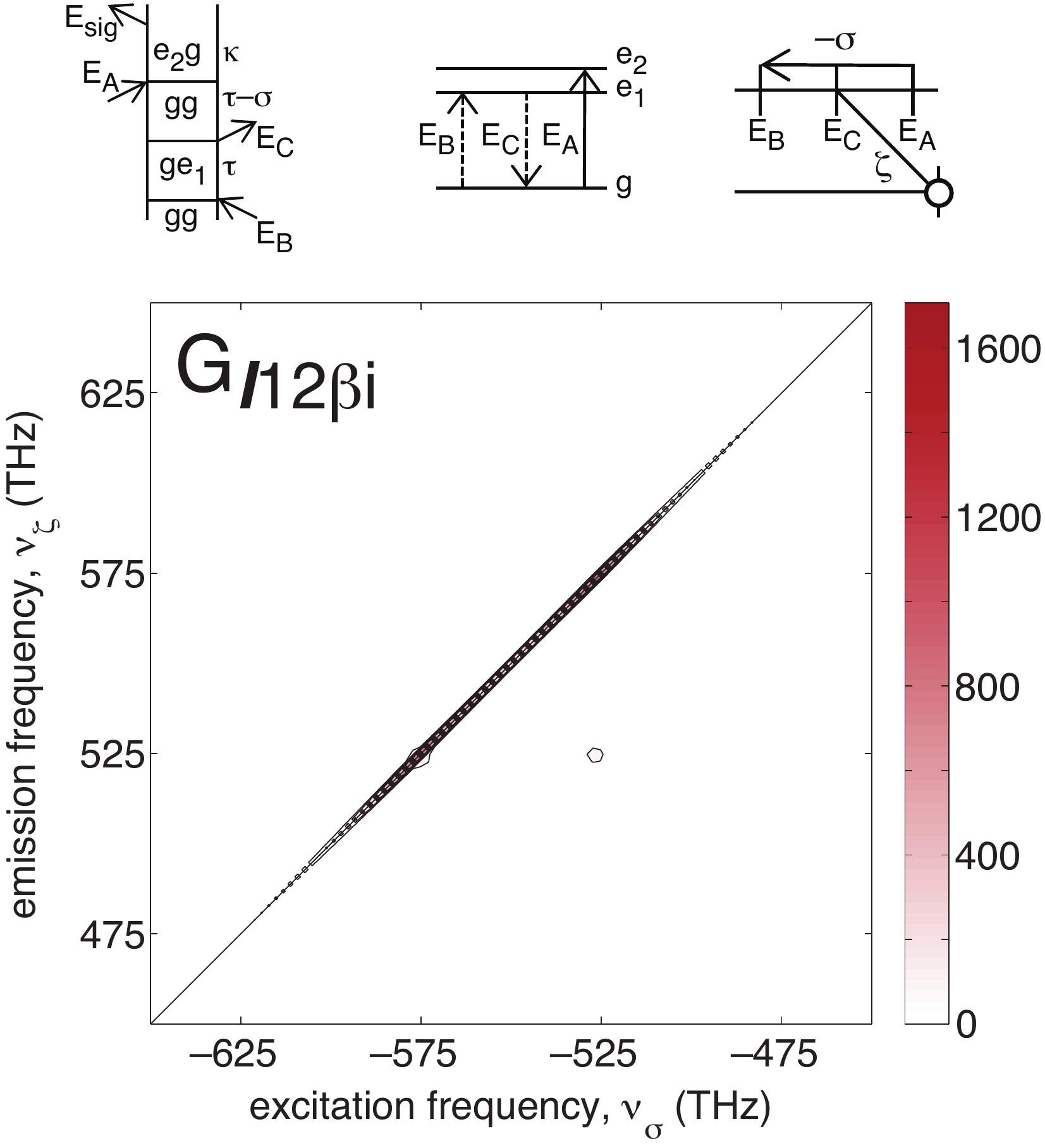} 
  \caption{The result for term $G_{I 12 \beta i}$. (top) The double-sided Feynman diagram, the WMEL diagram, and the FTC diagram for this term. (bottom) Absolute value I$^{(4)}$ 2D ES spectrum.} 
  \label{fig:spectrum_GI12bi}
\end{figure}

\begin{figure}
\centering
  \includegraphics[width=0.4\textwidth]{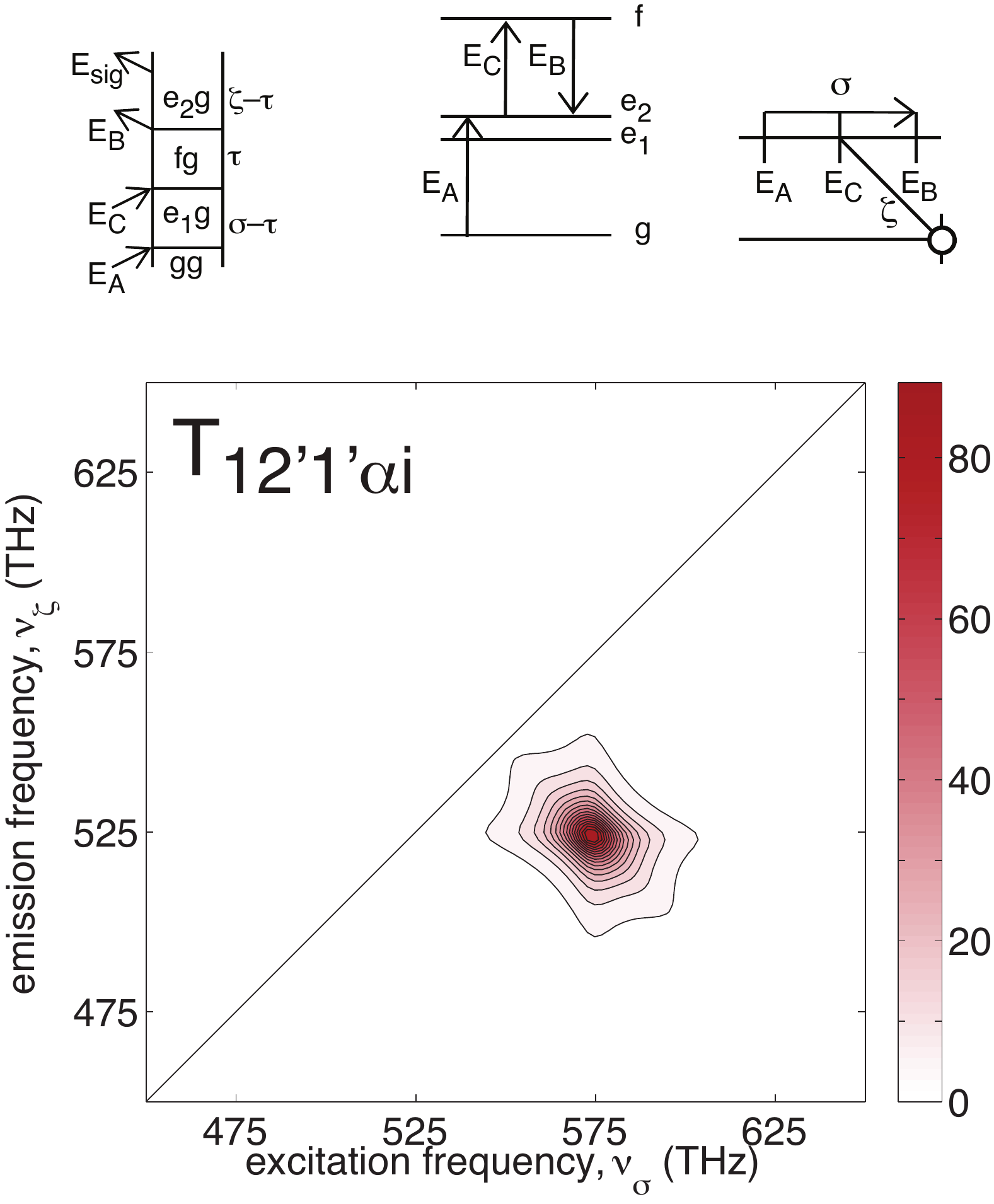} 
  \caption{The result for term $T_{12'1' \alpha i}$. (top) The double-sided Feynman diagram, the WMEL diagram, and the FTC diagram for this term. (bottom) Absolute value I$^{(4)}$ 2D ES spectrum.} 
  \label{fig:spectrum_T12'1'ai}
\end{figure}

The final example is term $T_{12'1' \alpha i}$, 
\begin{eqnarray}
I & = &  \frac{ \Theta[\zeta] \Theta[\sigma] \rho _{0}\vert \mu _{1} \vert^{2} \vert \mu _{2} \vert^{2} I_0^2}{\hbar ^{3}(\Omega_{e_{1} g} - \Omega_{fg} + \Omega_{e_{2} g})} \Bigg\{ \Theta[ \zeta - \sigma]  \nonumber \\
& & \times \bigg[e^{-i \Omega_{fg} \sigma} e^{-i \Omega_{e_{2} g} (\zeta - \sigma)} - e^{-i \Omega_{e_{1} g} \sigma} e^{-i \Omega_{e_{2} g} \zeta}\bigg] \nonumber \\
& & + \Theta[\sigma - \zeta] \nonumber \\
& & \times \bigg[e^{-i \Omega_{fg} \zeta} e^{-i \Omega_{e_{1} g} (\sigma - \zeta)}  -  e^{-i \Omega_{e_{1} g} \sigma} e^{-i \Omega_{e_{2} g} \zeta} \bigg] \Bigg\}. \qquad
\end{eqnarray} 
This term originates from a two-quantum Liouville pathway, and it leads to a small-amplitude cross peak in the 2D spectrum, see Fig. \ref{fig:spectrum_T12'1'ai}. As with the other similar terms, this term has no $\tau$ dependence, it is very small in amplitude, and its unusual peak shape is caused by the peculiar response in the time domain. 

\begin{figure} 
\centering 
  \includegraphics[width=0.5\textwidth]{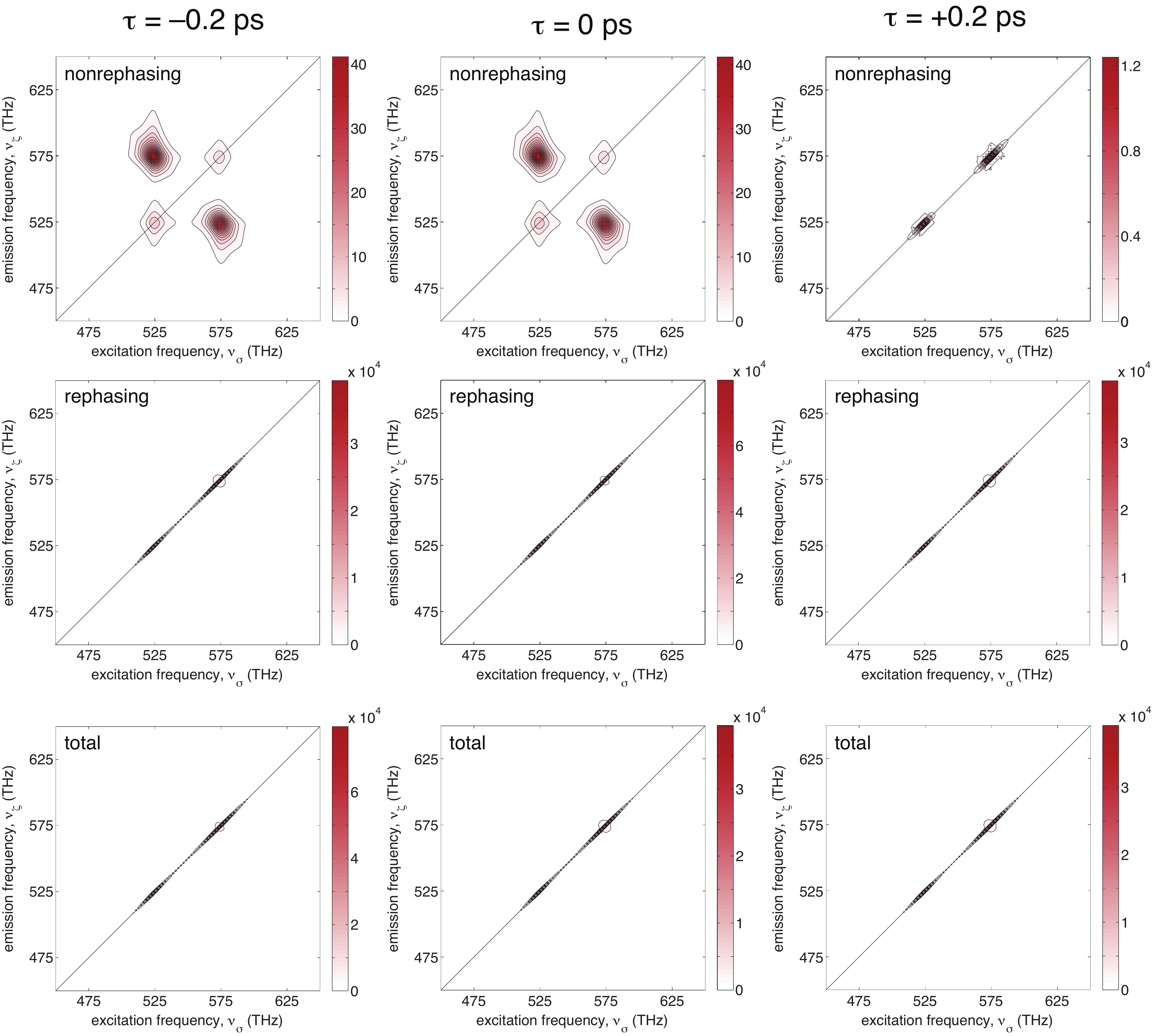}  
  \caption{The total result for the doubly restricted topological class for three values of $\tau$. (top) The nonrephasing contributions, meaning the contributions to the total signal under temporal scan conditions analogous to a femtosecond nonrephasing 2D measurement. (middle) The rephasing contributions flipped into the $(+,+)$ quadrant. (bottom) The total I$^{(4)}$ 2D ES.} 
  \label{fig:spectrum_doubly_restricted} 
\end{figure}

There are 40 nonzero terms in this class, and their contributions can be categorized again by the types of Heaviside restrictions.  In Fig. \ref{fig:spectrum_doubly_restricted} we show the total results under the same six different pulse-timing conditions used in the previous topological classes. The response under rephasing conditions is about an order of magnitude stronger than response under nonrephasing conditions (where the two-quantum pathways contribute), and thus the rephasing contributions dominate the total doubly-restricted response. 

Generally, terms in this topological class are much weaker than terms in the unrestricted class, and they can lead to diagonal and cross peaks. Most terms, including all the large-amplitude rephasing contributions, appear as narrow stripes along the diagonal. A few terms lead to peaks that oscillate due to true quantum beats, but they are again weak.  There are no one-quantum nonrephasing pathways in this topological class that contribute to the I$^{(4)}$ 2D ES signal.

\subsection{Total Signal}
An experiment would measure the total signal---meaning the sum of all 128 terms---regardless of topology. Thus we could present the total signal expected for typical rephasing and nonrephasing experiments under the various values of $\tau$ used above. However, inspection of the relative magnitudes shows that the entire class of singly restricted terms and the entire class of doubly restricted terms amounts to total signal on the order of $10^4$, whereas the unrestricted amplitudes are on the order of $10^8$. Thus the unrestricted terms completely dominate the response, and the total signal simply replicates those presented in Fig. \ref{fig:spectrum_unrestricted}.

\subsection{Two-level System I$^{(4)}$ 2D ES}
In the process of solving the Bloch four-level system, we have also solved a simpler problem, I$^{(4)}$ 2D ES for the two-level system. Many signal components for the Bloch four-level system do not exist for this system. In the end, only two $G$ type and two $P$ type Liouville pathways remain for $\alpha$ and $\beta$ types. This leads to a total of only sixteen FTC diagrams. For brevity, we do not present the expected signal.

\section{Conclusions}
Linear absorption spectroscopy is, formally, a coherent measurement technique. A linear absorption spectrum can be measured using a femtosecond laser and a suitable detector. Yet, it is well known that this measurement does not \emph{require} a coherent light source, and in fact incoherent light sources are extremely common for measurements in the visible and infrared portions of the spectrum. This is a powerful reminder that the coherence property of the light source plays a role less important than perhaps expected in coherent spectroscopy measurements. 

Our results suggest that there is value in attempting to measure a 2D spectrum using incoherent light. To be sure there will be a number of challenges to collecting the I$^{(4)}$ 2D ES data, especially if the goal is to observe coherent quantum beats. First, the interpretation of the experiment is not as straightforward as the femtosecond case because of the cw nature of noisy light. Second, methods to suppress the signals arising from terms in the unrestricted topological class will be needed to resolve the terms that produce quantum-beat signals clearly. Third, a noisy-light source with a significantly broader spectrum than any used to date will be needed. 
 
We hope that the connections between I$^{(4)}$ 2D ES and its femtosecond analogue will provide new insights into the interpretation of the corresponding femtosecond experiments.  The results presented here show that I$^{(4)}$ 2D ES should contain diagonal and cross peaks similar to femtosecond 2D ES, and they show that quantum beating persists in the third-order nonlinear response even under incoherent excitation. This work addresses \emph{in an experimentally verifiable manner} the important question of whether or not coherent quantum beating between exciton states is an artifact arising from the coherence properties of the light used to probe light-harvesting proteins in femtosecond measurements.

\newpage 
\mbox{ }

\section{Appendix}
The FTC diagram analysis presented in this work greatly reduces the amount of mathematical effort required to produce an expression for each component of the final signal. Although there is a formal isomorphism between the mathematical expressions and the FTC diagrams \cite{Biebighauser:2002aa,Biebighauser:2003aa}, it is informative to include selected analytic calculations in addition to the FTC diagram analysis. In this appendix we demonstrate the lengthy integration and algebra required for four of the 128 signal components, one of which is zero. The approach is to insert the response function and electric fields into Eqn. \ref{eqn:fullInt} and the goal is to arrive at the expression in the Tables that correspond to the terms  $G_{II 11 \alpha i}$, $G_{II 12 \beta i}$, $C_{I 1 2 \alpha ii}$, and $C_{I 1 2 \alpha i}$. We reiterate that while perhaps new to readers more familiar with femtosecond-pulse measurements, the assumptions and protocols used here are based in almost thirty years of experimental and theoretical work on noisy-light spectroscopy. 

\subsection{Term $G_{II 11 \alpha i}$}
This is one of the simplest terms. The response function for $G_{II 11 \alpha i}$---and in fact for all four $G_{II 11}$ terms---is
\begin{eqnarray}
R^{(3)}(t,t_1,t_2,t_3) & = & \left( \frac{i}{\hbar} \right)^3 \rho_{0} \vert \mu_1 \vert^4 e^{-i\Omega _{e_1 g}(t_{2} - t_{1})} \nonumber \\
& & \times e^{-i\Omega_{g g}(t_{3} - t_{2})}e^{-i\Omega _{e_1 g}(t - t_{3})}, 
\label{eqn:GII11_response} 
\end{eqnarray}
where the transition dipole $\mu_1$ corresponds to the label in Fig. \ref{fig:elevels}. In noisy-light spectroscopy, in addition to the response function, we must account for the quasi-cw and stochastic natures of the fields. There are four major steps. We first insert the response function into Eqn. \ref{eqn:fullInt}, 
\begin{eqnarray}
I(\sigma, \tau, \zeta) & = &\int_{-\infty }^{t}dt_{3}\int_{-\infty}^{t_{3}}dt_{2}\int_{-\infty }^{t_{2}}dt_{1} \langle E_{A}E_{B}^{\ast}E_{C}E_{LO}^{\ast }\rangle  \nonumber \\
& & \times \frac{\rho_{0} \vert \mu_1 \vert^4}{i \hbar^3} e^{-i\Omega _{e_1 g}(t_{2} - t_{1})} \nonumber \\
& & \times e^{-i\Omega_{g g}(t_{3} - t_{2})} e^{-i\Omega _{e_1 g}(t - t_{3})}.
\end{eqnarray}
Notice how the signal, $I$, is a function of the experimental time-delay variables $\sigma$, $\tau$, and $\zeta$. The interaction-time variables internal to the response function, $t_1$, $t_2$, $t_3$, and $t$, will disappear under integration. In femtosecond 2D ES there is a clear correlation between the experimental time-delay induced by, for example, a retroreflector, and the time period between field-matter interactions. In noisy-light, we must account for all possible interaction times for each experimental set of time-delay variables. While we must account for all possible interaction events, the time-delay variables are under experimental control. 

In the second step we reduce the four-point time correlator into a pair of two-point time correlators via the assumption that the fields obey circular complex Gaussian statistics. This leads to two separate terms: $G_{II 11 \alpha i}$ and $G_{II 11 \alpha ii}$. We select the two-point correlator pair that corresponds to $G_{II 11 \alpha i}$, 
\begin{eqnarray}
I(\sigma, \tau, \zeta) & = &\int_{-\infty }^{t}dt_{3}\int_{-\infty}^{t_{3}}dt_{2}\int_{-\infty }^{t_{2}}dt_{1} \langle E_{A}E_{B}^{\ast }\rangle \langle E_{C}E_{LO}^{\ast} \rangle \nonumber \\
& & \times \frac{\rho_{0} \vert \mu_1 \vert^4}{i \hbar^3} e^{-i\Omega _{e_1 g}(t_{2} - t_{1})} \nonumber \\
& & \times e^{-i\Omega_{gg}(t_{3} - t_{2})} e^{-i\Omega _{e_1 g}(t - t_{3})}.
\label{eqn:GII11ai_insertion}
\end{eqnarray}
Notice how the upper limits of integration are not $+ \infty$ but rather the next field-matter interaction because---for the particular Liouville pathway of interest---the field-matter interactions must have occurred in a specific order.  We then assign each field to a particular interaction time according to the time-ordered Liouville pathway depicted by the appropriate diagrams in Fig. \ref{fig:feynmans}. As a nonrephasing pathway with an $\alpha$ subscript, field $E_A(t_1)$, $E_B(t_2)$, and $E_C(t_3)$. Inserting the expression for each field according to Eqns. \ref{eqn:fields} into Eqn. \ref{eqn:GII11ai_insertion} leads to
\begin{eqnarray}
I & = &\int_{-\infty }^{t}dt_{3}\int_{-\infty}^{t_{3}}dt_{2}\int_{-\infty }^{t_{2}}dt_{1} \nonumber \\
& & \times \langle E_{0} p(t_{1}) e^{-i \omega t_{1}} E^{\ast}_{0} p^{\ast}(t_{2} - \sigma) e^{+i \omega (t_{2} - \sigma)}  \rangle \nonumber \\
& & \times \langle E_{0} p(t_{3} + \tau - \sigma) e^{-i \omega (t_{3} + \tau - \sigma)} E^{\ast}_{0} p^{\ast}(s - \kappa) e^{+i \omega (s - \kappa)}  \rangle   \nonumber \\
& & \times \frac{\rho_{0} \vert \mu_1 \vert^4}{i \hbar^3} e^{-i\Omega _{e_1 g}(t_{2} - t_{1})}  \nonumber \\
& & \times e^{-i\Omega_{gg}(t_{3} - t_{2})} e^{-i\Omega _{e_1 g}(t - t_{3})}.
\end{eqnarray}
Reorganization yields
\begin{eqnarray}
I & = & \frac{ \rho_{0} \vert \mu_1 \vert^4 I^{2}_{0}}{i \hbar^3} \int_{-\infty }^{t}dt_{3}\int_{-\infty}^{t_{3}}dt_{2}\int_{-\infty }^{t_{2}}dt_{1} \nonumber \\
& & \times \langle p(t_{1})  p^{\ast}(t_{2} - \sigma)  \rangle  \langle p(t_{3} + \tau - \sigma) p^{\ast}(s - \kappa)   \rangle  \nonumber \\
& & \times e^{-i \omega t_{1}} e^{+i \omega (t_{2} - \sigma)} e^{-i \omega (t_{3} + \tau - \sigma)} e^{+i \omega (s - \kappa)}  \nonumber \\
& & \times e^{-i\Omega _{e_1 g}(t_{2} - t_{1})} e^{-i\Omega_{gg}(t_{3} - t_{2})} e^{-i\Omega _{e_1 g}(t - t_{3})}, \quad
\end{eqnarray}
where $I_{0} = E_{0} E_{0}^{\ast}$. In the fourth major step, we approximate the noisy fields as being fully incoherent---their coherence times goes to zero---which allows us to replace the two-point correlators with $\delta$-functions, 
\begin{eqnarray}
I & = & \frac{\rho_{0} \vert \mu_1 \vert^4  I^{2}_{0}}{i \hbar^3} \int_{-\infty }^{t}dt_{3}\int_{-\infty}^{t_{3}}dt_{2}\int_{-\infty }^{t_{2}}dt_{1} \nonumber \\
& & \times \delta (t_{1} - t_{2} + \sigma ) \delta (t_{3} + \tau -\sigma - s + \kappa) \nonumber \\
& & \times e^{-i \omega t_{1}} e^{+i \omega (t_{2} - \sigma)} e^{-i \omega (t_{3} + \tau - \sigma)} e^{+i \omega (s - \kappa)} \nonumber \\
& & \times e^{-i\Omega _{e_1 g}(t_{2} - t_{1})} e^{-i\Omega_{gg}(t_{3} - t_{2})} e^{-i\Omega _{e_1 g}(t - t_{3})}. \quad
\end{eqnarray}
We are now almost in a position to carry out the integration. However, one must carefully inspect the arguments of the $\delta $-functions because blind analysis can lead to errors that are difficult to catch. Care must be taken to understand when the product of $\delta$-functions leads the integrand to be zero. This subtlety is better exposed by making a change of variables from absolute interaction times to interaction-time intervals, $T_{n}=t_{n+1}-t_{n}$:
\begin{eqnarray}
t_{3} &=&t-T_{3}, \\
t_{2} &=&t-T_{3}-T_{2}, \\
t_{1} &=&t-T_{3}-T_{2}-T_{1}.
\end{eqnarray}
This, in addition to setting $t = s$ in anticipation of the heterodyned I$^{(4)}$ 2D ES experiment, results in
\begin{eqnarray}
I(\sigma, \tau, \zeta) &=& \frac{\rho _{0}\vert \mu _{1} \vert^{4} I_0^2}{-i \hbar ^{3}} \int_{0}^{\infty }dT_{3}\int_{0}^{\infty }dT_{2}\int_{0}^{\infty }dT_{1} \nonumber \\
& & \times \delta (\sigma-T_{1} )\delta (\zeta - T_{3})  e^{-i\omega (t-T_{3}-T_{2}-T_{1})} \nonumber \\
& & \times e^{+i\omega (t-T_{3}-T_{2}-\sigma)}e^{-i\omega (t-T_{3}+\tau -\sigma )}e^{+i\omega (t-\kappa )}  \nonumber \\
& & \times e^{-i\Omega _{e_{1} g}T_{1}}e^{-i\Omega_{g g}T_{2}}e^{-i\Omega _{e_{1} g}T_{3}}.
\end{eqnarray}
The negative sign results from the change of integration limits. Simplification of the above expression yields 
\begin{eqnarray}
I(\sigma, \tau, \zeta) & = & \frac{\rho _{0}\vert \mu _{1} \vert^{4} I_0^2}{-i \hbar ^{3}} \int_{0}^{\infty }dT_{3}\int_{0}^{\infty }dT_{2}\int_{0}^{\infty }dT_{1}  \nonumber \\
& & \times \delta (\sigma-T_{1} )\delta (\zeta - T_{3})  \nonumber \\
& & \times e^{-i\omega (\tau + \kappa - T_{3} - T_{1})} \nonumber \\
& & \times e^{-i\Omega_{e_{1} g}T_{1}}e^{-i\Omega _{g g} T_{2}}e^{-i\Omega _{e_{1} g}T_{3}}.
\end{eqnarray}
What makes this term simple is that the three integrals are independent, allowing us to reorganize the expression to
\begin{eqnarray}
I(\sigma, \tau, \zeta) & = & \frac{\rho _{0}\vert \mu _{1} \vert^{4} I_0^2}{-i \hbar ^{3}} e^{-i\omega (\tau + \kappa)}   \nonumber \\
& & \times \int_{0}^{\infty }dT_{3} \delta (\zeta -  T_{3}) e^{-i (\Omega_{e_{1} g} - \omega) T_{3}} \nonumber \\
& & \times \int_{0}^{\infty }dT_{2} e^{-i\Omega _{g g} T_{2}} \nonumber \\
& & \times \int_{0}^{\infty }dT_{1} \delta (\sigma-T_{1}) e^{-i(\Omega_{e_{1} g} - \omega) T_{1}}.
\end{eqnarray}

We can work in any order to solve the integrals, and we choose to first solve the $T_{2}$ integral.  The indefinite integral is
\begin{equation}
\int dT_{2} e^{-i\Omega _{g g} T_{2}}  = \frac{i}{\Omega_{gg}} e^{-i \Omega_{gg} T_{2}}.
\end{equation}
Since $\Omega_{gg} = -i \Gamma_{gg}$, where $\Gamma_{gg}$ is a positive value, the function decays and the definite integral yields
\begin{equation}
 \frac{i}{\Omega_{gg}} e^{-i \Omega_{gg} T_{2}} \Bigg\vert_{0}^{\infty} =  \frac{-i}{\Omega_{g g}}.
\end{equation}
We now handle the $T_{1}$ and $T_{3}$ integrals using the generic expression,
\begin{equation}
\int_{0}^{\infty }dT_{x} \delta (z-T_{x}) e^{-i(\Omega - \omega) T_{x}}.
\end{equation}
Due to the $\delta$-function, the integrand---and therefore the integral as well---is nonzero only when $T_{x} = z$. Thus
\begin{equation}
\int_{-\infty}^{\infty }dT_{x} \delta (z-T_{x}) e^{-i(\Omega - \omega) T_{x}} = e^{-i(\Omega - \omega) z},
\end{equation}
which is also true when the lower limit of integration is $0$ instead of $-\infty$ as long as $z$ was a value inside the limits of integration: $0 \le z < \infty$.  If instead $z < 0$, the integration would yield $0$, not the given expression. In other words,
\begin{equation}
\int_{0}^{\infty }dT_{x} \delta (z-T_{x}) e^{-i(\Omega - \omega) T_{x}} = e^{-i(\Omega - \omega) z} \Theta[z],
\end{equation}
where $\Theta$ is the Heaviside step function. Using this result, the total expression for this term becomes
\begin{eqnarray}
I(\sigma, \tau, \zeta) & = & \frac{\rho _{0}\vert \mu _{1} \vert^{4} I_0^2}{ \hbar ^{3} \Omega_{gg}} e^{-i\omega (\sigma + \zeta)}   \nonumber \\
& & \times e^{-i(\Omega_{e_{1} g} - \omega) \zeta} \Theta[\zeta] \nonumber \\
& & \times e^{-i(\Omega_{e_{1} g} - \omega) \sigma} \Theta[\sigma],
\end{eqnarray}
where we used the identity $\tau + \kappa = \sigma + \zeta$. We can simplify this expression to its final form
\begin{equation}
I(\sigma, \tau, \zeta)  = \Theta[\sigma]  \Theta[\zeta] \frac{\rho _{0}\vert \mu _{1} \vert^{4} I_0^2}{\hbar ^{3} \Omega_{gg}}  e^{-i\Omega_{e_{1} g}  \zeta} e^{-i\Omega_{e_{1} g}  \sigma}.
\end{equation}

\subsection{Term $G_{II 12 \beta i}$}
Performing all initial steps, the change of variables, and setting $t = s$, the intensity for this term is 
\begin{eqnarray}
I(\sigma, \tau, \zeta) & = & e^{-i \omega (\sigma + \zeta)} \frac{\rho_{0} \vert \mu_1 \vert^2 \vert \mu_2 \vert^2 I_0^2}{-i \hbar^3} \nonumber \\
& & \times \int_{0}^{\infty}dT_{3}\int_{0}^{\infty}dT_{2}\int_{0}^{\infty}dT_{1} \nonumber \\
& & \times \delta \left(-T_{3} - T_{2} + \zeta - T_{1} \right) \delta \left( T_{2} + \sigma \right)   \nonumber \\
& & \times e^{+i\omega (T_{1} + T_{3})}   \nonumber \\
& & \times e^{-i\Omega_{e_{1} g} T_{1}}e^{-i\Omega_{g g} T_{2}}e^{-i\Omega_{e_{2} g} T_{3}}.
\end{eqnarray}
This term is slightly more complicated than the first term because one of the $\delta$-functions contains multiple integration variables. This means we need to consider the integrals in consecutive order. Using the generic expression above, the $T_{1}$ integral yields 
\begin{eqnarray}
& & \int_{0}^{\infty}dT_{1}  \delta \left(-T_{3} - T_{2} - T_{1} + \zeta \right)  e^{-i (\Omega_{e_1 g} - \omega) T_{1}} \nonumber \\
& = & e^{-i (\Omega_{e_1 g} - \omega) (-T_{3} - T_{2} + \zeta)} \Theta[-T_{3} - T_{2} + \zeta].
\end{eqnarray}
This leads to a $T_{2}$ integral which is slightly different than the previous terms. However, because of the second $\delta$-function, this term is also straightforward 
\begin{eqnarray}
& & \int_{0}^{\infty}dT_{2}  \delta \left(\sigma + T_{2} \right)  e^{-i (\Omega_{g g} + \omega - \Omega_{e_{1} g}) T_{2}}  \Theta[-T_{3} - T_{2} + \zeta] \nonumber \\
& = & e^{+i (\Omega_{g g} + \omega - \Omega_{e_{1} g}) \sigma}  \Theta[-T_{3} + \sigma + \zeta] \Theta[- \sigma]. 
\end{eqnarray}
After these two integration steps, the total intensity is
\begin{eqnarray}
I(\sigma, \tau, \zeta) & = & \Theta[- \sigma] e^{-i \Omega_{e_1 g} \zeta} e^{-i (\Omega_{e_1 g} - \Omega_{gg}) \sigma} \frac{\rho_{0} \vert \mu_1 \vert^2 \vert \mu_2 \vert^2 I_0^2}{-i \hbar^3}  \nonumber \\
& & \times \int_{0}^{\infty}dT_{3} \Theta[\sigma + \zeta - T_{3}]  e^{-i (\Omega_{e_{2} g} - \Omega_{e_1 g}) T_{3}}. \quad \nonumber \\
& &
\end{eqnarray}

The $T_{3}$ integral is unlike any we have encountered thus far due to the presence of the Heaviside function involving the integration variable. This Heaviside function has two consequences. First it forces the maximum integration limit of the $T_{3}$ integral to be $\zeta + \sigma$, because if $T_{3} > \sigma + \zeta$, then the argument of the Heaviside function would be negative and thus the result would be zero. Second, that change to the limit of integration requires that $\zeta + \sigma > 0$, which we can ensure by applying a Heaviside of $\sigma + \zeta$ to the resulting solution. Thus, 
\begin{eqnarray}
& &  \int_{0}^{\infty}dT_{3} \Theta[\sigma + \zeta - T_{3}]  e^{-i (\Omega_{e_{2} g} - \Omega_{e_1 g}) T_{3}} \nonumber \\
& = &  \Theta[\sigma + \zeta] \int_{0}^{\sigma + \zeta} dT_{3}  e^{-i (\Omega_{e_{2} g} - \Omega_{e_1 g}) T_{3}},
\end{eqnarray}
which can be solved to yield
\begin{eqnarray}
& = & \Theta[\sigma + \zeta]  \frac{-i}{\Omega_{e_2 g} - \Omega_{e_1 g}} \left(1 - e^{-i (\Omega_{e_{2} g} - \Omega_{e_1 g}) (\zeta + \sigma)} \right). \qquad
\end{eqnarray}
The total intensity for the term is
\begin{eqnarray}
I & = & \Theta[- \sigma] e^{-i \Omega_{e_1 g} \zeta} e^{-i (\Omega_{e_1 g} - \Omega_{gg}) \sigma} \frac{\rho_{0} \vert \mu_1 \vert^2 \vert \mu_2 \vert^2 I_0^2}{-i \hbar^3}   \\
& & \times  \Theta[\sigma + \zeta]  \frac{-i}{\Omega_{e_2 g} - \Omega_{e_1 g}} \left(1 - e^{-i (\Omega_{e_{2} g} - \Omega_{e_1 g}) (\zeta + \sigma)} \right), \nonumber
\end{eqnarray}
which simplifies to
\begin{eqnarray}
I(\sigma, \tau, \zeta) & = & \Theta[- \sigma] \Theta[\sigma + \zeta]  \frac{\rho_{0} \vert \mu_1 \vert^2 \vert \mu_2 \vert^2 I_0^2}{(\Omega_{e_2 g} - \Omega_{e_1 g}) \hbar^3}   e^{+i \Omega_{g g} \sigma}  \nonumber \\
& & \times \left( e^{-i \Omega_{e_1 g} (\sigma + \zeta)}  - e^{-i \Omega_{e_{2} g}  (\sigma + \zeta)} \right). 
\end{eqnarray}

\subsection{Term $C_{I 12 \alpha ii}$}
Performing all initial steps, the change of variables, and setting $t = s$, the intensity for this term is 
\begin{eqnarray}
I & = & e^{-i \omega (\tau + \kappa)} \frac{\rho _{0}\vert \mu _{1} \vert^{2} \vert \mu _{2} \vert^{2} I_0^2}{-i\hbar ^{3}} \nonumber \\
& & \times \int_{0}^{\infty }dT_{3}\int_{0}^{\infty }dT_{2}\int_{0}^{\infty }dT_{1}  \nonumber \\
& & \times \delta (\kappa -T_{3}-T_{2})\delta (\tau +T_{2}+T_{1})  \nonumber \\
& & \times e^{+i\omega (T_{3}-T_{1})} \nonumber \\
& & \times e^{-i\Omega_{ge_{1}}T_{1}}e^{-i\Omega _{e_{2}e_{1}}T_{2}}e^{-i\Omega _{e_{2}g}T_{3}}.
\end{eqnarray}
Like the term above, even though the integration steps must be performed consecutively, the $\delta$-functions make the first two integration steps straightforward. Integration over $T_{1}$ and $T_{2}$ in the usual manner yields
\begin{eqnarray}
I & = & \frac{\rho _{0}\vert \mu _{1} \vert^{2} \vert \mu _{2} \vert^{2} I_0^2}{-i\hbar ^{3}} e^{+i\Omega_{g e_{1}} \tau} e^{-i (\Omega_{e_{2} e_{1}} - \Omega_{g e_1} )\kappa} \nonumber \\
& & \times \int_{0}^{\infty }dT_{3}  e^{-i (\Omega_{e_{2} g} - \Omega_{e_{2} e_{1}} + \Omega_{g e_{1}} ) T_{3}} \nonumber \\
& & \times \Theta[\kappa - T_{3}] \Theta[-\tau - \kappa + T_{3}] .
\end{eqnarray}
At this stage we must again carefully consider how the Heaviside functions impact the range of integration and the total signal. We first look back at the original expression and observe that the final signal will acquire $\Theta[\kappa]$ due to the first $\delta$-function, which requires $\kappa  - T_3 - T_2 = 0$, where $T_3$ and $T_2$ are positive values.  Similar reasoning leads to the signal acquiring a factor of $\Theta[-\tau]$. 

We can make more progress by following the line of reasoning above for the first Heaviside function alone. It will act to change the upper limit of integration to $\kappa$ and to introduce the same factor of $\Theta[\kappa]$ we found above. Based on this analysis, the signal becomes
\begin{eqnarray}
I & = & \Theta[\kappa] \Theta[- \tau] \frac{\rho _{0}\vert \mu _{1} \vert^{2} \vert \mu _{2} \vert^{2} I_0^2}{-i\hbar ^{3}} e^{+i\Omega_{g e_{1}} \tau} e^{-i (\Omega_{e_{2} e_{1}} - \Omega_{g e_1} )\kappa} \nonumber \\
& & \times \int_{0}^{\kappa }dT_{3}  e^{-i (\Omega_{e_{2} g} - \Omega_{e_{2} e_{1}} + \Omega_{g e_{1}} ) T_{3}} \Theta[T_{3} -\tau - \kappa].
\end{eqnarray}

Finally, we treat the second Heaviside function. We recognize that $T_3 > \kappa + \tau$, but the lower limit of integration depends on whether or not $\kappa + \tau$ is greater than or less than zero. Thus we must consider two cases: $\kappa + \tau > 0$ and $\kappa + \tau < 0$. 

For the $\kappa + \tau > 0$ case, the lower limit of integration is what we initially expected, $\kappa + \tau$, and the signal gains $\Theta[\kappa + \tau]$. For the $\kappa + \tau < 0$ case, the lower limit is zero (because the original value of the limit was $T_3 = 0$) and the signal gains $\Theta[- (\kappa + \tau)]$. This yields a single expression that uses the Heaviside functions to treat the cases
\begin{eqnarray}
I & = & \Theta[\kappa] \Theta[-\tau]  \frac{\rho _{0}\vert \mu _{1} \vert^{2} \vert \mu _{2} \vert^{2} I_0^2}{-i\hbar ^{3}} e^{+i\Omega_{g e_{1}} \tau} e^{-i (\Omega_{e_{2} e_{1}} - \Omega_{g e_1} )\kappa} \nonumber \\
& & \times \Bigg( \Theta[\tau + \kappa] \int_{\tau + \kappa}^{\kappa }dT_{3}  e^{-i (\Omega_{e_{2} g} - \Omega_{e_{2} e_{1}} + \Omega_{g e_{1}} ) T_{3}} \nonumber \\
& & + \Theta[-\tau - \kappa] \int_{0}^{\kappa }dT_{3}  e^{-i (\Omega_{e_{2} g} - \Omega_{e_{2} e_{1}} + \Omega_{g e_{1}} ) T_{3}} \Bigg).
\end{eqnarray}
Integration over $T_{3}$ then yields
\begin{eqnarray}
I & = & \Theta[\kappa] \Theta[-\tau]  \frac{\rho _{0}\vert \mu _{1} \vert^{2} \vert \mu _{2} \vert^{2} I_0^2}{\hbar ^{3}(\Omega_{e_{2} g } - \Omega_{e_{2}e_{1}} + \Omega_{g e_{1}})} e^{+i\Omega_{g e_{1}} \tau}  \nonumber \\
& & \times e^{-i (\Omega_{e_{2} e_{1}} - \Omega_{g e_1} )\kappa} \Bigg\{ \Theta[\tau + \kappa] \nonumber \\
& & \times  \bigg[ e^{-i (\Omega_{e_{2} g } - \Omega_{e_{2}e_{1}} + \Omega_{g e_{1}})  (\kappa +\tau)}  - e^{-i (\Omega_{e_{2} g } - \Omega_{e_{2}e_{1}} + \Omega_{g e_{1}}) \kappa}\bigg] \nonumber \\
& & + \Theta[-\tau - \kappa] \bigg[1 - e^{-i (\Omega_{e_{2} g } - \Omega_{e_{2}e_{1}} + \Omega_{g e_{1}}) \kappa}\bigg] \Bigg\}.
\end{eqnarray} 
This simplifies a bit to
\begin{eqnarray}
I & = &  \frac{ \Theta[\kappa] \Theta[-\tau] \rho _{0}\vert \mu _{1} \vert^{2} \vert \mu _{2} \vert^{2} I_0^2}{\hbar ^{3}(\Omega_{e_{2} g} - \Omega _{e_{2}e_{1}} + \Omega_{g e_{1}})} \Bigg\{ \Theta[\tau + \kappa]  \nonumber \\
& & \times \bigg[e^{+i \Omega_{e_{2}e_{1}} \tau} e^{-i \Omega_{e_{2} g} (\tau + \kappa)} - e^{+i \Omega_{ge_{1}} \tau} e^{-i \Omega_{e_{2} g} \kappa}\bigg] \nonumber \\
& & + \Theta[-\tau - \kappa] \nonumber \\
& & \times \bigg[e^{-i \Omega_{e_2 e_1} \kappa} e^{+i \Omega_{ge_{1}} (\tau +\kappa)} -  e^{+i \Omega_{ge_{1}} \tau} e^{-i \Omega_{e_{2} g} \kappa} \bigg] \Bigg\}. \qquad
\end{eqnarray} 
In one case the $\Theta[\kappa]$ Heaviside is redundant, and in the other case the $\Theta[-\tau]$ Heaviside is redundant. For completeness we include both in each case. The Heaviside in the first term ensures ensures $\kappa > \vert \tau \vert $ and corresponds to column 2 in Table \ref{tab:tab3}, and the Heaviside in the second term $\kappa < \vert \tau \vert$ and corresponds to column 3 in Table \ref{tab:tab3}.

\subsection{Term $C_{I 12 \alpha i}$}
We turn now to term $C_{I12\alpha i}$, the partner FTC diagram to the previous. FTC diagram analysis predicts this term is zero due to color locking. The mathematical analysis is similar to the examples above but we choose to use a path that better exposes the effects of color locking. Like above, the intensity for this term is 
\begin{eqnarray}
I & = &\int_{-\infty }^{t}dt_{3}\int_{-\infty}^{t_{3}}dt_{2}\int_{-\infty }^{t_{2}}dt_{1} \langle E_{A}E_{B}^{\ast}E_{C}E_{LO}^{\ast }\rangle  \nonumber \\
& & \times \frac{\rho_{0} \vert \mu_1 \vert^2 \vert \mu_2 \vert^2}{-i \hbar^3} e^{-i\Omega _{ge_1}(t_{2} - t_{1})} \nonumber \\
& & \times e^{-i\Omega_{e_2 e_1}(t_{3} - t_{2})} e^{-i\Omega _{e_2 g}(t - t_{3})}.
\end{eqnarray}
Now though, when we reduce the four-point time correlator into a pair of two-point time correlators, we select the two-point correlator pair that corresponds to the $i$ subscript, which yields
\begin{eqnarray}
I & = & \frac{\rho_{0} \vert \mu_1 \vert^2 \vert \mu_2 \vert^2 I_0^2}{-i \hbar^3} \int_{-\infty}^{t}dt_{3}\int_{-\infty }^{t_{3}}dt_{2}\int_{-\infty }^{t_{2}}dt_{1} \nonumber \\
& & \times \langle p(t_{2})p^{\ast }(t_{1}-\sigma )\rangle \langle p(t_{3}+\tau -\sigma )p^{\ast }(s-\kappa )\rangle   \nonumber \\
& & \times e^{-i\omega t_{2}}e^{+i\omega (t_{1}-\sigma )}e^{-i\omega(t_{3}+\tau -\sigma )}e^{+i\omega (s-\kappa )}  \nonumber \\
& & \times e^{-i\Omega _{ge_{1}}(t_{2}-t_{1})}e^{-i\Omega_{e_{2}e_{1}}(t_{3}-t_{2})}e^{-i\Omega _{e_{2}g}(t-t_{3})},
\end{eqnarray}
At this point it is convenient to deviate from the path taken for $C_{I12\alpha ii}$ and consider $\langle p(t_{2})p^{\ast }(t_{1}-\sigma)\rangle$ which we can write as 
\begin{eqnarray}
\langle p(t_{2})p^{\ast }(t_{1}-\sigma )\rangle & = & \frac{1}{(2\pi )^{2}}\int_{-\infty }^{\infty }dx_{1}\int_{-\infty }^{\infty }dx_{2} \nonumber \\
& & \langle \tilde{p}(x_{2})\tilde{p}^{\ast }(x_{1})\rangle e^{-ix_{2}t_{2}}e^{ix_{1}(t_{1}-\sigma )}, \quad
\end{eqnarray}
where $\tilde{p}(x)$ is the Fourier Transform of $p(t)$. Now invoking the Weiner-Khintchine theorem, Eqn. \ref{eqn:colorlock}, we have 
\begin{eqnarray}
\langle p(t_{2})p^{\ast }(t_{1}-\sigma )\rangle & = & \frac{1}{(2\pi )^{2}} \int_{-\infty }^{\infty }dx_{1}\int_{-\infty }^{\infty }dx_{2}\nonumber \\
& & \Gamma (x_{1})\delta (x_{1}-x_{2})e^{-ix_{2}t_{2}}e^{ix_{1}(t_{1}-\sigma )}, \quad \quad
\end{eqnarray}
where $\Gamma (x)$ is the spectral density of the noisy light. The $\delta(x_{1}-x_{2})$ is the mathematical manifestation of color locking as it represents the $x-x$ correlated pair. Inspection of the Liouville pathway---the WMEL or double-sided Feynman diagrams---for this `C' term shows that $x_{1}\neq x_{2}$. Consequently, 
\begin{equation}
\langle p(t_{2})p^{\ast }(t_{1}-\sigma )\rangle = 0.
\end{equation}
Thus $I=0$ and this term vanishes.

\section{Acknowledgements}
D.B.T. was supported by the DARPA QuBE program and the National Science and Engineering Research Council of Canada. D.J.H., E.J.S., R.A.H., M.W.G., and D.J.U. were supported by Minnesota Space Grant A001584008, NSF grant DUE-0969568, and the Concordia College Chemistry Research Fund.

%


\end{document}